DEPARTMENT OF INFORMATICS

TECHNISCHE UNIVERSITÄT MÜNCHEN

Master's Thesis in Biomedical Computing

# Convolution Neural Networks for Semantic Segmentation: Application to Small Datasets of Biomedical Images

Vitaly Nikolaev

DEPARTMENT OF INFORMATICS

TECHNISCHE UNIVERSITÄT MÜNCHEN

Master's Thesis in Biomedical Computing

# Convolution Neural Networks for Semantic Segmentation: Application to Small Datasets of Biomedical Images

# Convolution Neuronale Netze für die semantische Segmentierung: Anwendung auf kleine Datensätze biomedizinischer Bilder

Author: Vitaly Nikolaev

Supervisor: Prof. Dr. Bjoern Menze

Advisor: Dr. Marie Piraud

Submission Date: 15.07.2018

I confirm that this master's thesis in Biomedical Computing is my own work and I have documented all sources and material used.

Munich, 15 July 2018 Vitaly Nikolaev

## Abstract

This thesis studies how the segmentation results, produced by convolutional neural networks (CNN), is different from each other when applied to small biomedical datasets. We use different architectures, parameters and hyper-parameters, trying to find out the better configurations for our task, and trying to find out underlying regularities. Two working datasets are from biomedical area of research. We conducted a lot of experiments with the two types of networks and the received results have shown the preference of some conditions of experiments and parameters of the networks over the others. All testing results are given in the tables and some selected resulting graphs and segmentation predictions are shown for better illustration.

# Contents

# 1 Introduction

From about 2012 Convolutional Neural Networks (CNNs) have become the gold standard for image classification [1]. While image classification task uses separate images as variables in studied datasets, CNNs can also be used for classification separate pixels in each image of the dataset. Such task of segmentation of various semantically grouped areas of pixels in the images is also of great interest in biomedical research.

On the contrary to conventional image segmentation tasks in computer vision, segmentation of biomedical images is more challenging in two aspects:

- Usually we have small number of available images;
- To make ground truth segmentation we need a help of professional from medicine (doctor or radiologist).

Results of image segmentation can be very different depending on

- The different types of network architectures;
- The chosen parameters and hyper-parameters of the network.

In this work we build different CNNs for biomedical semantic image segmentation and trying to find out which conditions of experiments give better results for segmentation of the given datasets.

# 2 Datasets

## 2.1 Light microscopy dataset

This dataset originally consists of 16 grayscale images with resolution of 1024 x 1024 (saved in RGBa). For each image we have black and white image images of:

- cell masks;
- nuclei masks.

Therefore, there are three classes of pixels: backgroung, cells, nuclei. We combined the both types of masks into single mask images and encoded each type of the pixel as:

0 – background;

1 – cells;

2 – nuclei.

Three examples from the dataset are shown in the figure 2.1.

From initial analysis it is obvious that the main problem for this type of segmentation will be the imbalance of classes. As a result, the usual accuracy as evaluation of the total number of correctly predicted pixels over the total number of pixels will not reveal a lot since a great imbalance between the classes of the pixels.

Two of the original images in bigger size (resolution) are shown in the figures 2.2 and 2.3.

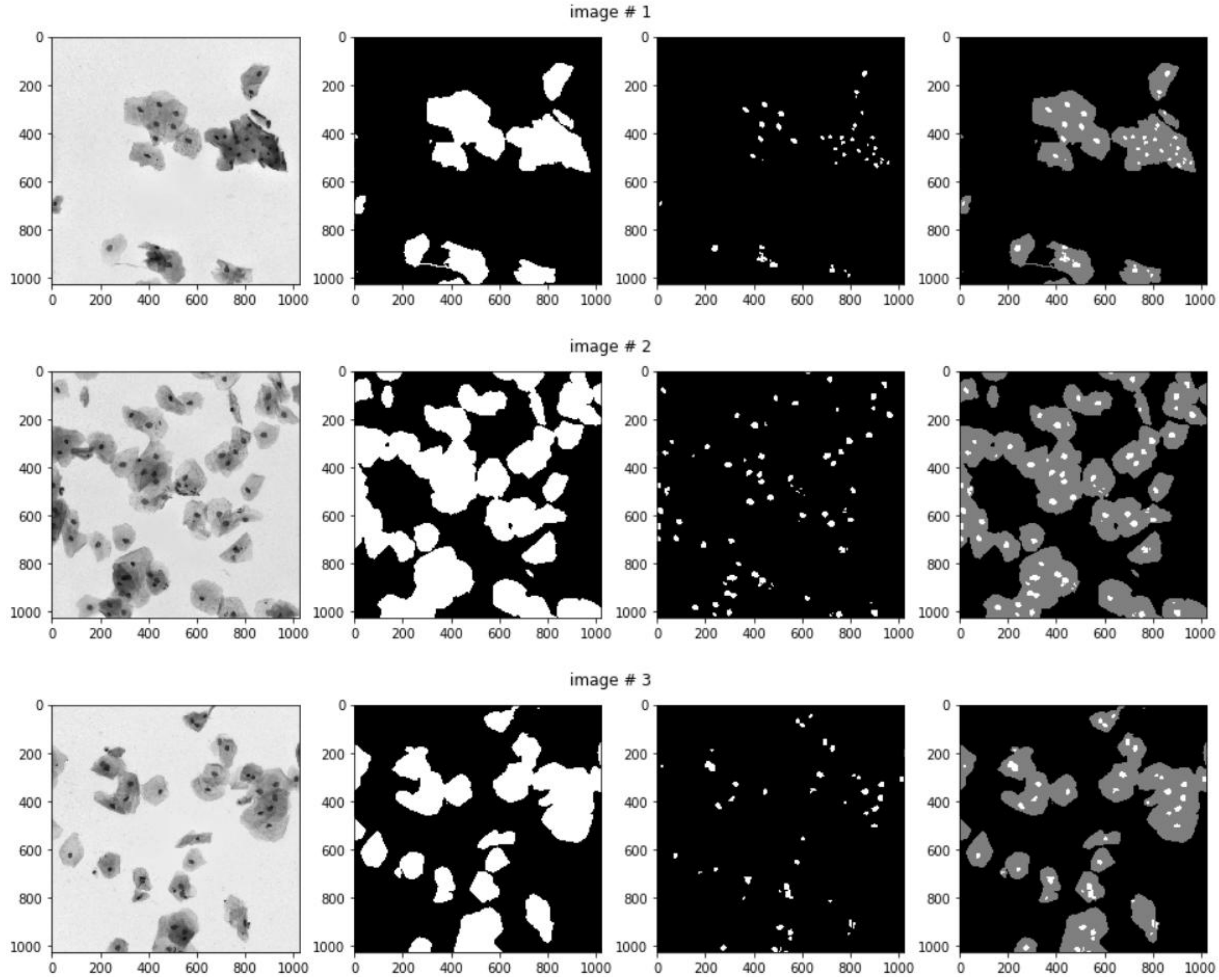

Figure 2.1: The examples of the original images (left), two given masks (middle) and the combined mask (right).

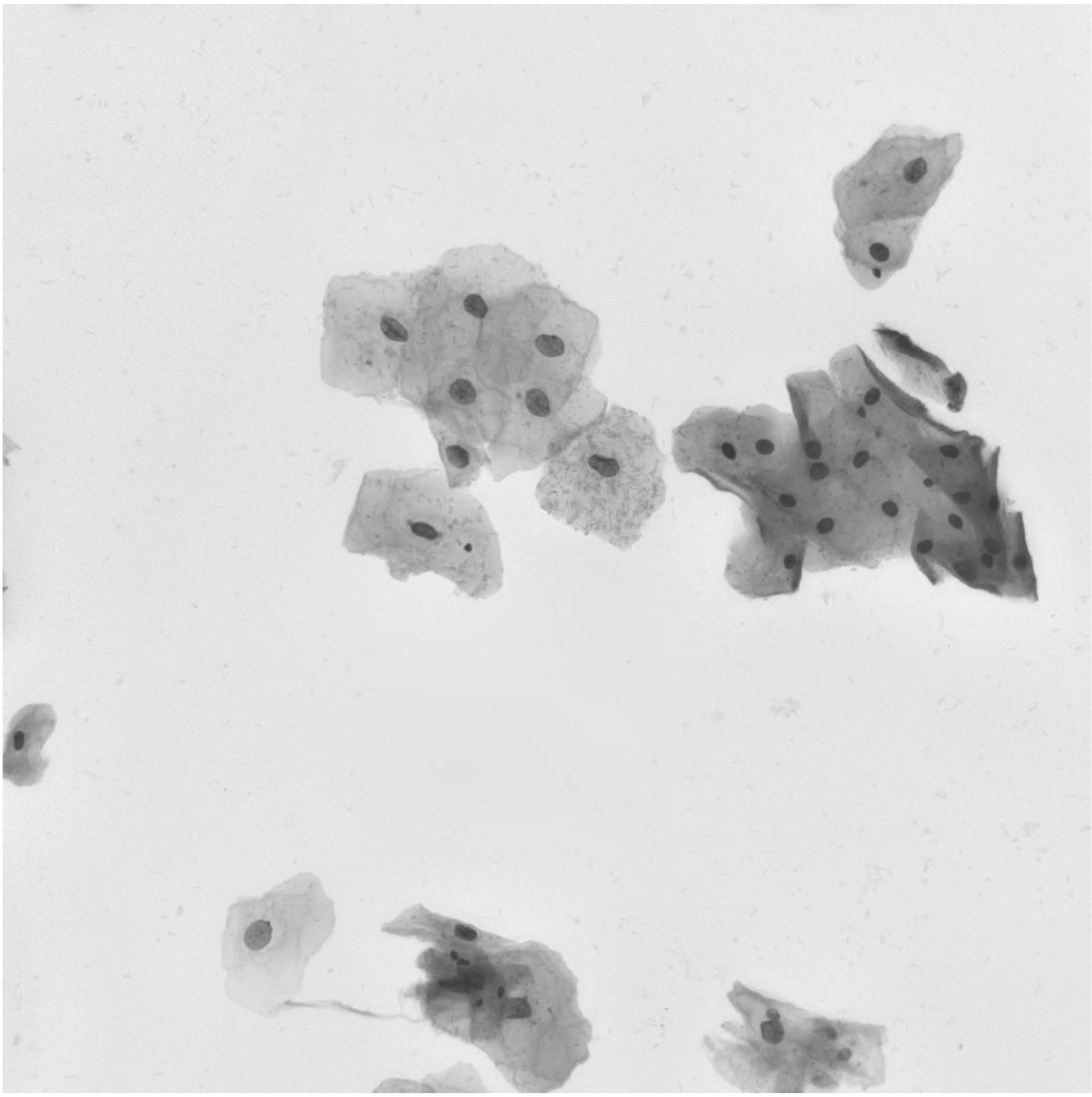

Figure 2.2: The example of the original image in a greater size (resolution).

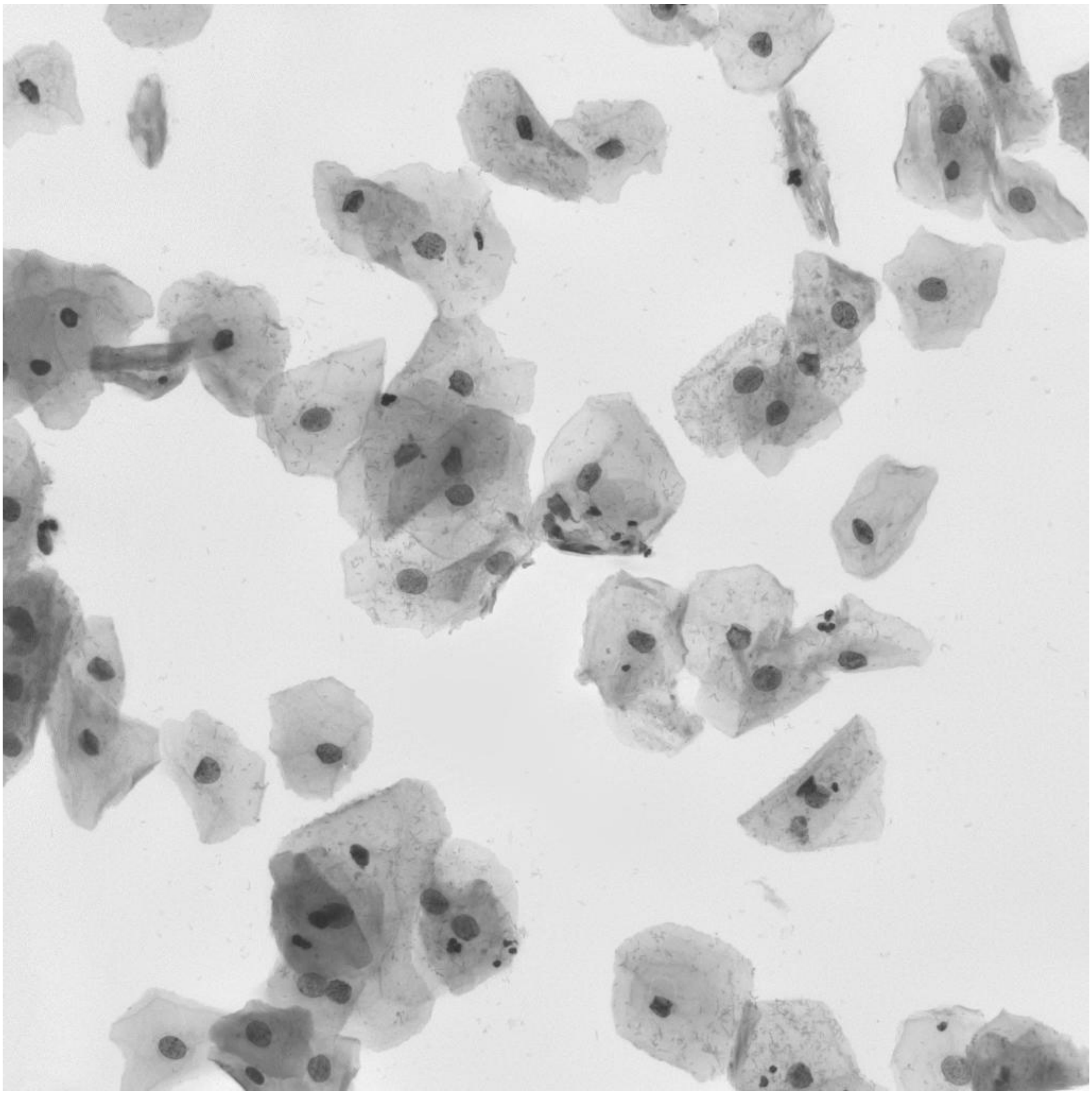

Figure 2.3: The example of the original image in a greater size (resolution).

## 2.2 BRATS dataset

The BRATS dataset consists of the data from the International Multimodal Brain Tumor Segmentation (BraTS) Challenge 2017 [6, 7, 8, 9]. There are the set of images, including ground truth segmentation labels. For each of 285 patients, which data is contained in the dataset, there are given 5 files, for example:

- Brats17_2013_24_1_flair.nii.gz;
- Brats17_2013_24_1_seg.nii.gz;
- Brats17_2013_24_1_t1.nii.gz;
- Brats17_2013_24_1_t1ce.nii.gz;
- Brats17_2013_24_1_t2.nii.gz.

In total we have 285 full 3d tomographic annotated scans (each for one patient) with four modalities of the same data and the ground truth segmentations for each.
The modalities are:

- T1;
- T1 CE: T1 contrast-enhanced;
- T2;
- T2 FLAIR: T2-weighted-fluid-attenuated inversion recovery.

For the each set of four modalities we have the segmentation in four classes (labels):

1 – Necrotic core (NCR);

2 – Peritumoral edema (ED);

4 – Enchancing tumor (ET);

0 – Else;

The sizes of the three projections are:

- Axial: 240 x 240, with total of 155 images per 3d scan;
- Saggital (lateral): 240 x 155, with total of 240 images per 3d scan;
- Coronal (frontal): 240 x 155, with total of 240 images per 3d scan.

The whole dataset is divided in two groups:

- HGG: 210 images of high-grade gliomas;
- LGG: 75 images of low-grade gliomas.

Each of the two groups have subdivisions into the following codifications:

- 2013;
- CBICA;
- TCIA,

which means that the scans from the different subdivisions were taken in different conditions and time.

The examples of the 2d scans taken from the dataset are given in the figures 2.4 and 2.5.

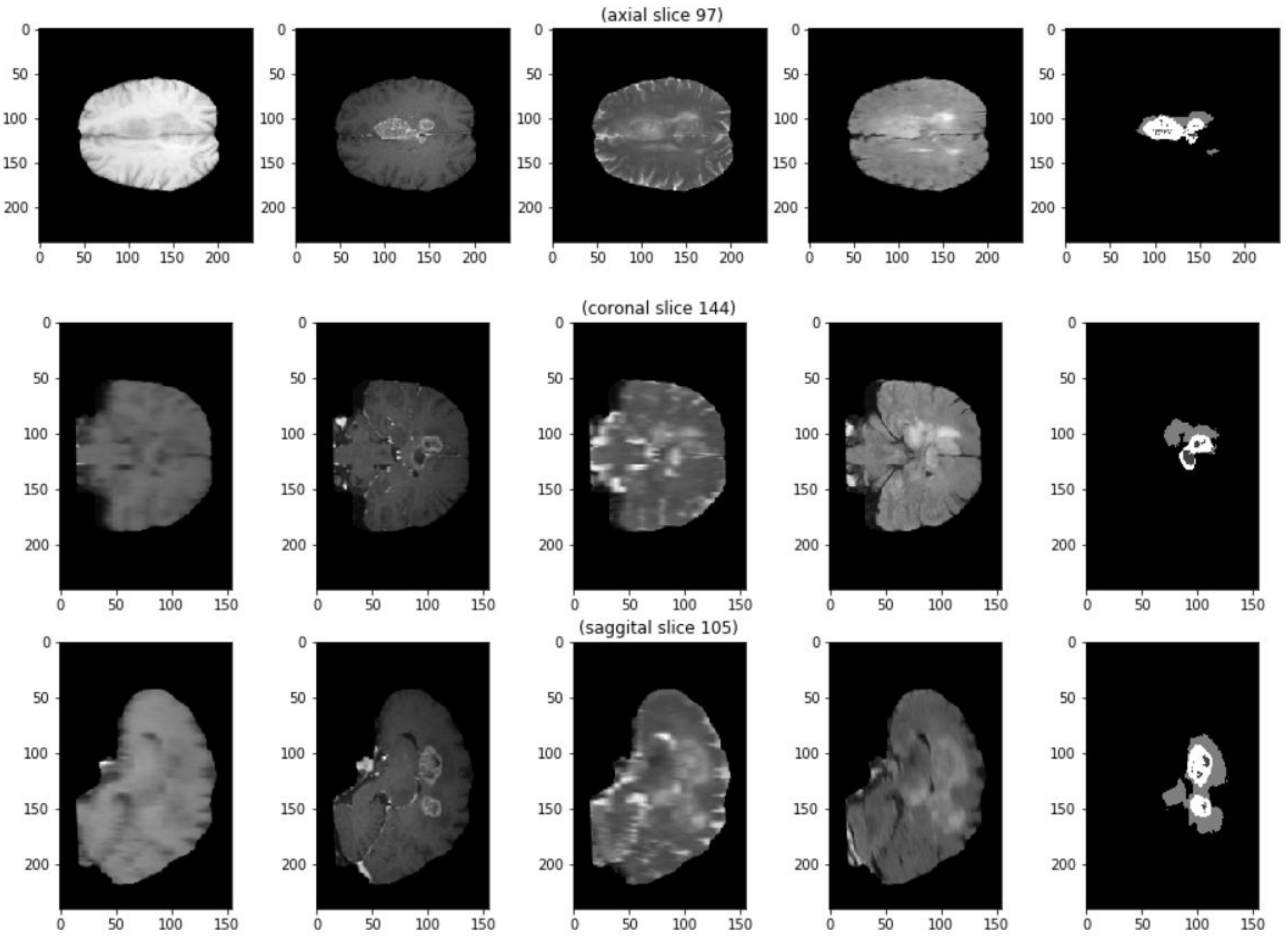

Figure 2.4: The examples of the axial, coronal and saggital slices in four modalities and segmentation of the patient encoded as 2013_2_1.

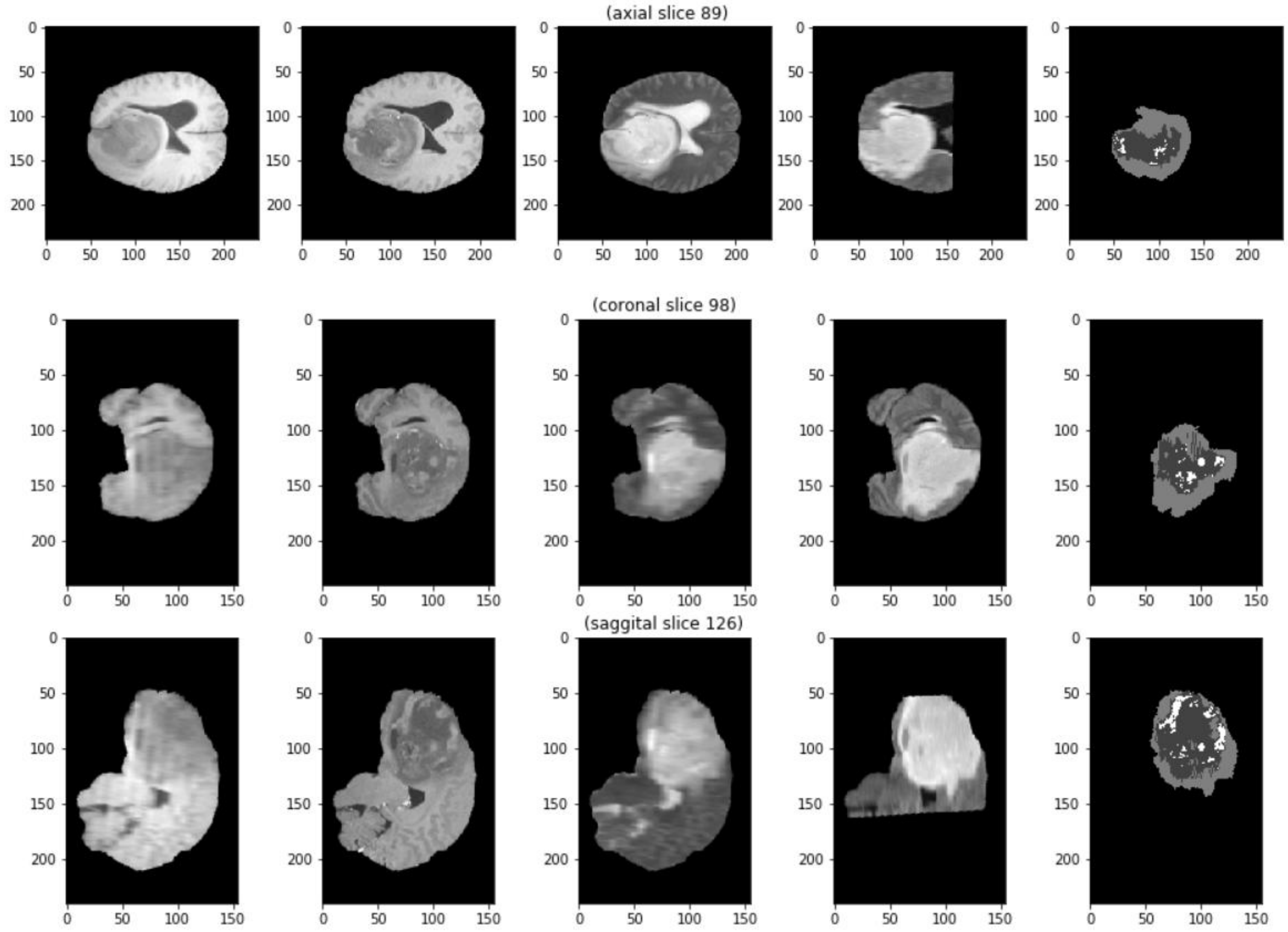

Figure 2.5: The examples of the axial, coronal and saggital slices in four modalities and segmentation of the patient encoded as 2013_6_1.

# 3 Architectures of the networks

To solve the task the two types of existing CNN’s architectures were used:

- conventional feed-forward fully convolutional network without downsampling (FCN) [2];
- more sofisticated type of CNNs, called U-Net [3].

The U-Net authors wrote that “using the same network trained on transmitted light microscopy images (phase contrastand DIC) we won the ISBI cell tracking challenge 2015” [3]. Also, this type of architecture is widely considered as very good for semantic segmentation of small medical datasets [11].

Since our datasets are also from the area of biomedical image processing, it would be interesting to learn how U-Net works, and compare its results with more traditional feedforward fully convolutional network.

In order to compare two types of architectures we will use different parameters and hyper-parameters for them, and then compare their results.

Parameters and hyper-parameters that will be used are:

- the structure of layers;
- parameters of convolution, deconvolution, and pooling kernels:
    - filter size;
    - number of filters;
    - stride;
    - padding;
    - etc.
- batch size;
- learning rate;

- activation functions;
- optimizers (optimization techniques);
- metrics;
- etc.

The typical architecture of fully convolutional network is shown on figure 3.1, and the architecture of U-Net is shown in figure 3.2.

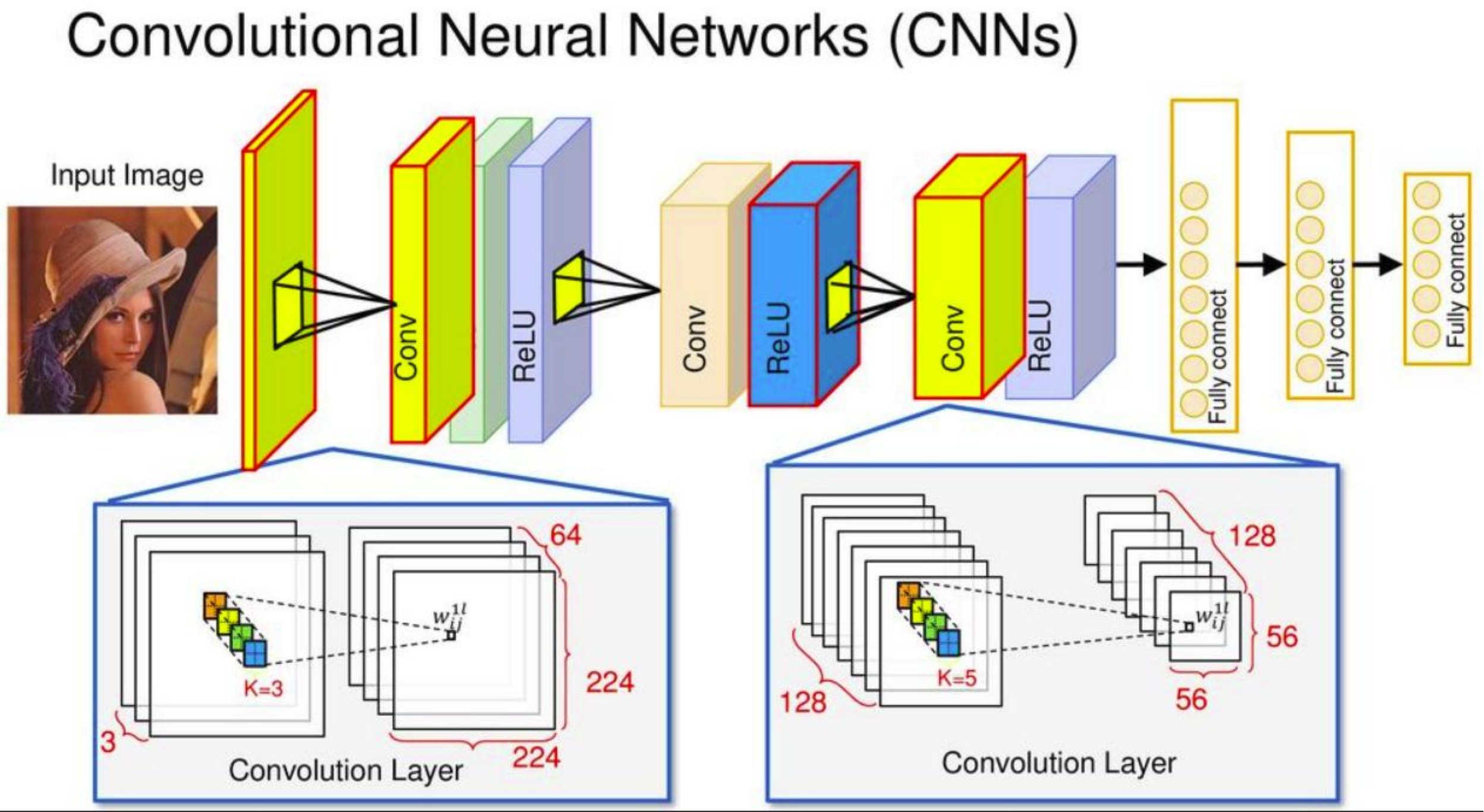

Figure 3.1: The architecture of a fully convolutional network.

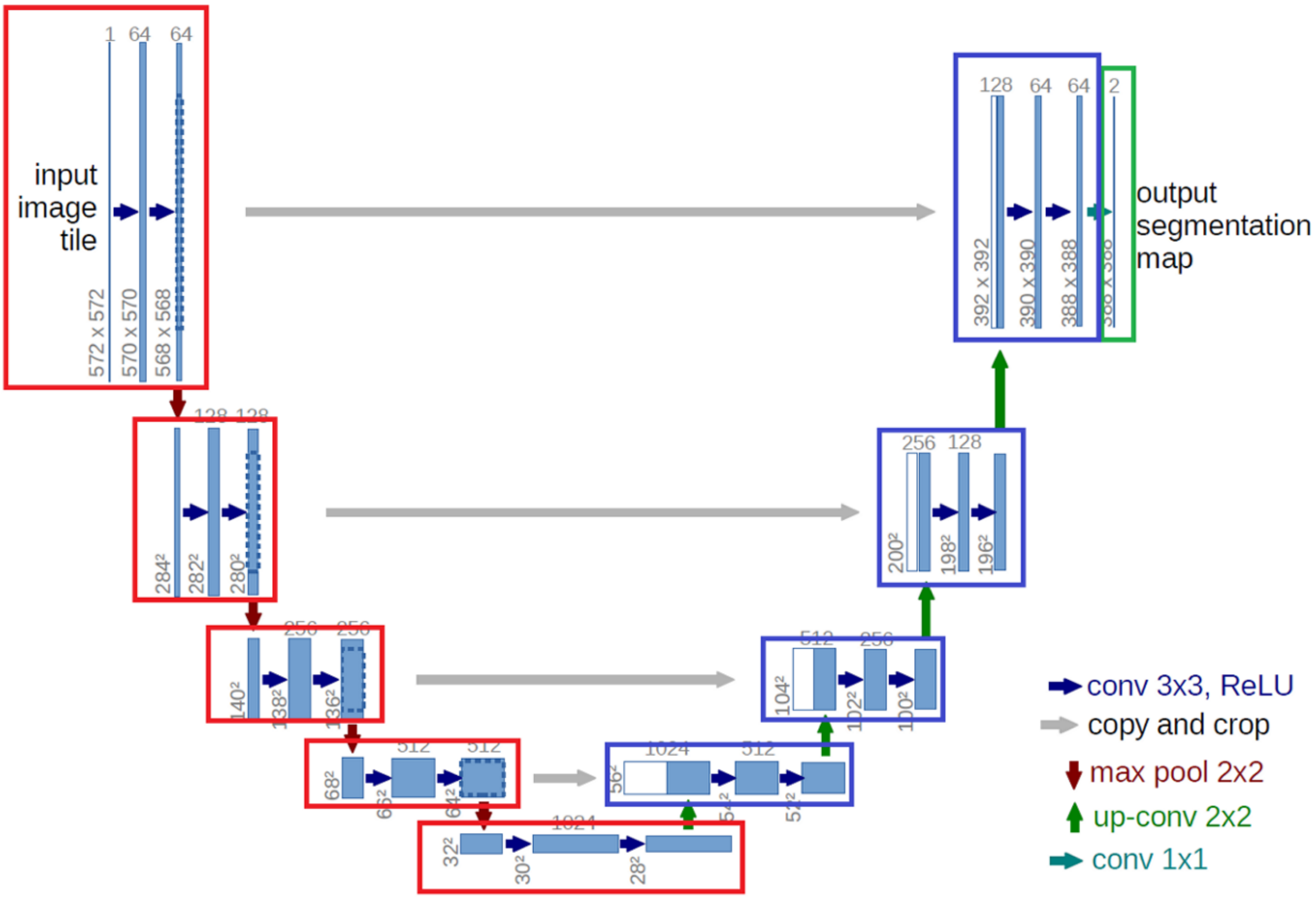

Figure 3.2: The architecture of a U-Net network.

# 4 Training of the networks

Before starting to train networks we need to divide our datasets into training, validation, and test sets, and then we need to preprocess and/or augment all the images according our plan. After, we need to choose the parameters and hyper-parameters of the networks we want to control and compare the results of. At the end, we have to train our networks with the chosen parameters and hyper-parameters, on the generated datasets.

## 4.1 Data preprocessing

Before we divide our datasets into different sets of training, validation, and test images, and choose how to preprocess and augment them, we describe the procedures that will be used for preprocessing and augmenting the images:

- Normalization of image values to the interval [0,1];
- Scaling to different sizes;
- Histogram equalization;
- Transformation with a certain probability:
    - Flipping: perform flips around horizontal and vertical axis with a certain probability;
    - Horizontal and vertical warping, based on sine function (points are mapped to points without changing the colors);
    - Rotation by a random angle;
    - Zooming by a random factor;
    - Cropping, if generated image is out of initial size.

Histogram equalization is a method in image processing of contrast adjustment using the image's histogram [10] (see figure 4.1).

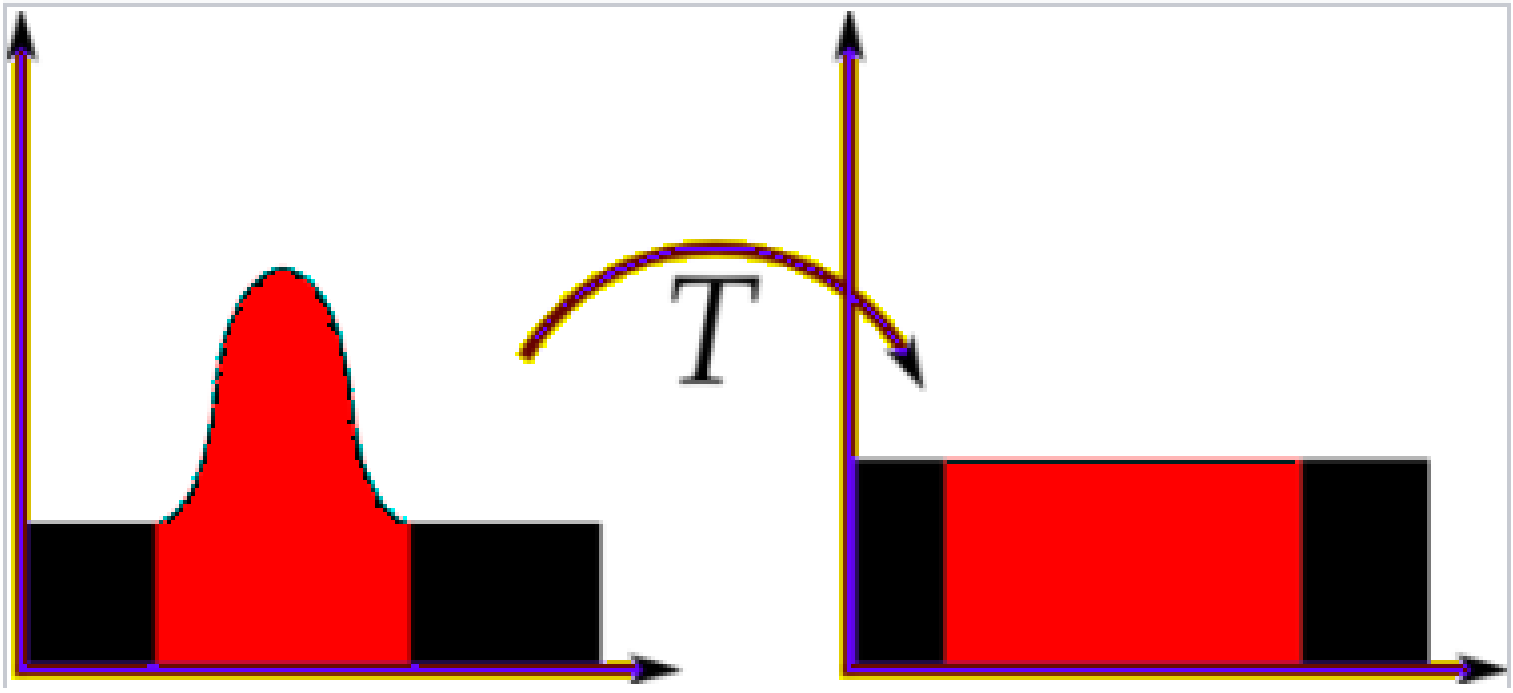

Figure 4.1: Histograms of an image before and after equalization.

For warping we used the formula:

int(warping_amplitude * (sin(warping_frequency * Pi * i / 180) + 1) / 2),

where i is a row or column number depending on what type of warping is produced, horizontal or vertical.

Warping amplitude and frequency are chosen randomly from the intervals (10, 50) and (0.5, 2).

The probability of warping is 0.5.

The probability of flipping is also 0.5.

The maximum rotation angle is 60 degrees for light microscopy images, and 20 degrees for BRATS images. The interval for zooming factor is (0.8, 1.2) for light microscopy images, and (0.9, 1.1) for BRATS images.

The whole (random) transform procedure, applied to an image, is:

transform: (scale) -> flip -> warp -> rotate -> zoom -> (crop),

where procedures in parentheses are applied only if necessary.

Also, for BRATS dataset the label “4” (“Enchancing tumor”) was changed in the all dataset to “3” for the sake of convenience during training and evaluation.

### 4.1.1 Light microscopy dataset

First, we divide our dataset randomly into two parts: 12 training and 4 testing images. Then, to have more data for experiments, we generate six training datasets:

- four main datasets:
    - 12 images of 1024 x 1024 (12 original images without rescaling);
    - 96 images 1024 x 1024, augmented (transformed) from 12 original;
    - 384 images of 512 x 512;
    - 1536 images of 256 x 256;
- two additional sets (without histogram equalization):
    - 12 images of 1024 x 1024;
    - 1536 images of 256 x 256.

We decided to generate two datasets without histogram equalization applied during preprocessing since some images after histogram equalization looked like some useful information has been lost. The example of the image from the figure 2.2 after histogram equalization is shown in the figure 4.2. The example of the same image after transformation procedure is shown in the figure 4.3.

The two samples of preprocessed first two test images are shown in the figure 4.4.
The third and fourth test images are in the figure 4.5.

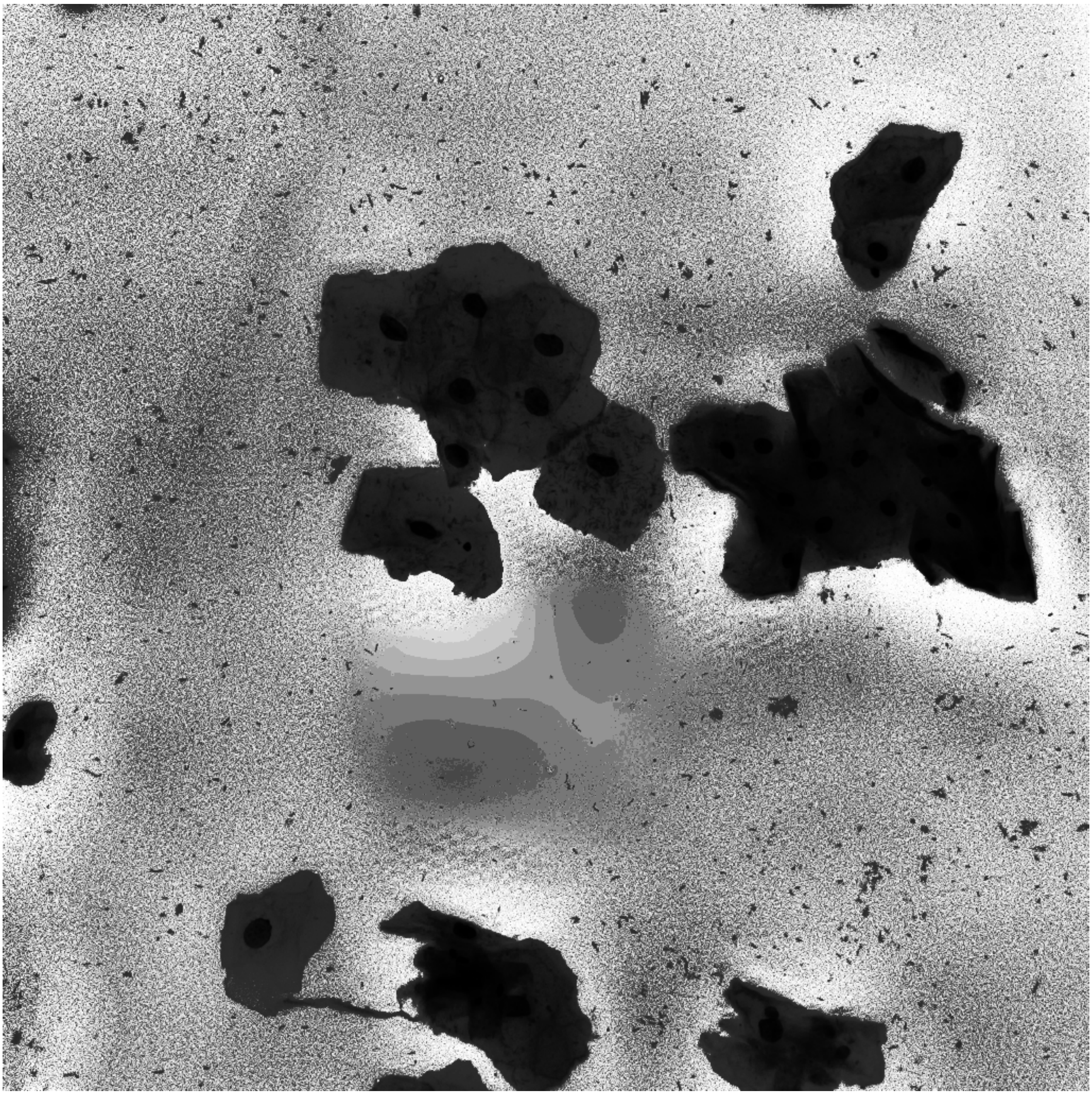

Figure 4.2: The image from figure 2.2 after histogram equalization.

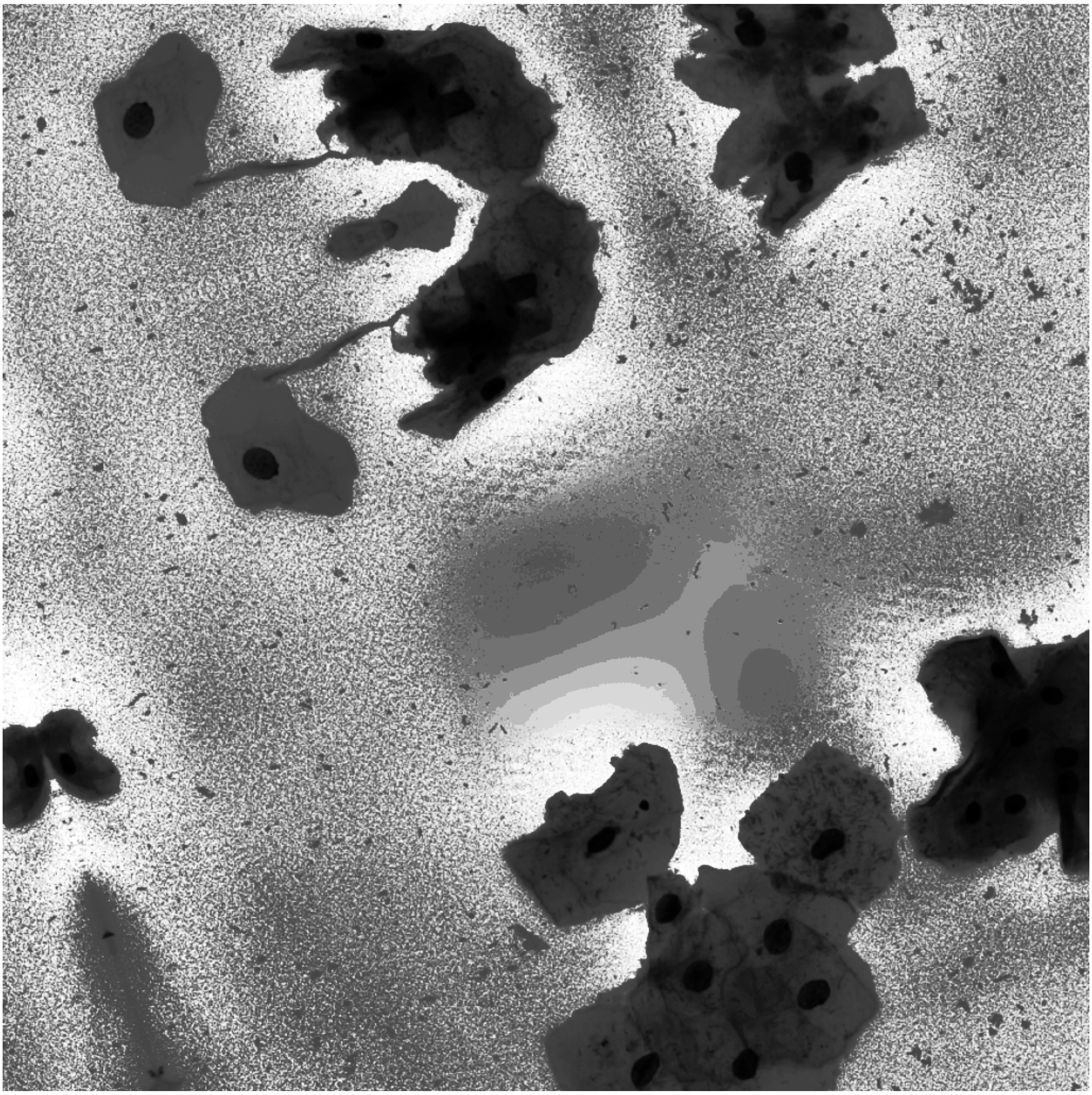

Figure 4.3: The image from figure 2.2 after transformation.

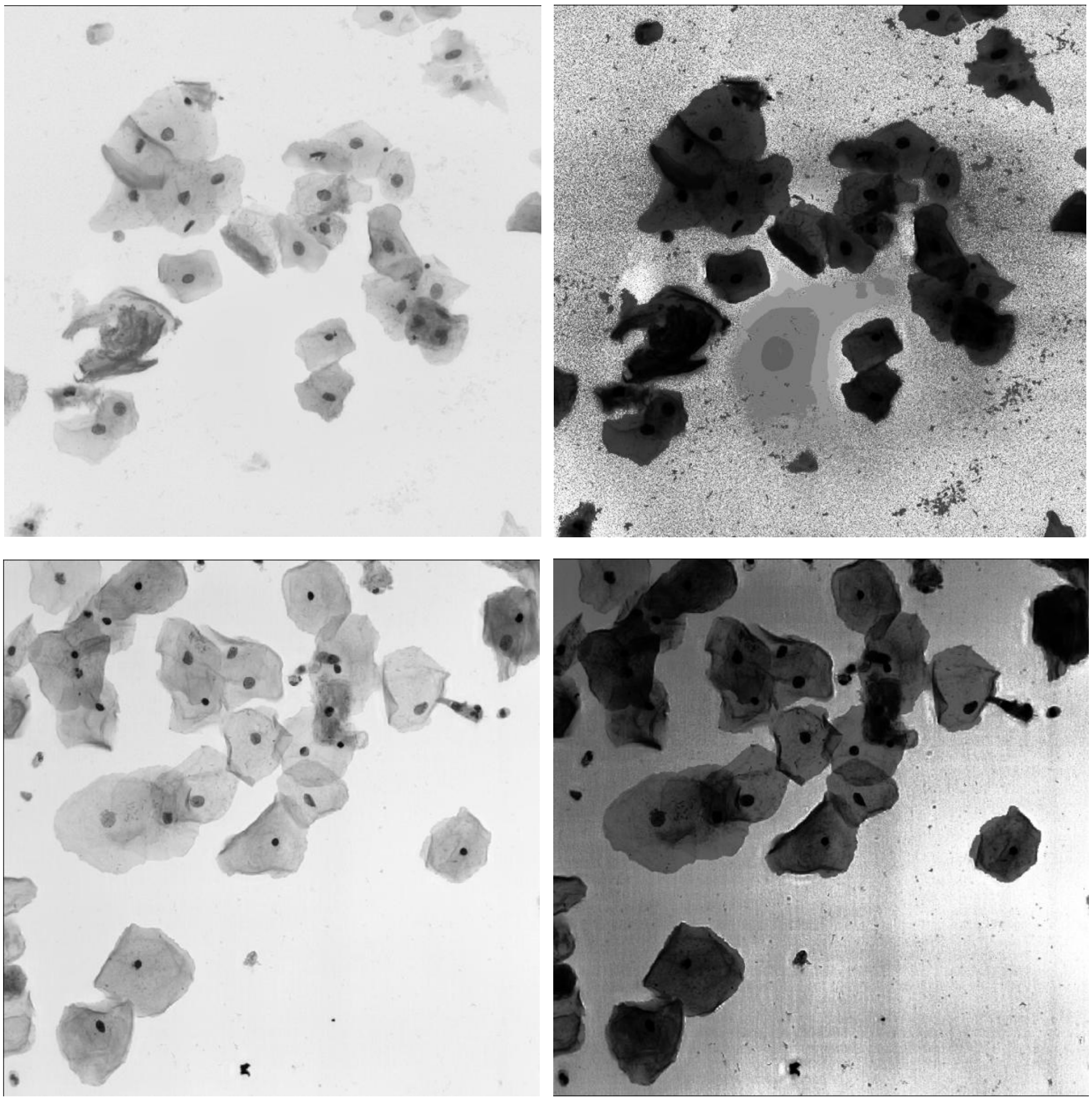

Figure 4.4: The examples of two preprocessed test images.

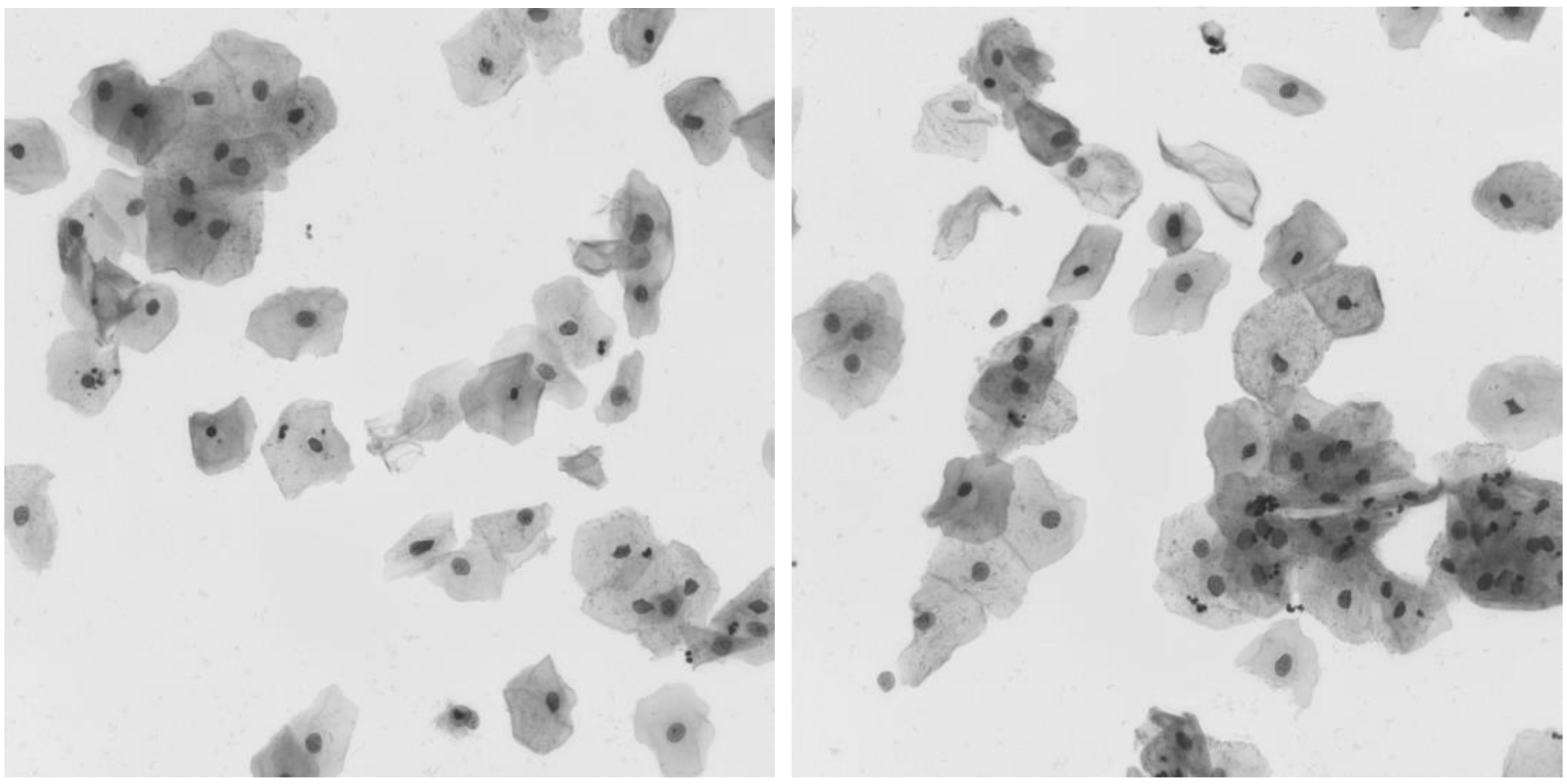

Figure 4.5: The third and fourth test images.

### 4.1.2 BRATS dataset

We will use 2d slices from the BRATS dataset. So, for each 3d volume data of a patient we will use one or more 2d slices from each modality and segmentation data.

We divide the dataset into 270 training patients' data and 15 test patients' data. 270 training data has the following quantities in the groups:

- HGG – 201;
- GG – 69.

From 270 training data we will form three different training sets:

- 270 2d slices by one, with the most intensity of segmentation area, from each patient;
- 15 2d slices by one, with the most intensity of segmentation area, from 15 patients, afterwards augmented 18 times each image to get 270 images in total;
- all non-empty slices (each modality has to be non-empty) from two patients (so, in total we will have about 300 training images).

15 train data are taken from patients with the codes:

- HGG: Brats17_2013_2_1, Brats17_2013_3_1, Brats17_2013_4_1, Brats17_CBICA_AAB_1, Brats17_CBICA_AAG_1, Brats17_CBICA_AAL_1, Brats17_TCIA_105_1, Brats17_TCIA_111_1, Brats17_TCIA_113_1;
- LGG: Brats17_2013_0_1, Brats17_2013_1_1, Brats17_2013_6_1, Brats17_TCIA_101_1, Brats17_TCIA_103_1, Brats17_TCIA_109_1.

All non-empty slices are taken from the patients with the codes:

- HGG: Brats17_2013_2_1;
- LGG: Brats17_TCIA_109_1.

After all preprocessing and augmenting we got the following datasets:

- AXIAL_IMG_2_TRAIN: slices from two patients;
- AXIAL_IMG_15_TRAIN: 270 slices from 15 patients;
- AXIAL_IMG_270_TRAIN: 270 slices from 270 patients.
- AXIAL_SEG_2_TRAIN: segmentations of slices from two patients;
- AXIAL_SEG_15_TRAIN: segmentations of 270 slices from 15 patients;
- AXIAL_SEG_270_TRAIN: segmentations of 270 slices from 270 patients.

And similarly the training data were taken also in coronal and saggital projections, six for each.

From 15 test data we will take by one slice from each patient (with the most intensity of segmentation area). We will use the following files:

- 9 from group HGG: 3 x 2013, 3 x CBICA, 3 x TCIA;
- 6 from group LGG: 3 x 2013, 3 x TCIA.

They are namely:

- HGG: Brats17_2013_25_1, Brats17_2013_26_1, Brats17_2013_27_1, Brats17_CBICA_BHB_1, Brats17_CBICA_BHK_1, Brats17_CBICA_BHM_1, Brats17_TCIA_606_1, Brats17_TCIA_607_1, Brats17_TCIA_608_1;
- LGG: Brats17_2013_24_1, Brats17_2013_28_1, Brats17_2013_29_1, Brats17_TCIA_650_1, Brats17_TCIA_653_1, Brats17_TCIA_654_1.

We got the datasets:

- AXIAL_IMG_15_TEST: 15 axial test slices;
- AXIAL_SEG_15_TEST: segmentations of 15 axial test slices;
- CORONAL_IMG_15_TEST: 15 coronal test slices;
- CORONAL_SEG_15_TEST: segmentations of 15 coronal test slices;
- SAGGITAL_IMG_15_TEST: 15 saggital test slices;
- SAGGITAL_SEG_15_TEST: segmentations of 15 saggital test slices.

Also, we formed second test data from the full data (i.e. all slices) of two patients from the dataset of 15 test data. They are namely: HGG: Brats17_2013_27_1 and LGG: Brats17_TCIA_606_1.

From each ot the two we got the datasets:

- AXIAL_IMG_1_TEST: all axial slices from one patient;
- AXIAL_SEG_1_TEST: segmentations for all axial slices from one patient;
- CORONAL_IMG_1_TEST: all coronal slices from one patient;
- CORONAL_SEG_1_TEST: segmentations for all coronal slices from one patient;
- SAGGITAL_IMG_1_TEST: all saggital slices from one patient;
- SAGGITAL_SEG_1_TEST: segmentations for all saggital slices from one patient.

The examples of preprocessed training images of the patient encoded as Brats17_2013_2_1 are shown in figure 4.6.

The examples of preprocessed and transformed (augmented) training images of the same patient are shown in figure 4.7.

The examples of slices from the all non-empty slices which are taken from the patients with the code Brats17_2013_2_1 are shown in the figure 4.8.

The examples of the preprocessed test slices are given in the figures 4.9 and 4.10.

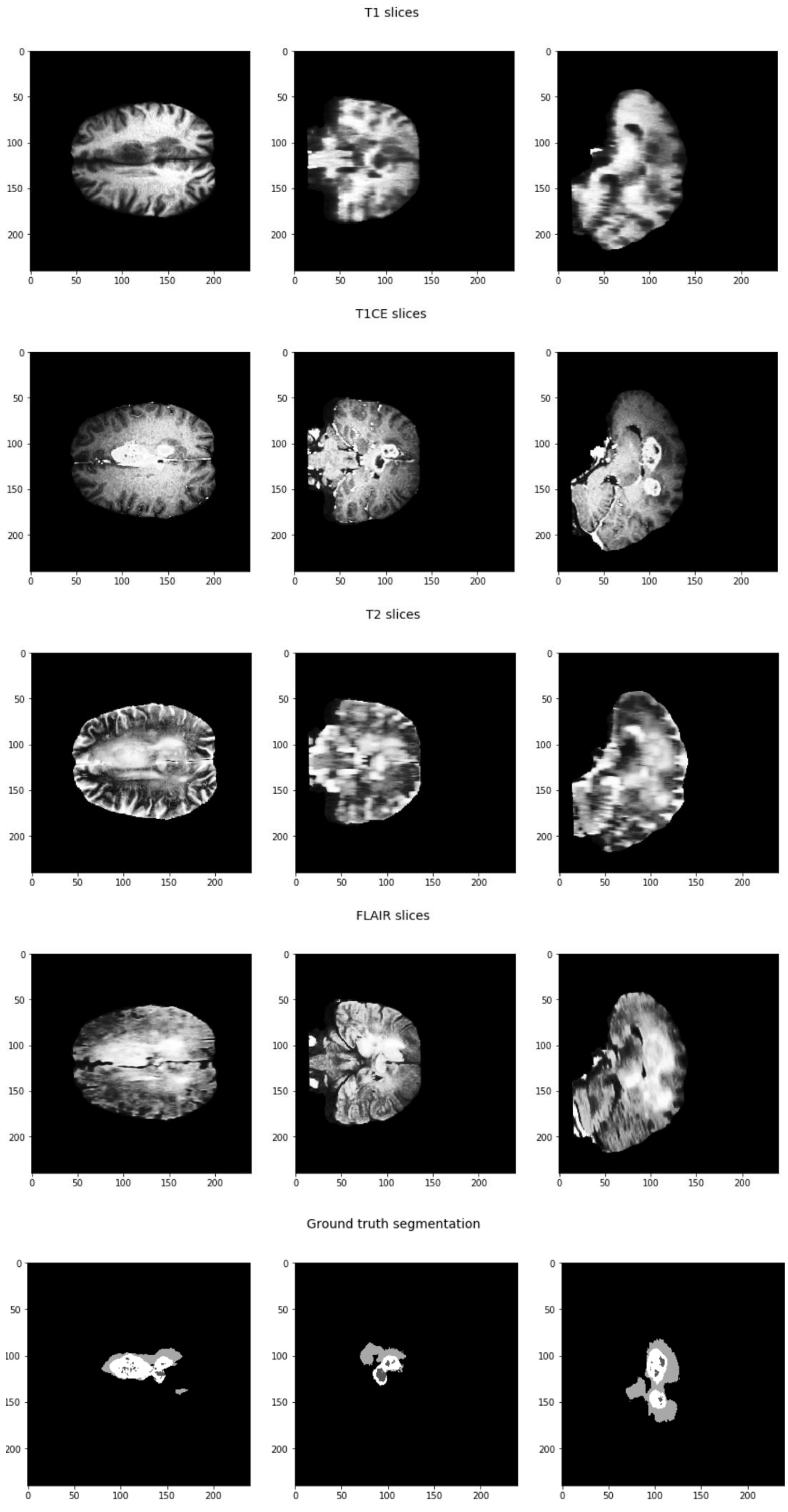

Figure 4.6: The examples of preprocessed training images of the patient encoded as Brats17_2013_2_1

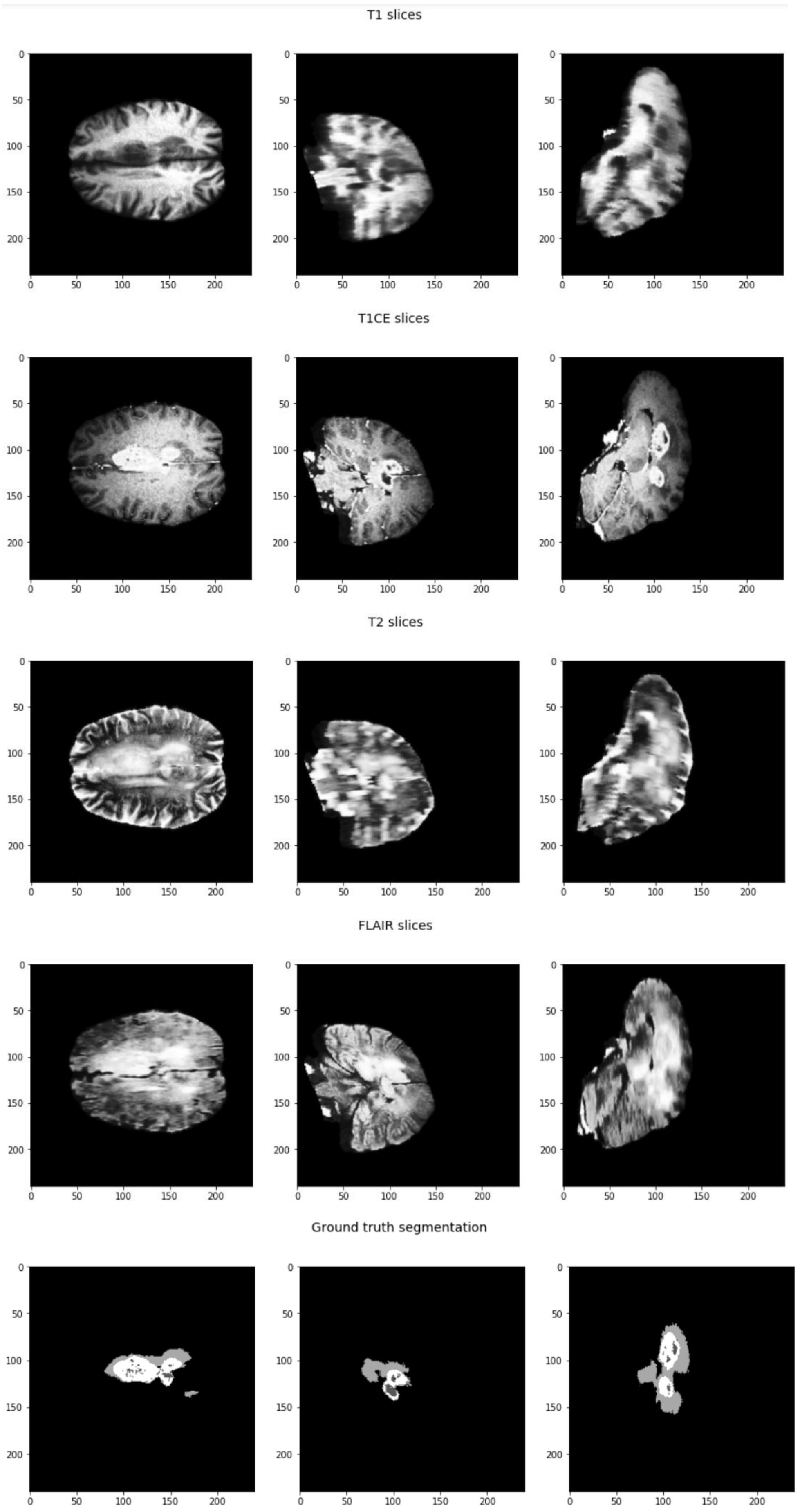

Figure 4.7: The examples of preprocessed and transformed training images of the patient encoded as Brats17_2013_2_1

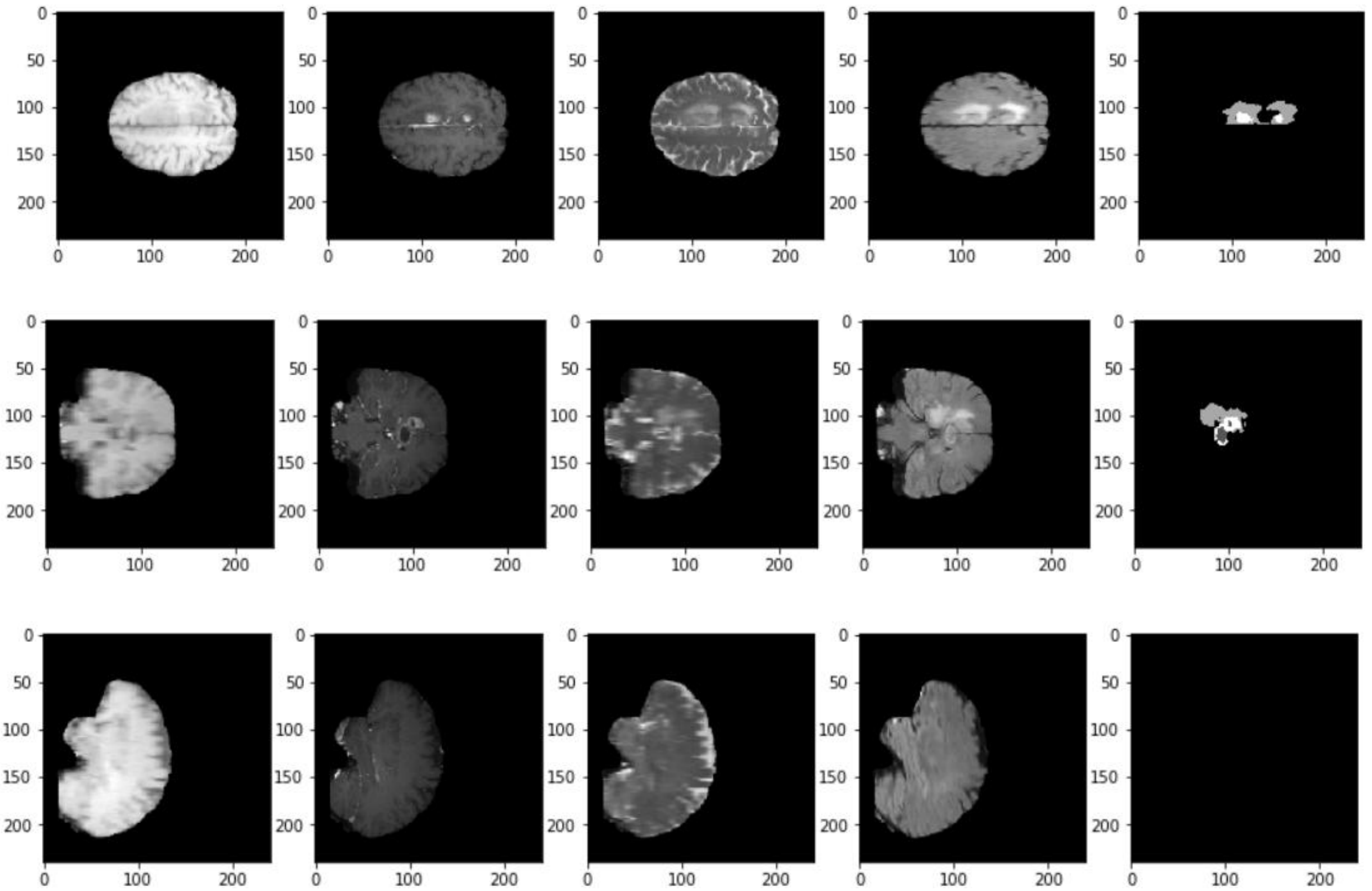

Figure 4.8: The examples from all non-empty slices which are taken from the patients with the code Brats17_2013_2_1.

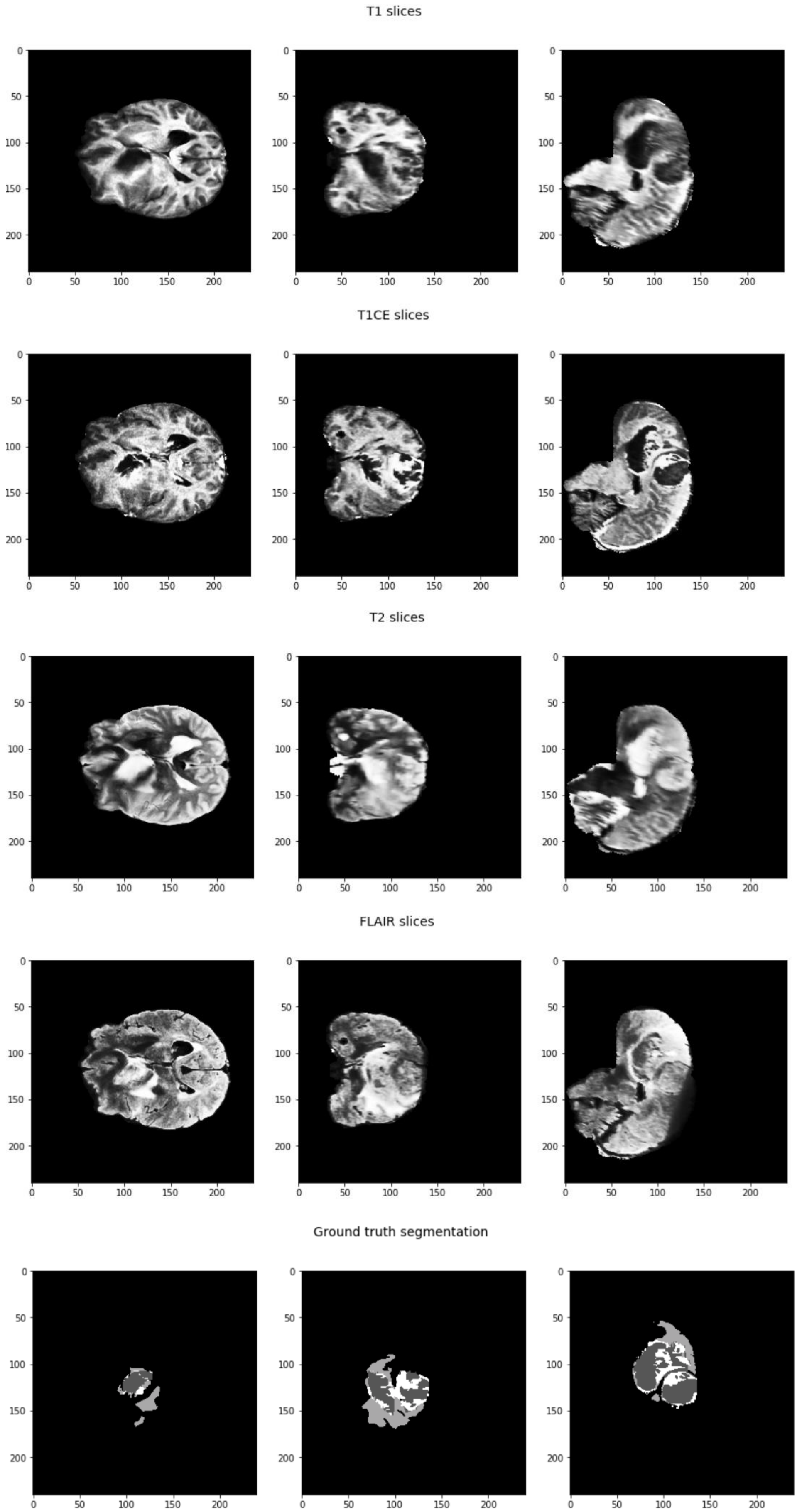

Figure 4.9: The examples of the slices of the test data form full scans of one patient.

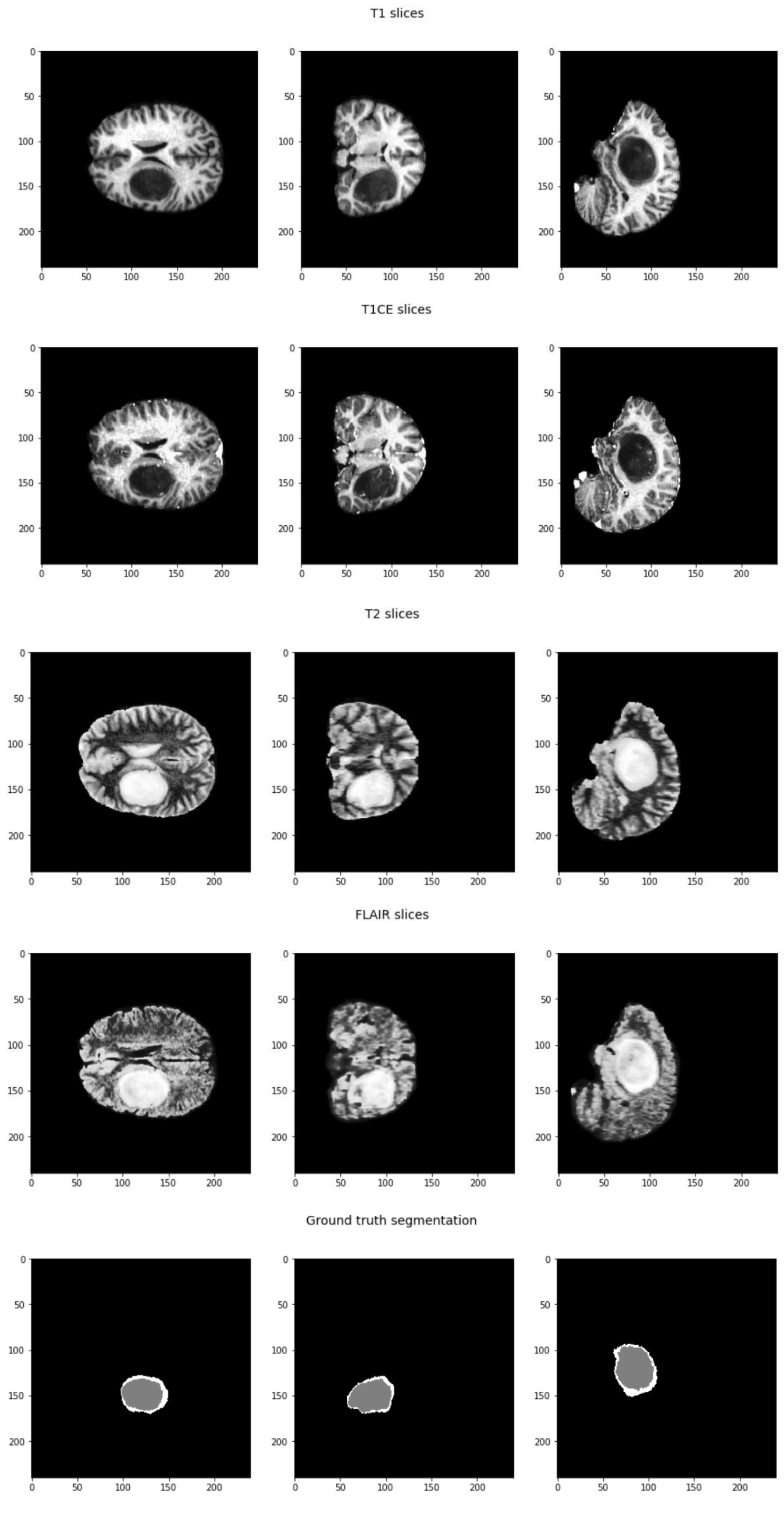

Figure 4.10: The examples of the slices of the test data form full scans of one patient.

## 4.2 Networks parameters and hyper-parameters

Since we are implementing semantic segmentation (or multi-class classification of each pixel), we need to classifying each pixel in the image, and so we will use a cross-entropy function, i.e. categorical cross entropy in our case. To use this we need the softmax activation in the final layer to predict multiple mutually-exclusive classes. Softmax classifier minimizes cross-entropy between estimated class probabilities and true probabilities.

Regarding optimization algorithms, we will try different, but initially stick to one, which is often regarded as the best "out of the box" optimiser. Initial learning rates and learning rate decay schedule will be set before training.

While training a classifier we will be controlling training and validation accuracy, as well as training and validation loss.

We will use 0.1 validation split, i.e. 10% of the initial training data will be used as validation data.

For convolution layers we will use:

- activation function: ReLU;
- padding: “same”;
- strides: (1, 1).

For pooling layers we will use:

- pool size: 2;
- padding: “same”.

For transposed convolution layers (deconvolution layers) we will use:

- strides: (2, 2);
- padding: “same”;
- activation: none.

For the final (convolution) layer we will use:

- padding: “same”;
- activation: softmax.

The optimizers and learning rates we will try:

- Adam with initial learning rate lr = 0.001;
- SGD with lr = 0.01;
- SGD with lr = 0.01 and Nesterov momentum = 0.9;
- RMSprop with lr = 0.001 and rho=0.9;
- Adagrad with lr = 0.01;
- Adadelta with lr = 1.0 and rho=0.95;
- Adamax with lr = 0.002, beta_1=0.9 and beta_2=0.999;
- Nadam with lr = 0.002, beta_1=0.9 and beta_2=0.

The architecture of the FCN we will use is:

input: (height, width, num_channels)

conv_1: (num_filters_1, conv_size, padding='same', activation='relu')

conv_2: (num_filters_2, conv_size, padding='same', activation='relu')

conv_3: (num_filters_3, conv_size, padding='same', activation='relu')

conv_4: (num_filters_4, conv_size, padding='same', activation='relu')

conv_5: (num_filters_5, conv_size, padding='same', activation='relu')

out: (num_classes, out_conv_size, padding='same', activation='softmax'),

where:

height and width: height and width of images;

num_channels is 1 for light microscopy images and 4 for BRATS images;

conv_size and out_conv_size: the sizes of filters which will be set during training;

num_filters: number of filters in layers which will be set during training.

The architecture of the U-Net we will use is:

conv_1: (num_filters_1, conv_size, padding='same', activation='relu')(input)

conv_1: (num_filters_1, conv_size, padding='same', activation='relu')(conv_1)

pool_1: (pool_size=(ps, ps), padding='same')(conv_1)

conv_2: (num_filters_2, conv_size, padding='same', activation='relu')(pool_1)

conv_2: (num_filters_2, conv_size, padding='same', activation='relu')(conv_2)

pool_2: (pool_size=(ps, ps), padding='same')(conv_2)

conv_3: (num_filters_3, conv_size, padding='same', activation='relu')(pool_2)

conv_3: (num_filters_3, conv_size, padding='same', activation='relu')(conv_3)

pool_3: (pool_size=(ps, ps), padding='same')(conv_3)

conv_4: (num_filters_4, conv_size, padding='same', activation='relu')(pool_3)

conv_4: (num_filters_4, conv_size, padding='same', activation='relu')(conv_4)

pool_4: (pool_size=(ps, ps), padding='same')(conv_4)

conv_5: (num_filters_5, conv_size, padding='same', activation='relu')(pool_4)

conv_5: (num_filters_5, conv_size, padding='same', activation='relu')(conv_5)

up_6: (num_filters_4, deconv_size, strides=deconv_strides, padding='same')(conv_5)

up_6: concatenate(up_6, conv_4)

conv_6: (num_filters_4, conv_size, padding='same', activation='relu')(up_6)

conv_6: (num_filters_4, conv_size, padding='same', activation='relu')(conv_6)

up_7: (num_filters_3, deconv_size, strides=deconv_strides, padding='same')(conv_6)

up_7: concatenate(up_7, conv_3)

conv_7: (num_filters_3, conv_size, padding='same', activation='relu')(up_7)

conv_7: (num_filters_3, conv_size, padding='same', activation='relu')(conv_7)

up_8: (num_filters_2, deconv_size, strides=deconv_strides, padding='same')(conv_7)

up_8: concatenate(up_8, conv_2)

conv_8: (num_filters_2, conv_size, padding='same', activation='relu')(up_8)

conv_8: (num_filters_2, conv_size, padding='same', activation='relu')(conv_8)

up_9 : (num_filters_1, deconv_size, strides=deconv_strides, padding='same')(conv_8)

up_9: concatenate(up_9, conv_1)

conv_9: (num_filters_1, conv_size, padding='same', activation='relu')(up_9)

conv_9: (num_filters_1, conv_size, padding='same', activation='relu')(conv_9)

out: (num_classes, out_conv_size, padding='same', activation='softmax')(conv_9),

where:
up: transposed convolution layers (deconvolution layers);
ps: size of pooling, will be set to 2 during training;
deconv_strides: parameter of deconvolution layers, which will be set to (2, 2) during training;
deconv_size: parameter of deconvolution layers, which will be set during training;
and all other parameters are the same meaning as in FCN.

## 4.3 Networks training and evaluation

Training a neural networks requires experiments to find the best hyperparameters, preprocessing for the data (which was already done), choosing an optimiser, loss function etc.
The gap between training and validation accuracy will be indicating the amount of overfitting.
To stop training appropriately we will monitor the validation loss, and when it will not lower on 0.0001 during 12 epochs, we will stop.
To get better results we will decrease the learning rate by the factor 0.2 after 8 epochs during which the validation loss will not lower on 0.0001.
We will also be saving the best current weights of the model to hard drive.

For the purpose of evaluation of the accuracy of prediction of segmentation labels we will use the metric called Dice coefficient, since it is widely used in evaluation of image segmentation, especially for imbalanced classes [5]:

$$Dice(A,B) = \frac{2|A \cdot B|}{|A| + |B|},$$

where A and B are binary vectors (with values of 1 for elements inside a group and 0 otherwise), one signify the ground truth and the other signify the classification result. When applied to boolean data, using the definition of true positive (TP), false positive (FP), and false negative (FN), it can be written as

$$DSC = \frac{2TP}{2TP + FP + FN}.$$

Also, we will use conventional accuracy metric.

# 5 Results

We did several dozens of experiments with different networks and different parameters and hyper-parameters.

## 5.1 Light microscopy dataset

All evaluations of trained networks have been summarized in the table 5.1.
Best of achieved results are op par with best known results, which are:
- cells: 0.977 – 0.979;
- nuclei: 0.83 – 0.87;

The table contains the following parameters:
- #: number of dataset;
- data info: short information about the dataset;
- data tr/val: number of training/validation images;
- # exp.: the number of the current experiment;
- net arch.: net architecture;
- ## epoch: total number of epochs spent for training;
- batch size;
- kernel: the size of one of the two sides of kernel, e.g. 3 for (3, 3);
- ## kernels: the number of kernels in the layers;
- optimizer: the type of optimization algorithm used;
- LR begin/end: initial and ending values of the learning rate;
- ## param.: the total number of learnable parameters in the network;
- acc.: the maximum of training accuracy;
- val. acc.: the maximum of validation accuracy;
- time: the amount of time spent on training;

- test acc.: the test accuracy on four test samples;
- cells: the value of Dice coefficient for the cells;
- nuclei: the value of Dice coefficient for the nuclei;
- comment: comment for specific feature of the current result.

The training graphs of training and validation loss, and training and validation accuracy are shown in the figure 5.1.

The results of predictions for four test images for the best received results of training and evaluation are presented in the following figures:

- figures 5.2 – 5.5: Experiment # 11 on the dataset # 1 (U-Net);
- figures 5.6 – 5.9: Experiment # 30 on the datset # 1 (FCN);
- figures 5.10 – 5.13: Experiment # 3 on the dataset # 3 (U-Net);
- figures 5.14 – 5.17: Experiment # 14 on the dataset # 3 (FCN).

Table 5.1: The results of evaluation of the CNNs on the test data of light microscopy images.

| # | data info | data tr/val | # exp. | net arch. | ## epoch | batch size | conv. layers kernel | conv. layers ## kernels | deconv. layers kernel | deconv. layers ## kernels | out l. kernel | optimizer | LR begin/end | ## param. | training max acc. | training max val.acc. | time | test acc. | Dice coeff. cells | Dice coeff. nuclei | comment |
|---|---|---|---|---|---|---|---|---|---|---|---|---|---|---|---|---|---|---|---|---|---|
| 1 | **1024 x 1024** | 86/10 | 1 | **U-Net** | 59 | 2 | 3 | 16/32/64/128/256 | 2 | 128/64/32/16 | 1 | Adam | 0.001/2e-4 | 1940851 | 0.9888 | 0.9845 | 1625 | 0.9837 | 0.9728 | 0.8654 | >0.865 |
| | **96 samples** | | 2 | | 90 | 2 | 5 | 16/32/64/128/256 | 2 | 128/64/32/16 | 1 | Adam | 0.001/4e-5 | 5078643 | 0.9855 | 0.9832 | 5625 | 0.9822 | 0.9706 | 0.8614 | |
| | (12 * 8) | | 3 | | 66 | 2 | 3 | 16/32/64/128/256 | 3 | 128/64/32/16 | 1 | Adam | 0.001/2e-4 | 2158451 | 0.9868 | 0.9842 | 1960 | 0.9847 | 0.9744 | 0.8633 | |
| | | | 4 | | 75 | 2 | 3 | 16/32/64/128/256 | 3 | 128/64/32/16 | 3 | Adam | 0.001/4e-5 | 2158835 | 0.9897 | 0.9854 | 2307 | 0.9838 | 0.9732 | 0.8618 | |
| | | | 5 | | 87 | 2 | 1 | 16/32/64/128/256 | 2 | 128/64/32/16 | 1 | Adam | 0.001/8e-6 | 371955 | 0.9733 | 0.9709 | 1627 | 0.972 | 0.9544 | 0.6906 | bad |
| | | | 6 | | 75 | 2 | 3 | 16/32/64/128/256 | 4 | 128/64/32/16 | 1 | Adam | 0.001/8e-6 | 2463091 | 0.9866 | 0.9847 | 2292 | 0.9833 | 0.9726 | 0.8599 | <0.86 |
| | | | 7 | | 65 | 2 | 3 | 16/32/64/128/256 | 1 | 128/64/32/16 | 1 | Adam | 0.001/4e-5 | 1810291 | 0.9883 | 0.9849 | 1749 | 0.9838 | 0.9724 | 0.8643 | |
| | | | 8 | | 58 | 2 | 3 | 16/32/64/128/256 | 2 | 128/64/32/16 | 3 | Adam | 0.001/4e-5 | 1941235 | 0.9877 | 0.9848 | 1648 | 0.9824 | 0.971 | 0.8677 | >0.865 |
| | | | 9 | | 67 | 2 | 3 | 16/32/64/128/256 | 1 | 128/64/32/16 | 3 | Adam | 0.001/2e-4 | 1810675 | 0.988 | 0.984 | 1865 | 0.9833 | 0.9722 | 0.8562 | <0.86 |
| | | | 32 | | 81 | 2 | 3 | 16/32/64/128/256 | 2 | 128/64/32/16 | 5 | Adam | 0.001/8e-6 | 1942003 | 0.9879 | 0.9846 | 2501 | 0.9838 | 0.9732 | 0.8617 | |
| | | | 33 | | 114 | 2 | 5 | 16/32/64/128/256 | 2 | 128/64/32/16 | 5 | Adam | 0.001/8e-6 | 5079795 | 0.985 | 0.9831 | 4460 | 0.9824 | 0.9708 | 0.8461 | <0.85 |
| | | | 10 | | 67 | 2 | 5 | 16/32/64/128/256 | 2 | 128/64/32/16 | 3 | Adam | 0.001/2e-4 | 5079027 | 0.9854 | 0.9841 | 4245 | 0.9815 | 0.9697 | 0.869 | >0.865 |
| | | | 11 | | 44 | 1 | 3 | 16/32/64/128/256 | 2 | 128/64/32/16 | 1 | Adam | 0.001/2e-4 | 1940851 | 0.986 | 0.9847 | 1241 | 0.984 | 0.9736 | 0.8755 | >0.875 |
| | | | 12 | | 70 | 1 | 3 | 16/32/64/128/256 | 2 | 128/64/32/16 | 3 | Adam | 0.001/4e-5 | 1941235 | 0.987 | 0.9845 | 2038 | 0.9831 | 0.972 | 0.8652 | |
| | | | 13 | | 124 | 1 | 5 | 16/32/64/128/256 | 2 | 128/64/32/16 | 3 | Adam | 0.001/1.6e-6 | 5079027 | 0.9846 | 0.9831 | 8116 | 0.9803 | 0.9678 | 0.8648 | |
| | | | 14 | | 148 | 1 | 3 | 16/32/64/128/256 | 2 | 128/64/32/16 | 1 | SGD | 0.01/8e-5 | 1940851 | 0.9561 | 0.9511 | 4110 | 0.9621 | 0.9416 | 0 | very bad |
| | | | 15 | | 137 | 1 | 3 | 16/32/64/128/256 | 2 | 128/64/32/16 | 1 | SGD Nest. | 0.01/8e-5 | 1940851 | 0.9761 | 0.9735 | 3822 | 0.9771 | 0.9632 | 0.7904 | bad |
| | | | 16 | | 62 | 1 | 3 | 16/32/64/128/256 | 2 | 128/64/32/16 | 1 | RMSprop | 0.001/4e-5 | 1940851 | 0.989 | 0.9845 | 1737 | 0.9828 | 0.9717 | 0.8469 | <0.85 |
| | | | 17 | | 54 | 1 | 3 | 16/32/64/128/256 | 2 | 128/64/32/16 | 1 | Adagrad | 0.01/4e-5 | 1940851 | 0.9882 | 0.9835 | 1511 | 0.9821 | 0.9697 | 0.8541 | <0.86 |
| | | | 18 | | 90 | 1 | 3 | 16/32/64/128/256 | 2 | 128/64/32/16 | 1 | Adadelta | 1.0/8e-3 | 1940851 | 0.9849 | 0.9841 | 2559 | 0.9823 | 0.9707 | 0.8551 | <0.86 |
| | | | 19 | | 53 | 1 | 3 | 16/32/64/128/256 | 2 | 128/64/32/16 | 1 | Adamax | 0.002/4e-4 | 1940851 | 0.9894 | 0.9845 | 1538 | 0.9831 | 0.9721 | 0.8627 | |
| | | | 20 | | 44 | 1 | 3 | 16/32/64/128/256 | 2 | 128/64/32/16 | 1 | Nadam | 0.002/4e-4 | 1940851 | 0.9742 | 0.9729 | 1279 | 0.9739 | 0.9569 | 0.7395 | bad |
| | | | 24 | | 53 | 1 | 3 | 8/16/32/64/128 | 2 | 64/32/16/8 | 1 | Adam | 0.001/4e-5 | 485691 | 0.9862 | 0.983 | 949 | 0.9822 | 0.9703 | 0.8348 | <0.84 |
| | | | 21 | | 108 | 4 | 3 | 8/16/32/64/128 | 2 | 64/32/16/8 | 1 | Adam | 0.001/4e-5 | 485691 | 0.9863 | 0.983 | 1834 | 0.9824 | 0.971 | 0.8335 | <0.84 |
| | | | 22 | **FCN** | 75 | 1 | 3 | 16/32/64/32/16 | | | 1 | Adam | 0.001/1.6e-6 | 46435 | 0.9734 | 0.9701 | 1975 | 0.971 | 0.9526 | 0.6795 | bad |
| | | | 31 | | 60 | 1 | 3 | 16/32/64/32/16 | | | 1 | Adamax | 0.002/8e-5 | 46435 | 0.9718 | 0.9695 | 1628 | 0.9706 | 0.9525 | 0.6688 | bad |
| | | | 23 | | 76 | 2 | 3 | 8/16/32/16/8 | | | 1 | Adam | 0.001/4e-5 | 11699 | 0.9709 | 0.9683 | 1184 | 0.9792 | 0.9501 | 0.6475 | bad |
| | | | 25 | | 39 | 1 | 3 | 8/16/32/16/8 | | | 1 | Adam | 0.001/2e-4 | 11699 | 0.9711 | 0.9699 | 628 | 0.9806 | 0.9554 | 0.6626 | bad |
| | | | 34 | | 32 | 1 | 3 | 8/16/32/16/8 | | | 5 | Adam | 0.001/4e-5 | 12275 | 0.971 | 0.9678 | 578 | 0.9698 | 0.9509 | 0.6553 | bad |
| | | | 35 | | 69 | 1 | 5 | 8/16/32/16/8 | | | 1 | Adam | 0.001/8e-6 | 32307 | 0.9752 | 0.9719 | 2156 | 0.9723 | 0.9548 | 0.7028 | >0.70 |
| | | | 36 | | 42 | 1 | 5 | 8/16/32/16/8 | | | 3 | Adam | 0.001/2e-4 | 32499 | 0.9745 | 0.9729 | 1337 | 0.9742 | 0.9582 | 0.7082 | >0.70 |
| | | | 37 | | 78 | 1 | 5 | 8/16/32/16/8 | | | 5 | Adam | 0.001/8e-6 | 32883 | 0.978 | 0.9771 | 2578 | 0.9741 | 0.9582 | 0.6658 | |
| | | | 26 | | 113 | 1 | 3 | 2/4/8/16/32/16/8/4/2 | | | 1 | Adam | 0.001/8e-6 | 12359 | 0.9746 | 0.974 | 2268 | 0.9772 | 0.9508 | 0.6738 | bad |
| | | | 27 | | 97 | 1 | 3 | 4/8/16/32/64/32/16/8/4 | | | 1 | Adam | 0.001/8e-6 | 49195 | 0.9743 | 0.9724 | 3156 | 0.9812 | 0.9565 | 0.677 | bad |
| | | | 28 | | 52 | 1 | 3 | 16/32/64/32/16 | | | 3 | Adam | 0.001/8e-6 | 46819 | 0.9723 | 0.9695 | 1431 | 0.9803 | 0.9526 | 0.6722 | bad |
| | | | 29 | | 92 | 1 | 5 | 16/32/64/32/16 | | | 1 | Adam | 0.001/1.6e-6 | 128611 | 0.9768 | 0.9743 | 5895 | 0.9728 | 0.9564 | 0.7491 | >0.74 |
| | | | 30 | | 86 | 1 | 5 | 16/32/64/32/16 | | | 1 | Adamax | 0.002/8e-5 | 128611 | 0.9783 | 0.9771 | 5462 | 0.9757 | 0.96 | 0.7733 | >0.77 |
| | | | 38 | | 36 | 1 | 5 | 16/32/64/32/16 | | | 3 | Adam | 0.001/2e-4 | 128995 | 0.9743 | 0.971 | 2318 | 0.974 | 0.958 | 0.7279 | |

Table 5.1: Continued.

| | | | | | | | | | | | | | | | | | | | | | |
|---|---|---|---|---|---|---|---|---|---|---|---|---|---|---|---|---|---|---|---|---|---|
| 2 | **1024 x 1024** | 10/2 | 1 | **U-Net** | 66 | 1 | 3 | 16/32/64/128/256 | 2 | 128/64/32/16 | 1 | Adam | 0.001/4e-5 | 1940851 | 0.9788 | 0.9821 | 227 | 0.9765 | 0.9622 | 0.7772 | bad |
| | **12 samples** | | 2 | | 68 | 1 | 3 | 16/32/64/128/256 | 2 | 128/64/32/16 | 3 | Adam | 0.001/8e-6 | 1941235 | 0.978 | 0.9819 | 241 | 0.9751 | 0.9604 | 0.7791 | bad |
| | (original) | | 3 | | 78 | 1 | 3 | 16/32/64/128/256 | 2 | 128/64/32/16 | 3 | Adam | 0.001/8e-6 | 5079027 | 0.9803 | 0.9829 | 622 | 0.9782 | 0.9647 | 0.8088 | cross-val |
| | | | 4 | | 80 | 1 | 3 | 16/32/64/128/256 | 2 | 128/64/32/16 | 1 | Adamax | 0.002/1.6e-5 | 1940851 | 0.9808 | 0.9796 | 274 | 0.9765 | 0.9616 | 0.6557 | bad |
| | | | 5 | | 102 | 1 | 3 | 16/32/64/128/256 | 2 | 128/64/32/16 | 3 | Adamax | 0.002/1.6e-5 | 1941235 | 0.9819 | 0.9824 | 358 | 0.9802 | 0.9678 | 0.7459 | bad |
| | | | 6 | | 75 | 1 | 5 | 16/32/64/128/256 | 2 | 128/64/32/16 | 3 | Adamax | 0.002/1.6e-5 | 5079027 | 0.9805 | 0.9811 | 588 | 0.9764 | 0.9622 | 0.7431 | bad |
| | | | 7 | **FCN** | 70 | 1 | 3 | 16/32/64/32/16 | | | 1 | Adam | 0.001/8e-6 | 46435 | 0.9709 | 0.9783 | 231 | 0.9801 | 0.9506 | 0.6359 | bad |
| | | | 8 | | 86 | 1 | 5 | 16/32/64/32/16 | | | 1 | Adam | 0.001/1.6e-6 | 128611 | 0.973 | 0.9777 | 660 | 0.9799 | 0.9532 | 0.6481 | bad |
| | | | 9 | | 94 | 1 | 3 | 16/32/64/32/16 | | | 1 | Adamax | 0.002/1.6e-5 | 46435 | 0.9717 | 0.979 | 301 | 0.9695 | 0.9512 | 0.6396 | bad |
| | | | 10 | | 84 | 1 | 5 | 16/32/64/32/16 | | | 1 | Adamax | 0.002/8e-5 | 128611 | 0.9736 | 0.979 | 639 | 0.9709 | 0.9535 | 0.6537 | bad |
| | **w/o hist eq** | | 11 | **U-Net** | 115 | 1 | 5 | 16/32/64/128/256 | 2 | 128/64/32/16 | 3 | Adam | 0.001/1.6e-6 | 5079027 | 0.9764 | 0.977 | 916 | 0.9756 | 0.9584 | 0.7612 | worse |
| | | | 12 | **FCN** | 106 | 1 | 5 | 16/32/64/32/16 | | | 1 | Adamax | 0.002/3.2e-6 | 128611 | 0.9742 | 0.9759 | 1010 | 0.9768 | 0.9606 | 0.7088 | better |
| 3 | **512 x 512** | 345/39 | 1 | **U-Net** | 56 | 1 | 3 | 16/32/64/128/256 | 2 | 128/64/32/16 | 1 | Adam | 0.001/2e-4 | 1940851 | 0.9887 | 0.9849 | 1890 | 0.9833 | 0.972 | 0.8767 | >0.875 |
| | **384 samples** | | 2 | | 58 | 8 | 3 | 16/32/64/128/256 | 2 | 128/64/32/16 | 1 | Adam | 0.001/2e-4 | 1940851 | 0.9872 | 0.9837 | 1596 | 0.9808 | 0.9685 | 0.8466 | |
| | (12 * 32) | | 3 | | 67 | 1 | 3 | 16/32/64/128/256 | 2 | 128/64/32/16 | 3 | Adam | 0.001/2e-4 | 1941235 | 0.9888 | 0.9852 | 2334 | 0.9836 | 0.9725 | 0.8815 | >0.88 |
| | | | 15 | | 57 | 1 | 3 | 16/32/64/128/256 | 2 | 128/64/32/16 | 5 | Adam | 0.001/4e-5 | 1942003 | 0.9854 | 0.9828 | 1870 | 0.9812 | 0.9686 | 0.8407 | |
| | | | 4 | | 93 | 1 | 5 | 16/32/64/128/256 | 2 | 128/64/32/16 | 3 | Adam | 0.001/4e-5 | 5079027 | 0.9843 | 0.9827 | 7030 | 0.9797 | 0.9668 | 0.8564 | |
| | | | 8 | | 48 | 1 | 3 | 16/32/64/128/256 | 1 | 128/64/32/16 | 1 | Adam | 0.001/4e-5 | 1810291 | 0.9882 | 0.9841 | 1865 | 0.9827 | 0.9711 | 0.8628 | |
| | | | 9 | | 60 | 1 | 3 | 16/32/64/128/256 | 1 | 128/64/32/16 | 3 | Adam | 0.001/2e-4 | 1810675 | 0.9875 | 0.9847 | 2333 | 0.9832 | 0.9721 | 0.8789 | >0.875 |
| | | | 10 | | 54 | 1 | 3 | 16/32/64/128/256 | 3 | 128/64/32/16 | 1 | Adam | 0.001/2e-4 | 2158451 | 0.989 | 0.985 | 2264 | 0.9836 | 0.9726 | 0.8536 | |
| | | | 11 | | 57 | 1 | 3 | 16/32/64/128/256 | 3 | 128/64/32/16 | 3 | Adam | 0.001/2e-4 | 2158835 | 0.9864 | 0.9844 | 2456 | 0.9839 | 0.973 | 0.8692 | |
| | | | 5 | | 45 | 1 | 3 | 64/128/256/512/1024 | 2 | 256/128/64/32 | 3 | Adam | 0.001/4e-5 | 31032259 | 0.9896 | 0.9849 | 9264 | 0.9837 | 0.9728 | 0.8684 | |
| | | | 6 | **FCN** | 103 | 1 | 3 | 64/128/256/128/64 | | | 1 | Adam | 0.001/1.6e-6 | 738691 | 0.9827 | 0.979 | 20634 | 0.9804 | 0.9674 | 0.8453 | |
| | | | 7 | | 94 | 1 | 3 | 16/32/64/32/16 | | | 1 | Adam | 0.001/8e-6 | 46435 | 0.9783 | 0.9785 | 3111 | 0.9776 | 0.9628 | 0.8171 | |
| | | | 12 | | 103 | 1 | 3 | 16/32/64/32/16 | | | 1 | Adamax | 0.002/8e-5 | 46435 | 0.9789 | 0.979 | 4013 | 0.9746 | 0.9581 | 0.8281 | |
| | | | 13 | | 102 | 1 | 5 | 16/32/64/32/16 | | | 1 | Adam | 0.001/8e-6 | 128611 | 0.9818 | 0.9817 | 8113 | 0.9819 | 0.97 | 0.8652 | >0.865 |
| | | | 14 | | 85 | 1 | 5 | 16/32/64/32/16 | | | 1 | Adamax | 0.002/1.6e-5 | 128611 | 0.9812 | 0.9814 | 6390 | 0.9796 | 0.9664 | 0.8698 | >0.865 |
| 4 | **256 x 256** | 1382/ | 1 | **U-Net** | 46 | 16 | 3 | 16/32/64/128/256 | 2 | 128/64/32/16 | 1 | Adam | 0.001/2e-4 | 1940851 | 0.9814 | 0.977 | 1485 | 0.9789 | 0.9649 | 0.8067 | |
| | **1536 sampl.** | /154 | 2 | | 51 | 16 | 3 | 16/32/64/128/256 | 2 | 128/64/32/16 | 3 | Adam | 0.001/2e-4 | 1941235 | 0.9809 | 0.9775 | 1707 | 0.9805 | 0.9676 | 0.8206 | |
| | (12 * 128) | | 3 | | 29 | 16 | 3 | 16/32/64/128/256 | 1 | 128/64/32/16 | 1 | Adam | 0.001/2e-4 | 1810291 | 0.974 | 0.9739 | 930 | 0.9761 | 0.9605 | 0.7667 | bad |
| | | | 4 | | 46 | 16 | 3 | 16/32/64/128/256 | 1 | 128/64/32/16 | 3 | Adam | 0.001/2e-4 | 1810675 | 0.9808 | 0.9765 | 1544 | 0.9808 | 0.9648 | 0.7957 | bad |
| | | | 5 | | 51 | 16 | 3 | 16/32/64/128/256 | 3 | 128/64/32/16 | 1 | Adam | 0.001/2e-4 | 2158451 | 0.9817 | 0.9765 | 1839 | 0.9784 | 0.9639 | 0.8016 | |
| | | | 6 | | 53 | 16 | 3 | 16/32/64/128/256 | 3 | 128/64/32/16 | 3 | Adam | 0.001/2e-4 | 2158835 | 0.9811 | 0.9772 | 2013 | 0.9794 | 0.9656 | 0.8006 | |
| | | | 7 | | 65 | 16 | 5 | 16/32/64/128/256 | 2 | 128/64/32/16 | 3 | Adam | 0.001/2e-4 | 5079027 | 0.9807 | 0.9772 | 3892 | 0.98 | 0.967 | 0.8254 | |
| | | | 9 | **FCN** | 102 | 1 | 3 | 16/32/64/32/16 | | | 1 | Adamax | 0.002/1.6e-5 | 46435 | 0.9747 | 0.9743 | 3333 | 0.9745 | 0.9579 | 0.8622 | |
| | | | 10 | | 166 | 16 | 3 | 16/32/64/32/16 | | | 1 | Adamax | 0.002/1.6e-5 | 46435 | 0.9727 | 0.9724 | 9467 | 0.9735 | 0.9484 | 0.848 | |
| | | | 11 | | 111 | 16 | 5 | 16/32/64/32/16 | | | 1 | Adam | 0.001/8e-6 | 128611 | 0.9746 | 0.9744 | 7300 | 0.9784 | 0.954 | 0.8567 | |
| | | | 12 | | 165 | 16 | 5 | 16/32/64/32/16 | | | 1 | Adamax | 0.002/1.6e-5 | 128611 | 0.9758 | 0.9753 | 11430 | 0.9822 | 0.9612 | 0.863 | >0.86 |
| | **w/o hist eq** | | 13 | **U-Net** | 111 | 16 | 5 | 16/32/64/128/256 | 2 | 128/64/32/16 | 3 | Adam | 0.001/4e-5 | 5079027 | 0.9799 | 0.9768 | 6768 | 0.9798 | 0.966 | 0.8256 | equal |
| | | | 14 | **FCN** | 127 | 1 | 3 | 16/32/64/32/16 | | | 1 | Adamax | 0.002/3.2e-6 | 46435 | 0.9776 | 0.9773 | 4099 | 0.9883 | 0.974 | 0.8645 | equal |

## Experiment # 11 on the dataset # 1 (U-Net):

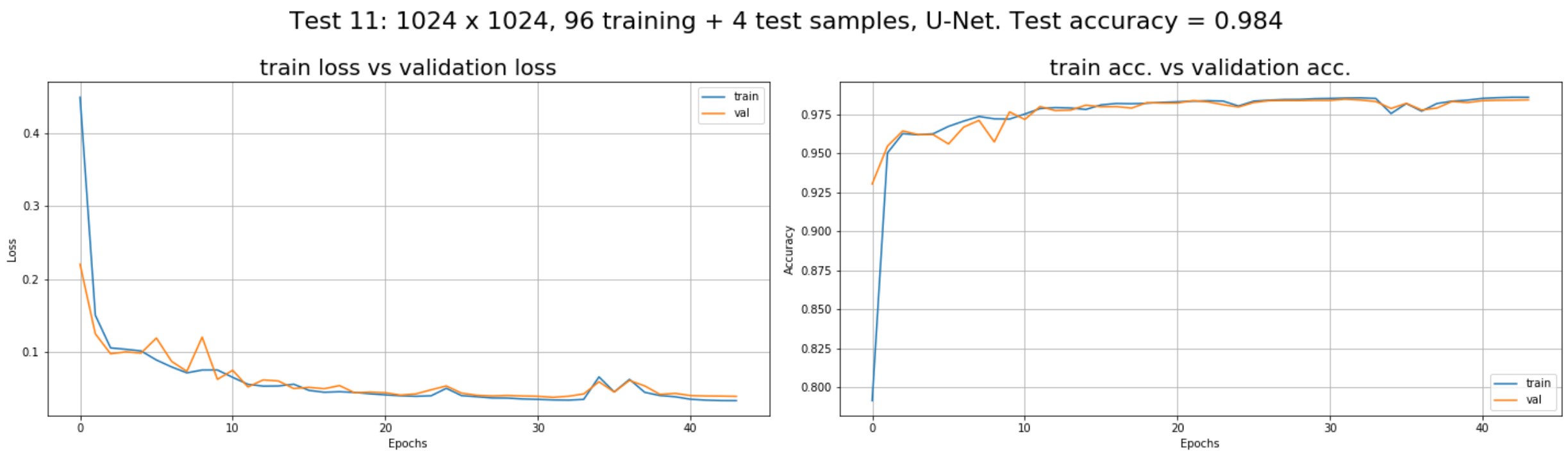

## Experiment # 30 on the datset # 1 (FCN):

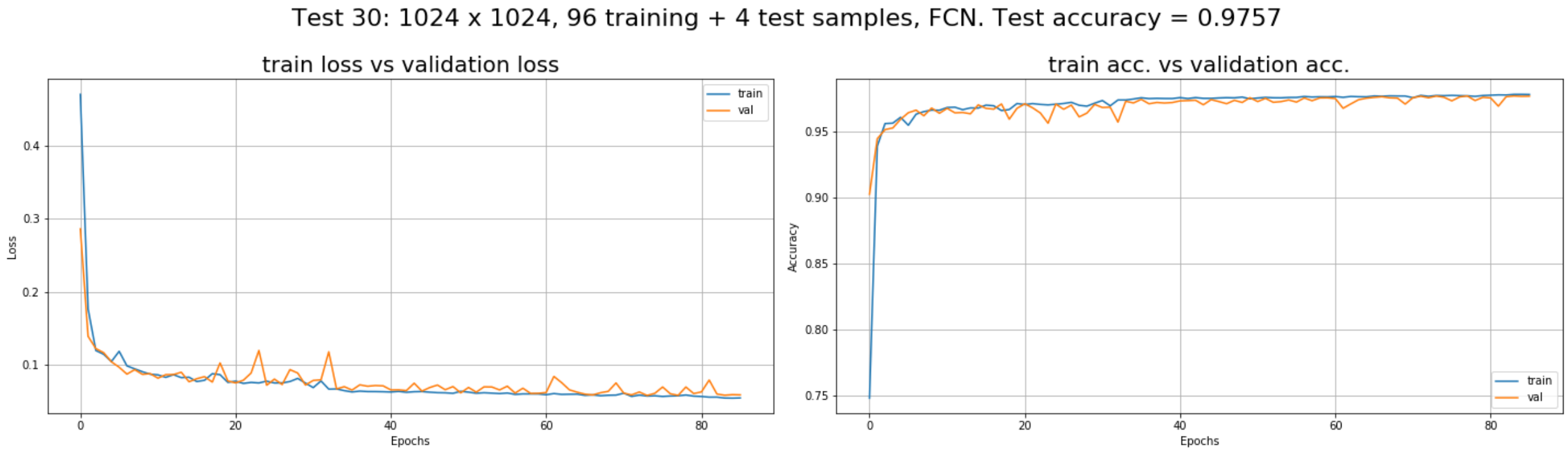

## Experiment # 3 on the dataset # 3 (U-Net):

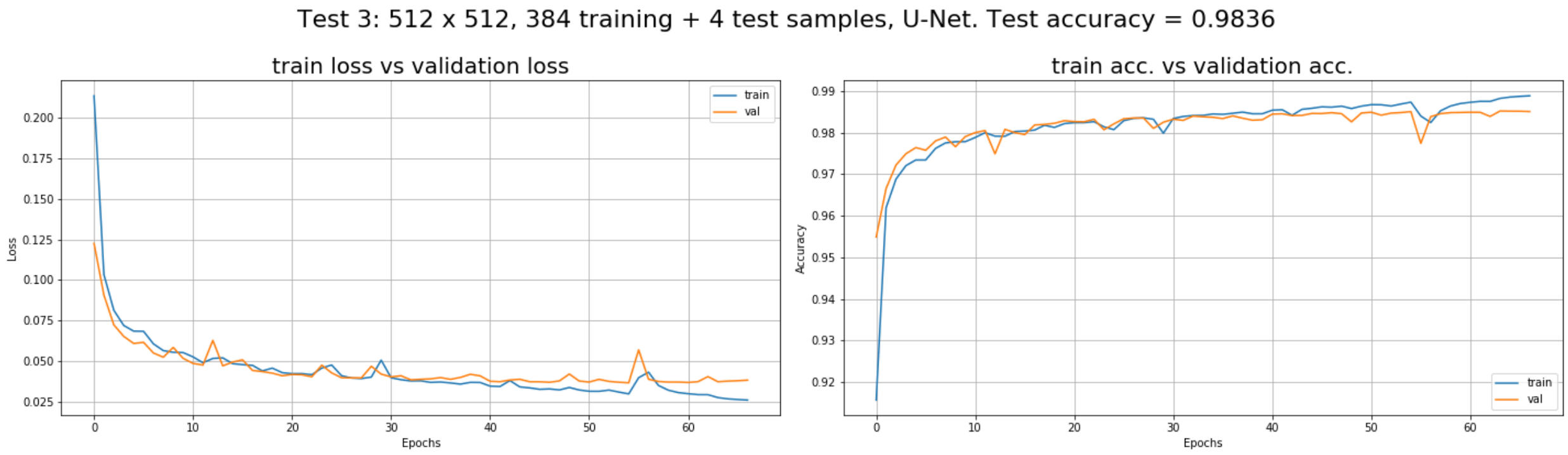

## Experiment # 14 on the dataset # 3 (FCN):

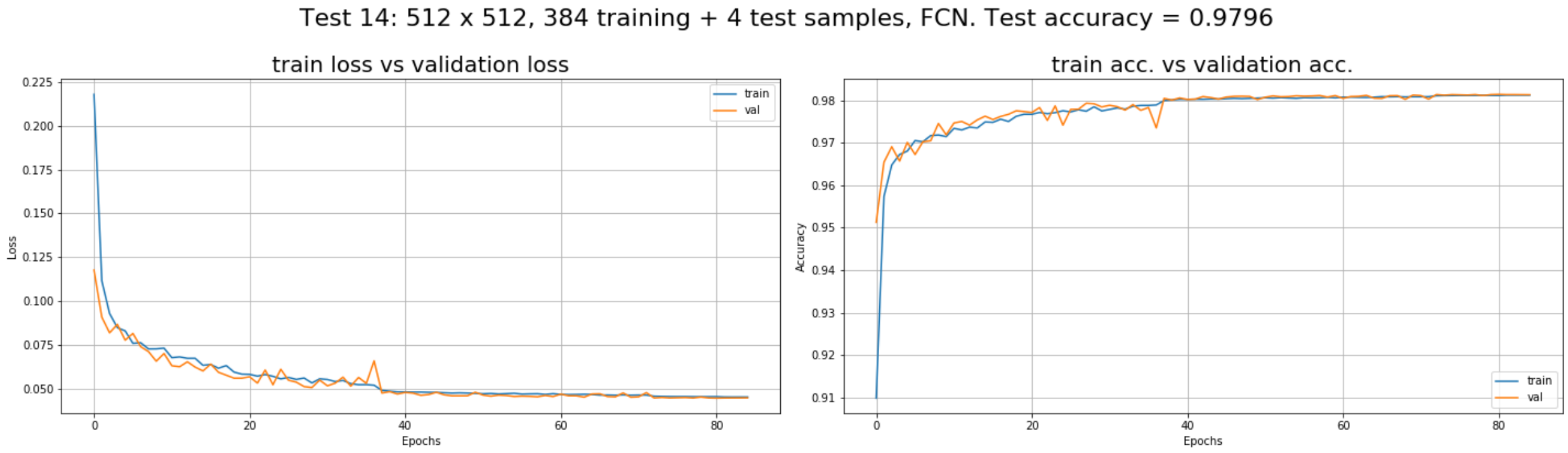

Figure 5.1: Graphs of training and validation loss and accuracy for selected experiments.

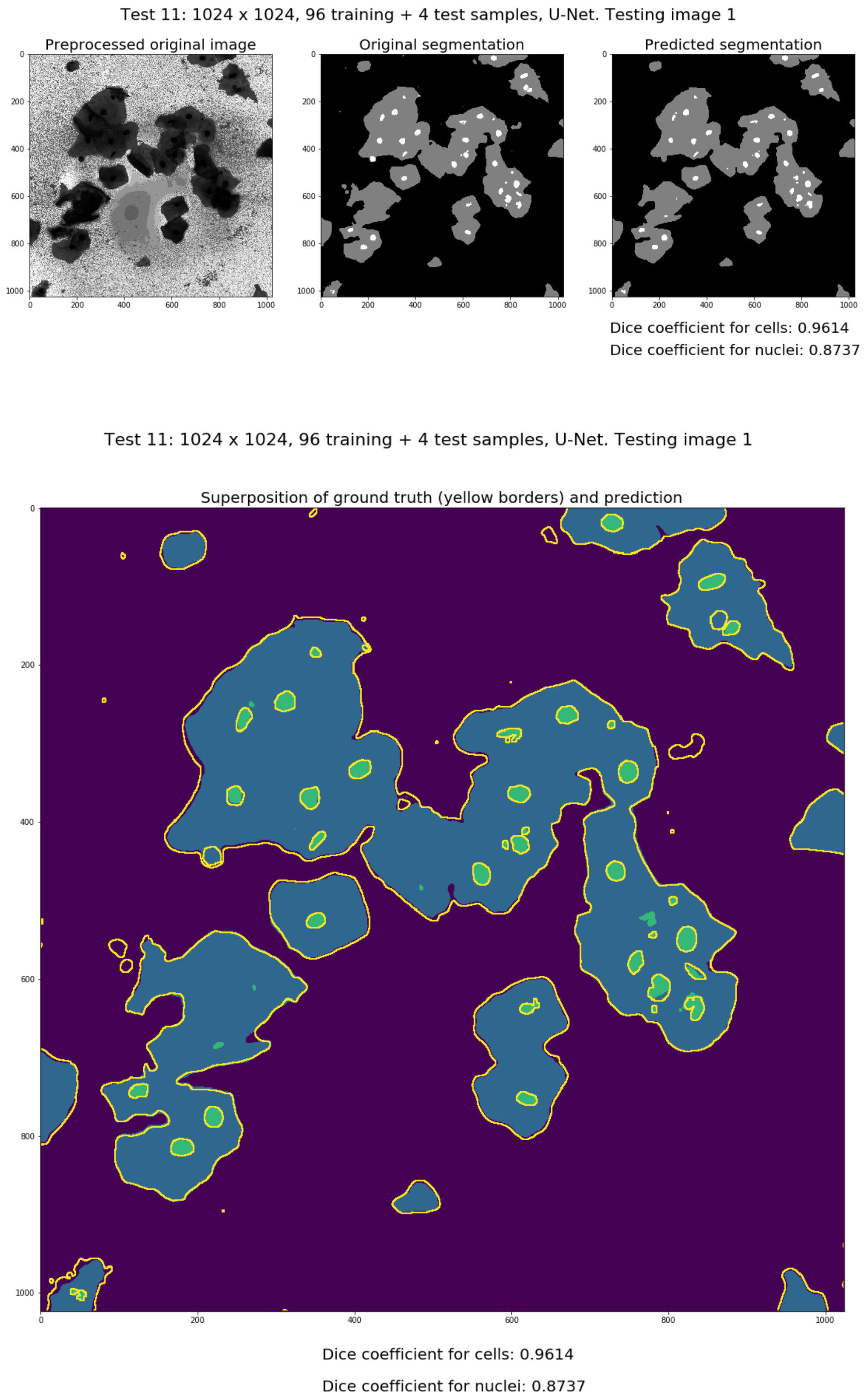

Figure 5.2: Experiment # 11 on the dataset # 1 (U-Net), test image # 1.

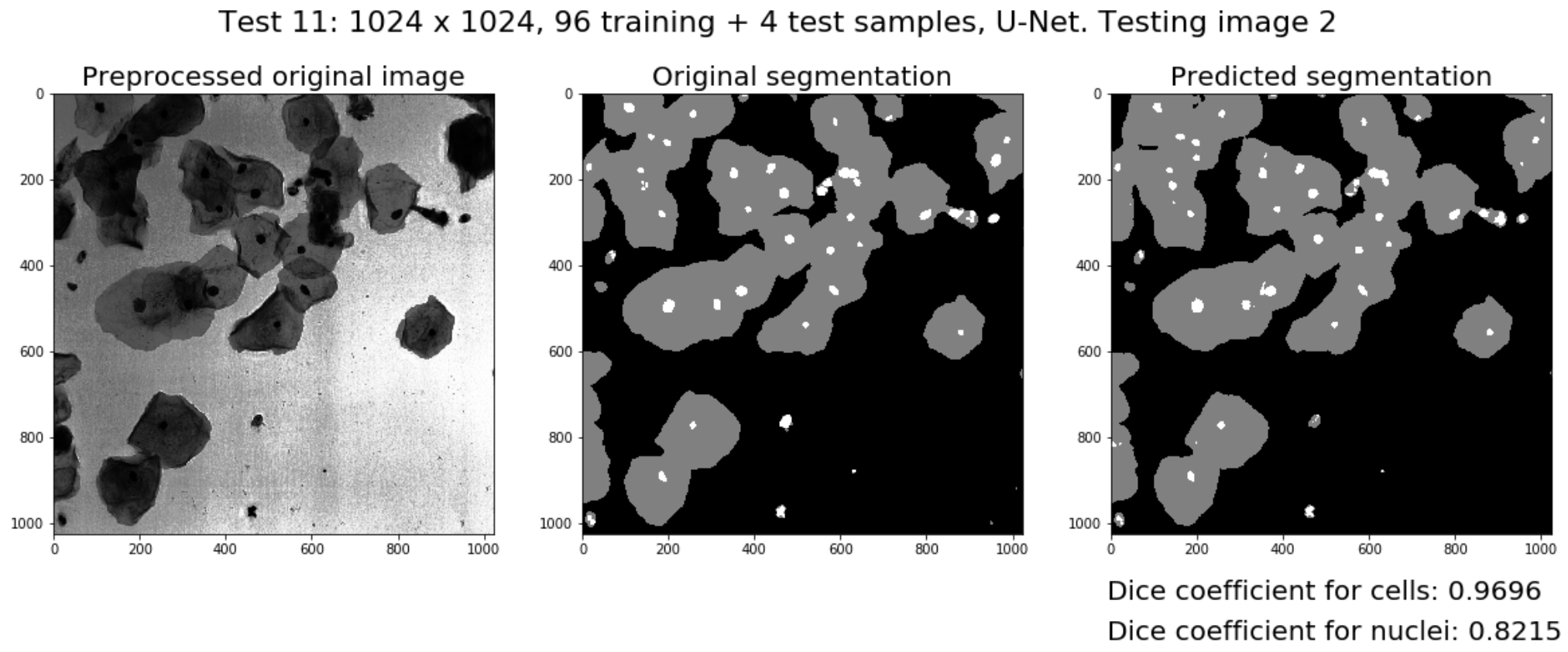

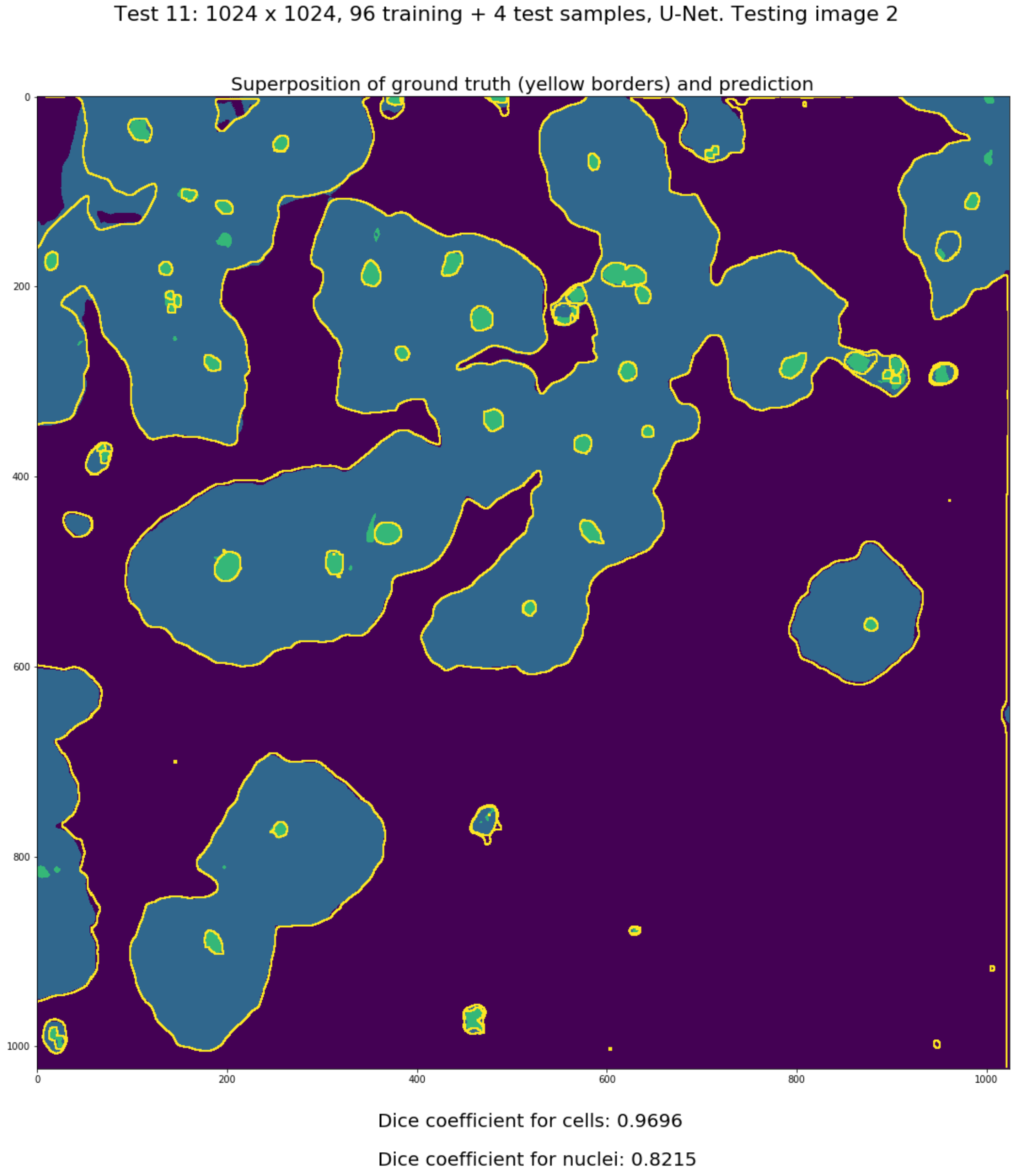

Figure 5.3: Experiment # 11 on the dataset # 1 (U-Net), test image # 2.

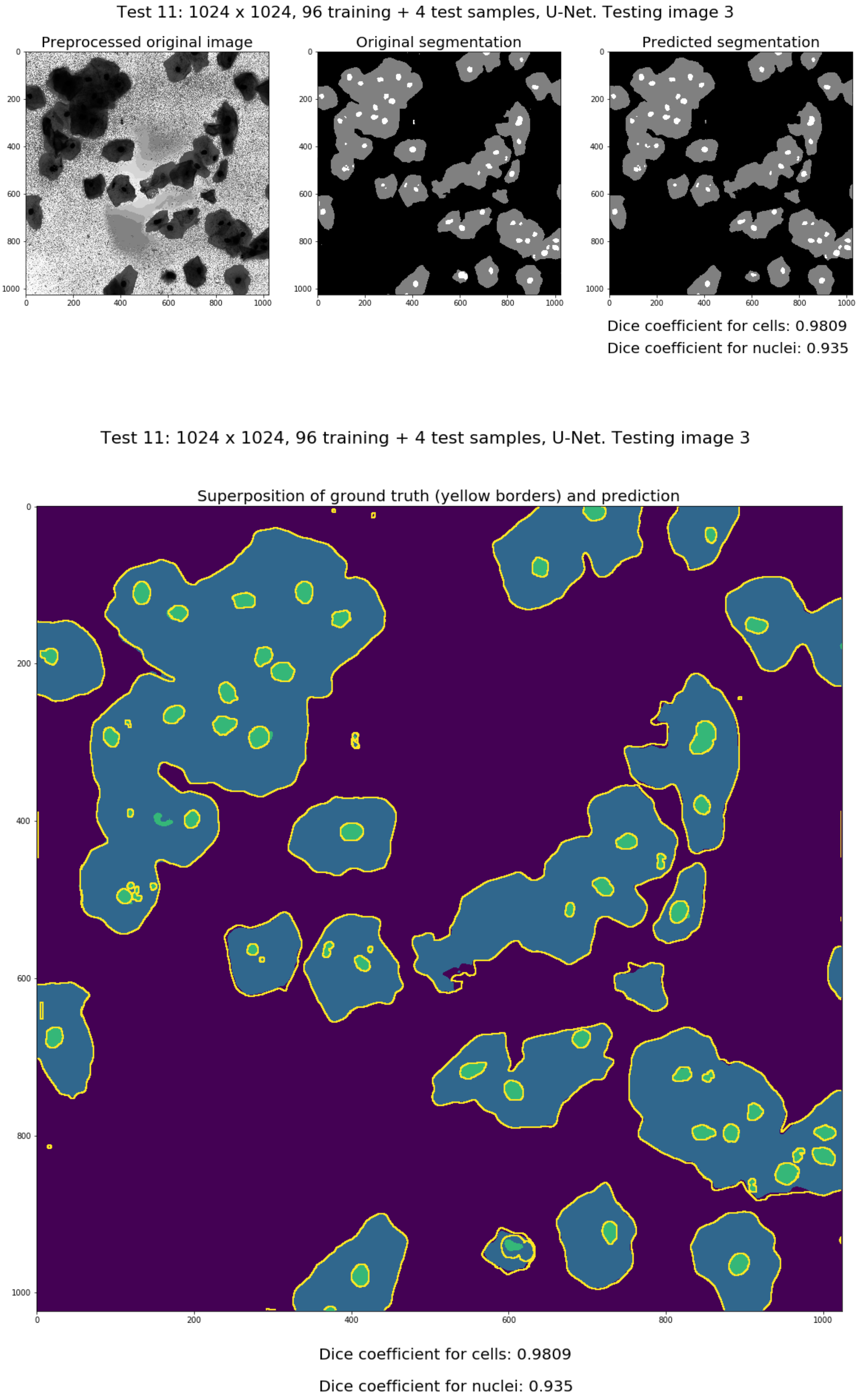

Figure 5.4: Experiment # 11 on the dataset # 1 (U-Net), test image # 3.

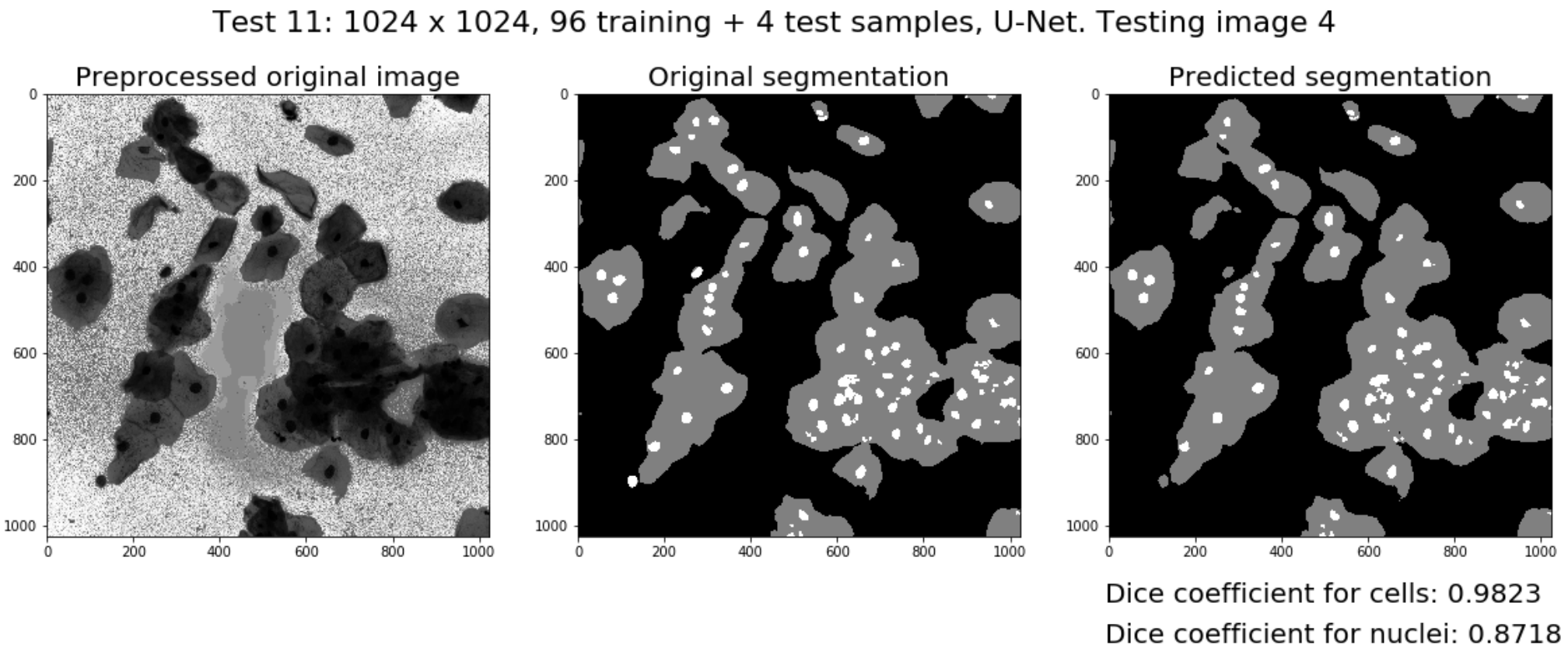

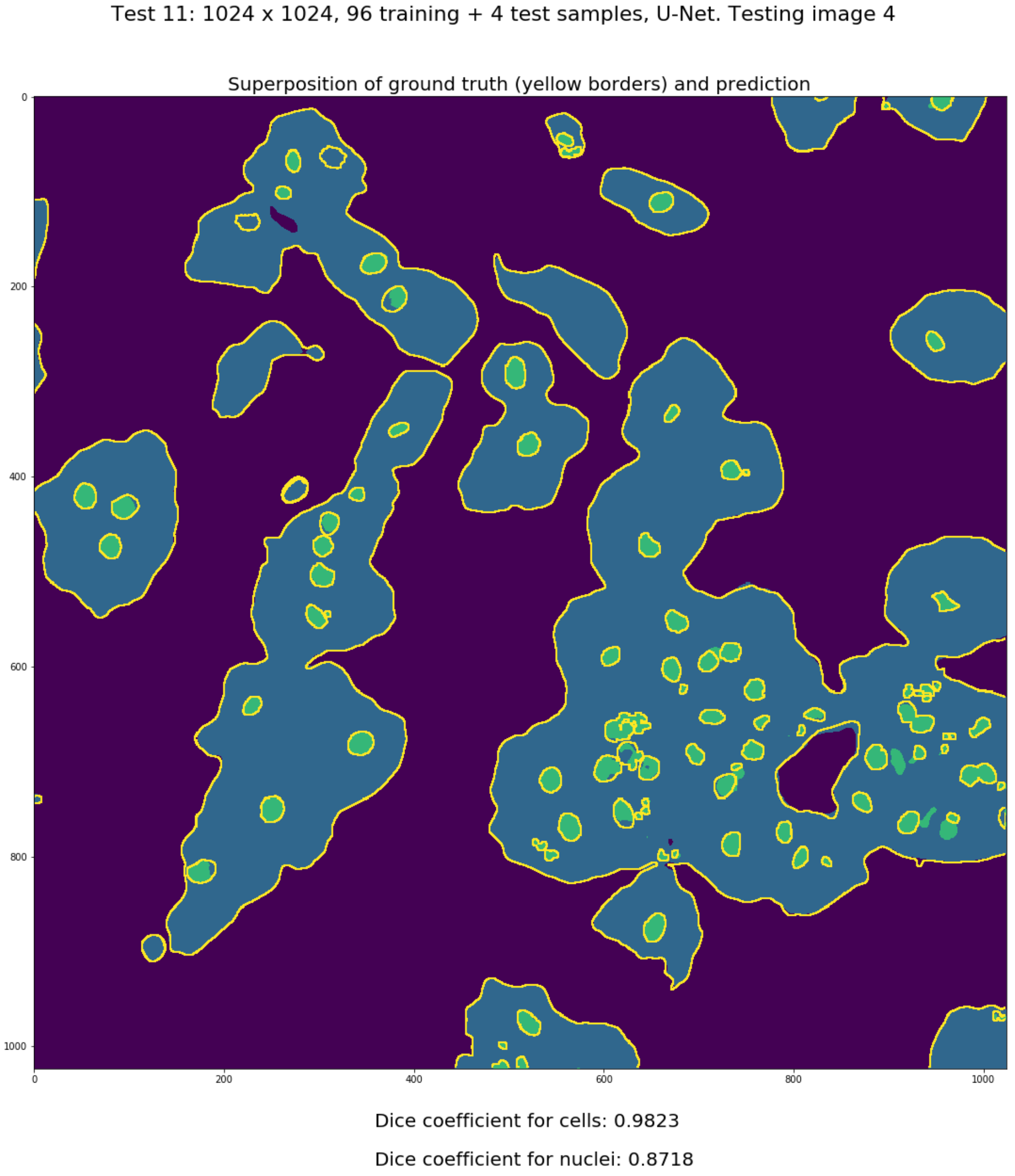

Figure 5.5: Experiment # 11 on the dataset # 1 (U-Net), test image # 4.

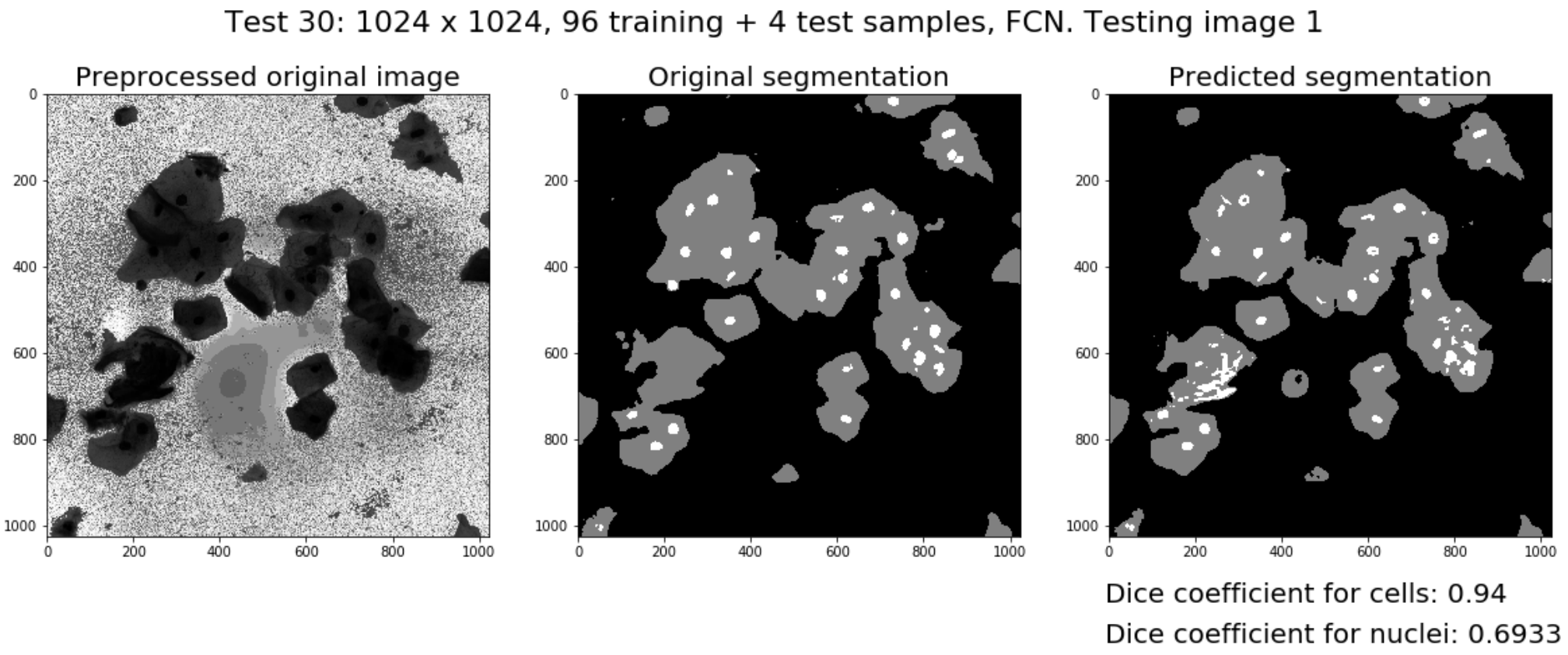

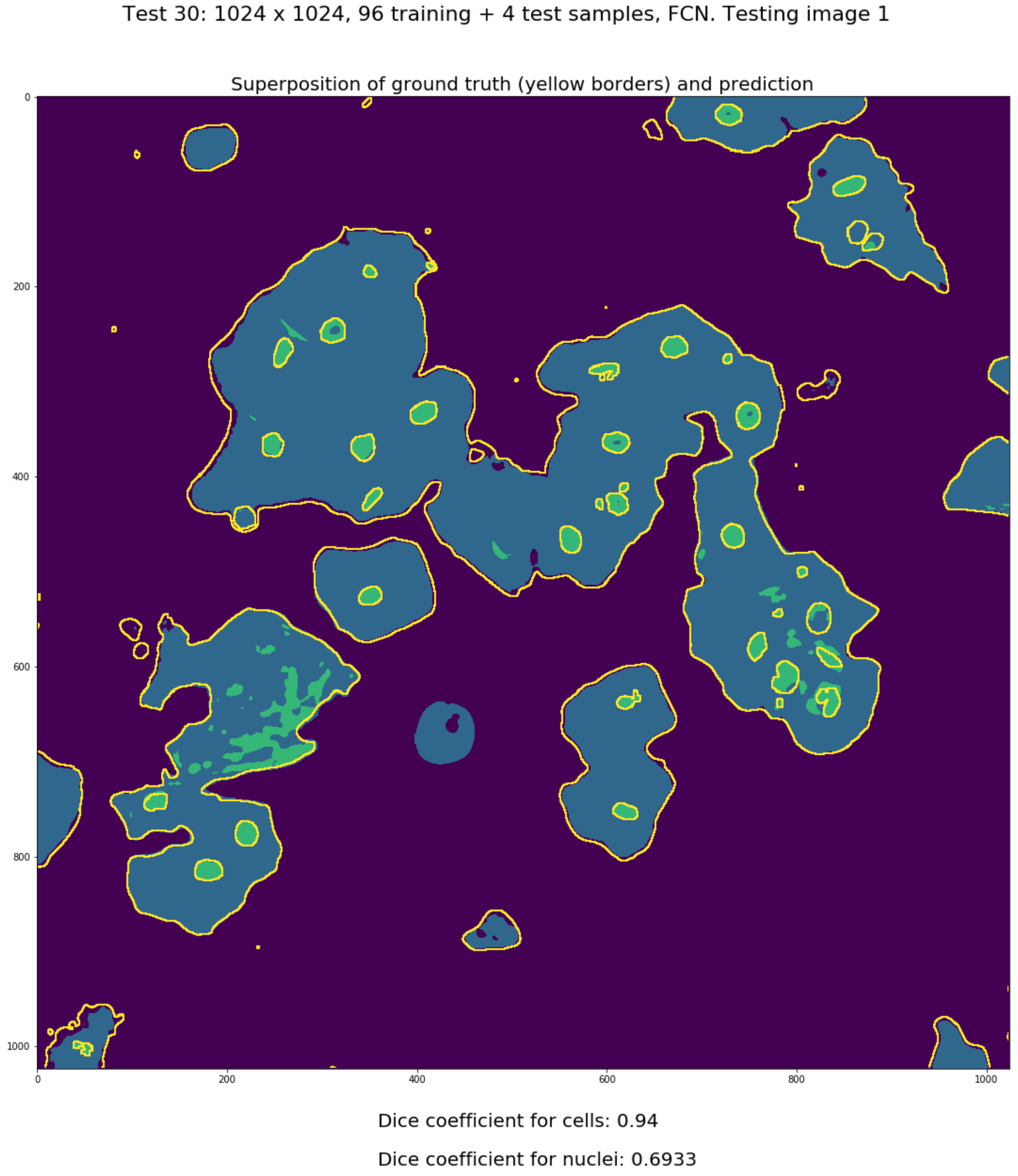

Figure 5.6: Experiment # 30 on the datset # 1 (FCN), test image # 1.

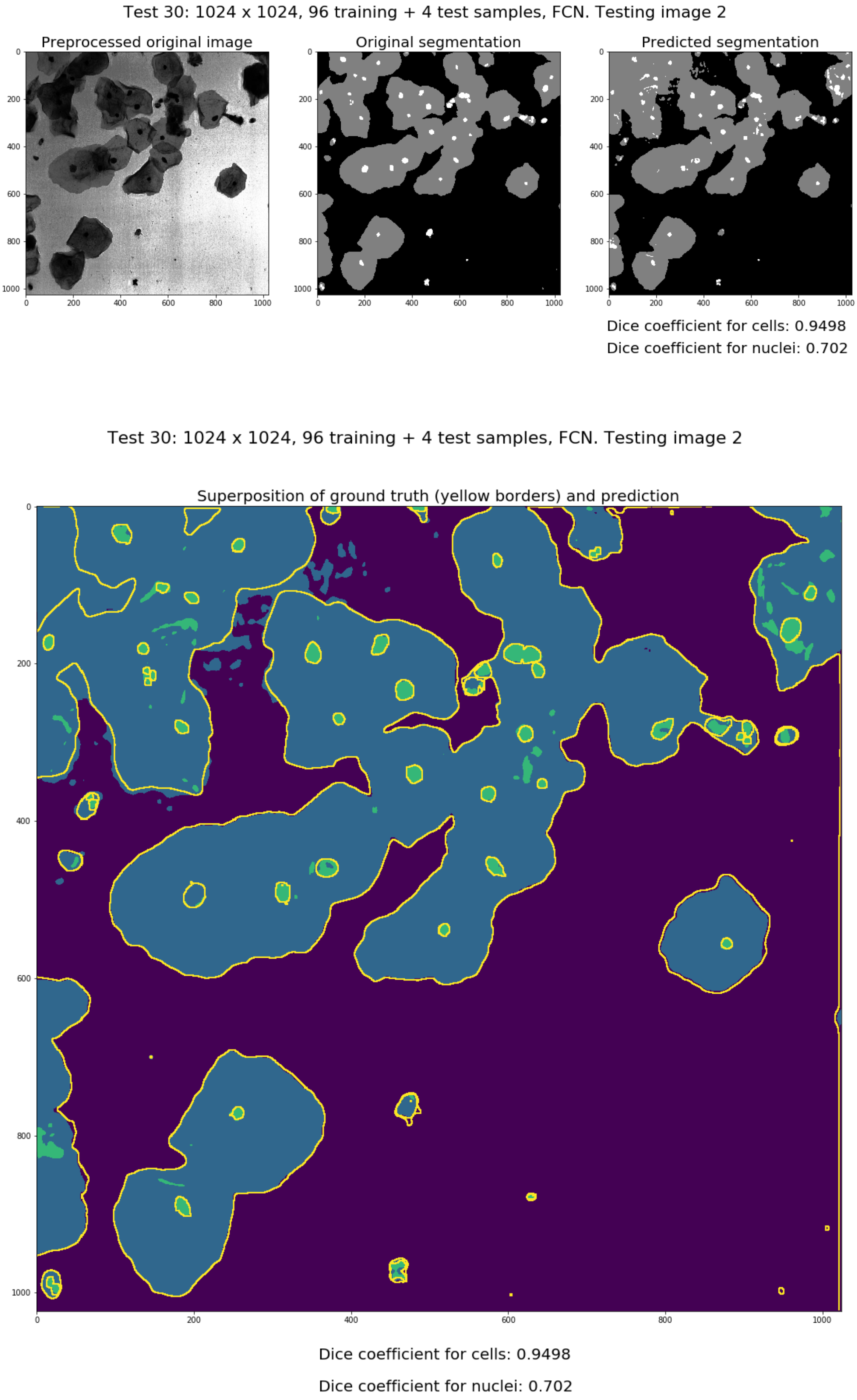

Figure 5.7: Experiment # 30 on the datset # 1 (FCN), test image # 2.

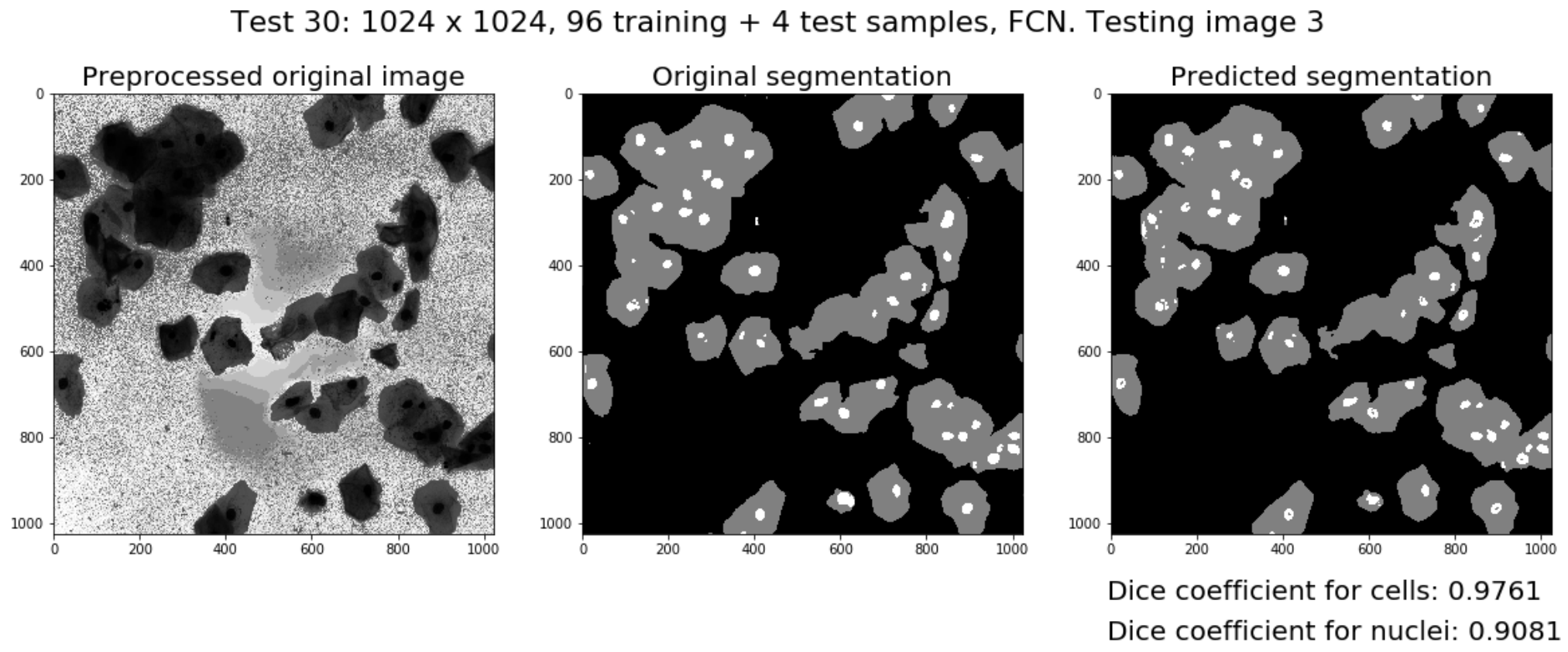

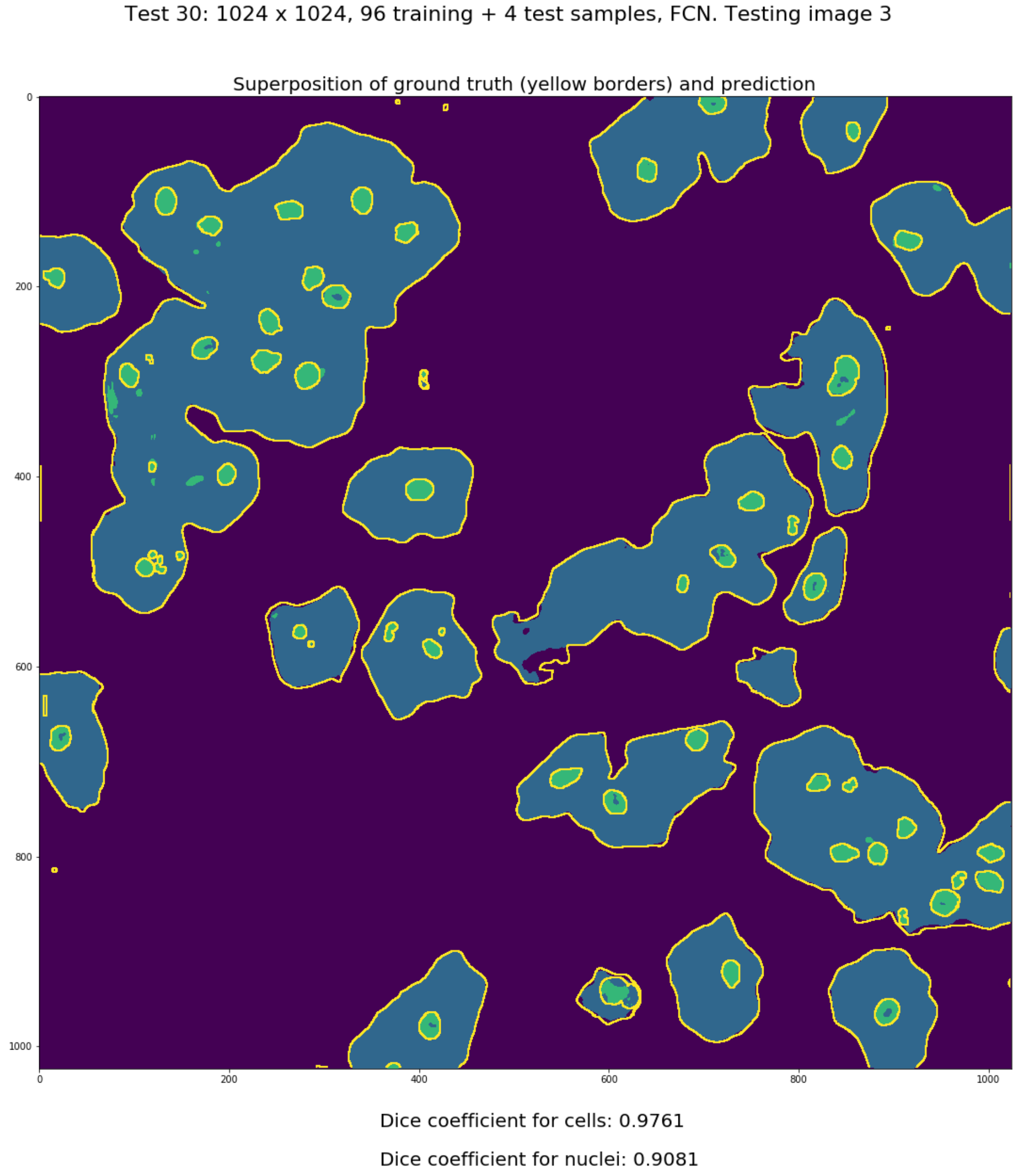

Figure 5.8: Experiment # 30 on the datset # 1 (FCN), test image # 3.

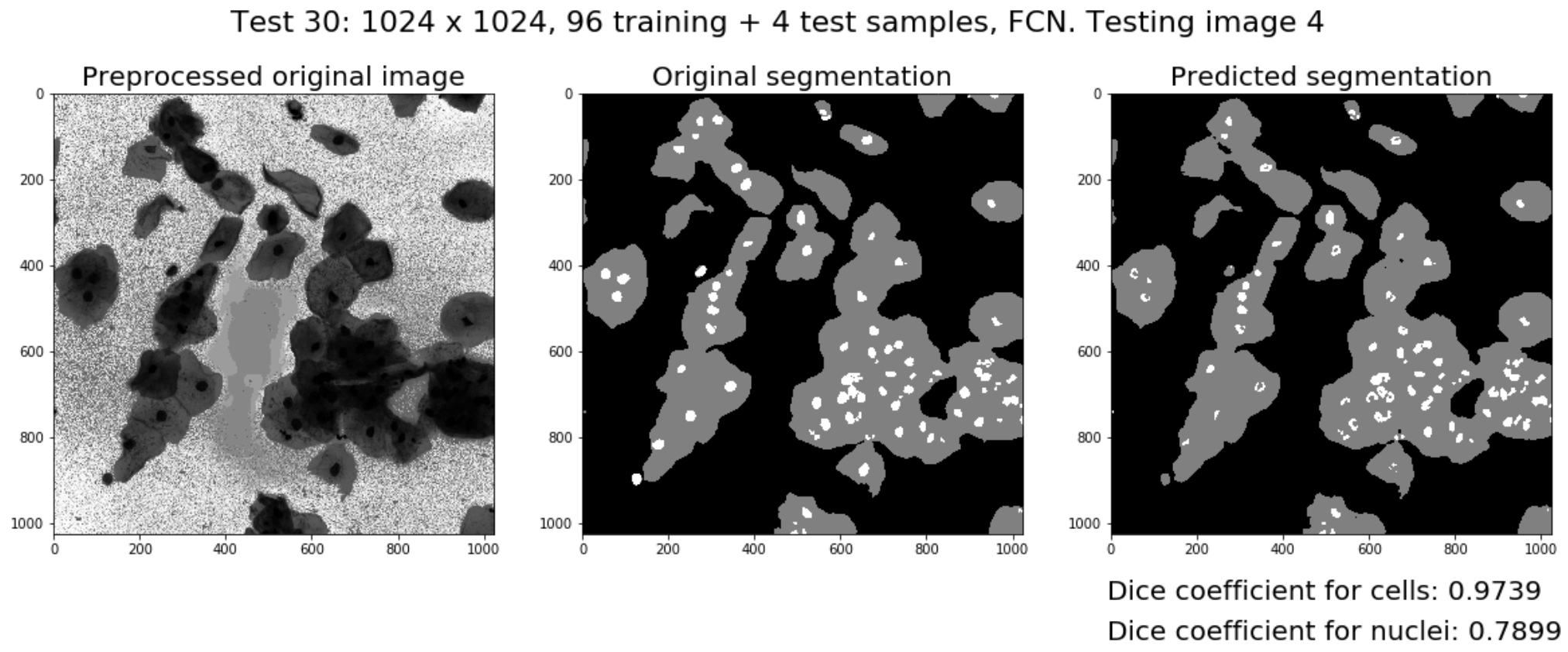

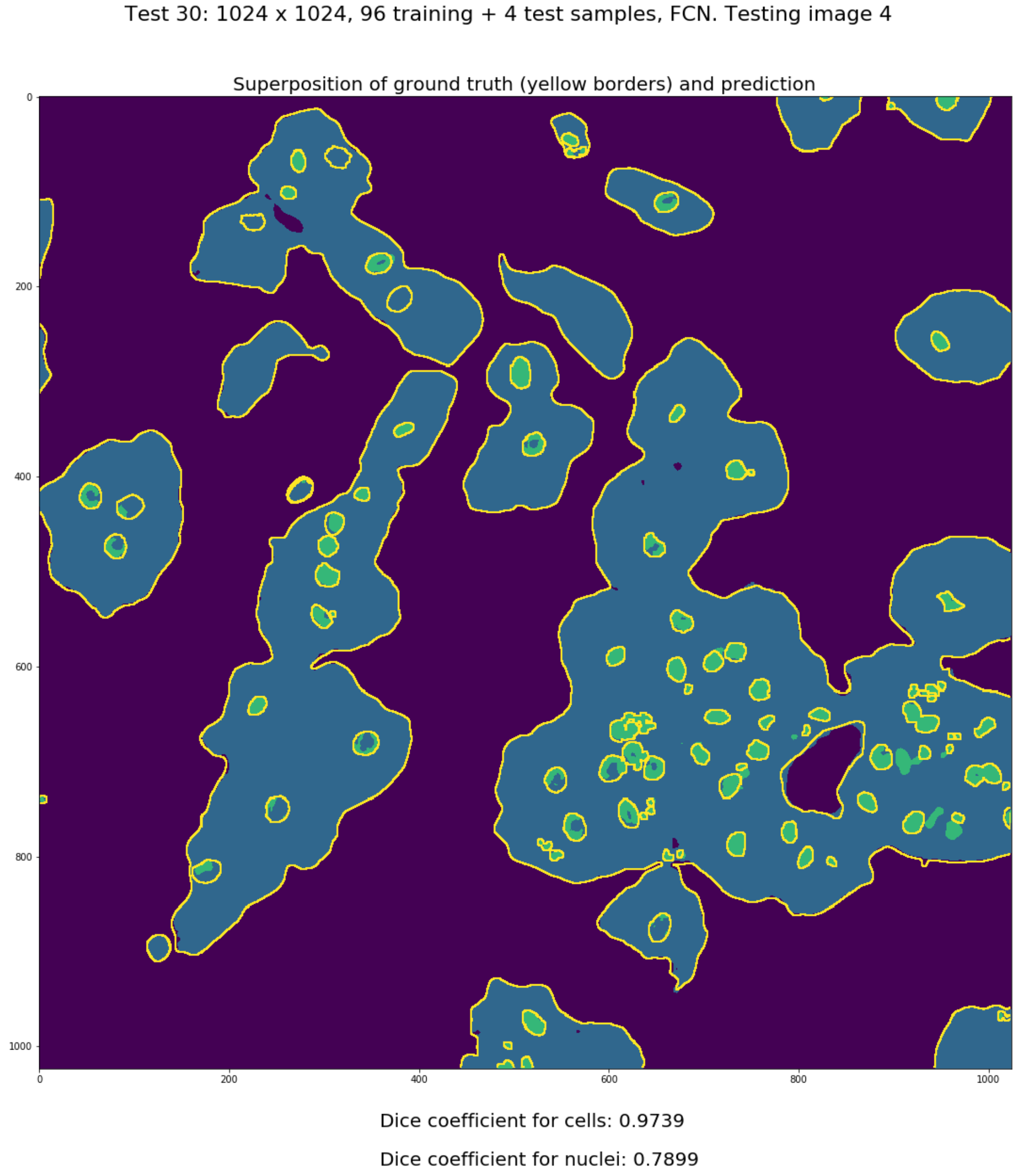

Figure 5.9: Experiment # 30 on the datset # 1 (FCN), test image # 4.

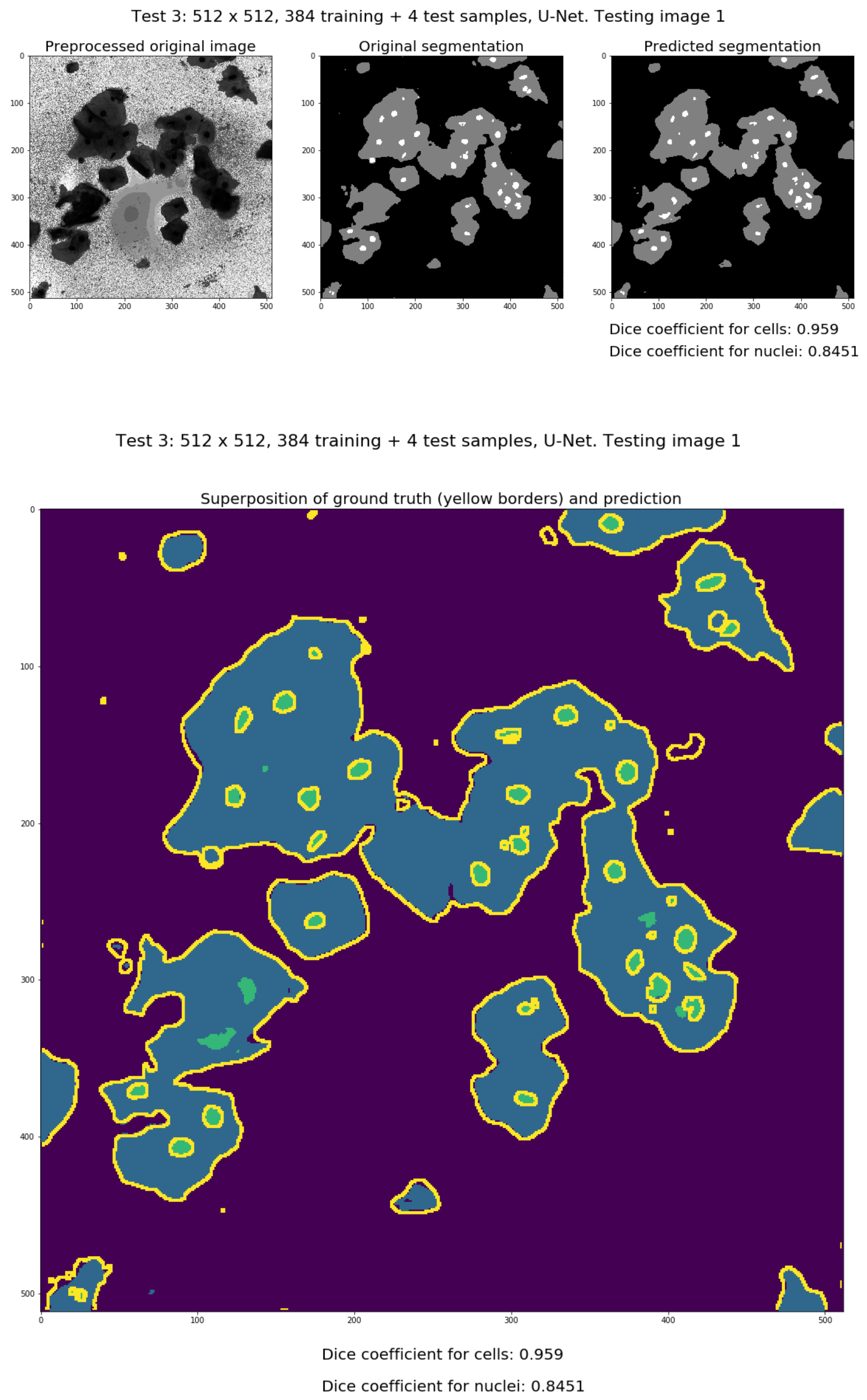

Figure 5.10: Experiment # 3 on the dataset # 3 (U-Net), test image # 1.

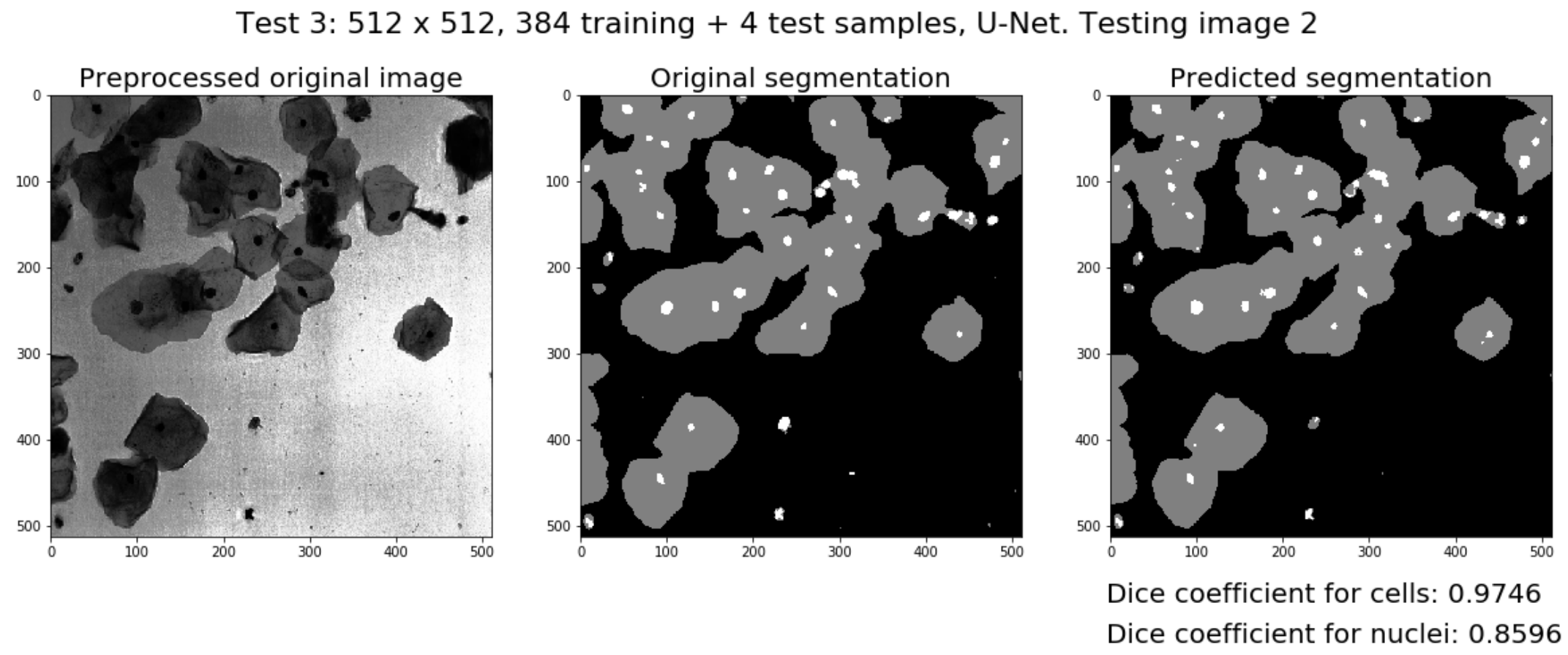

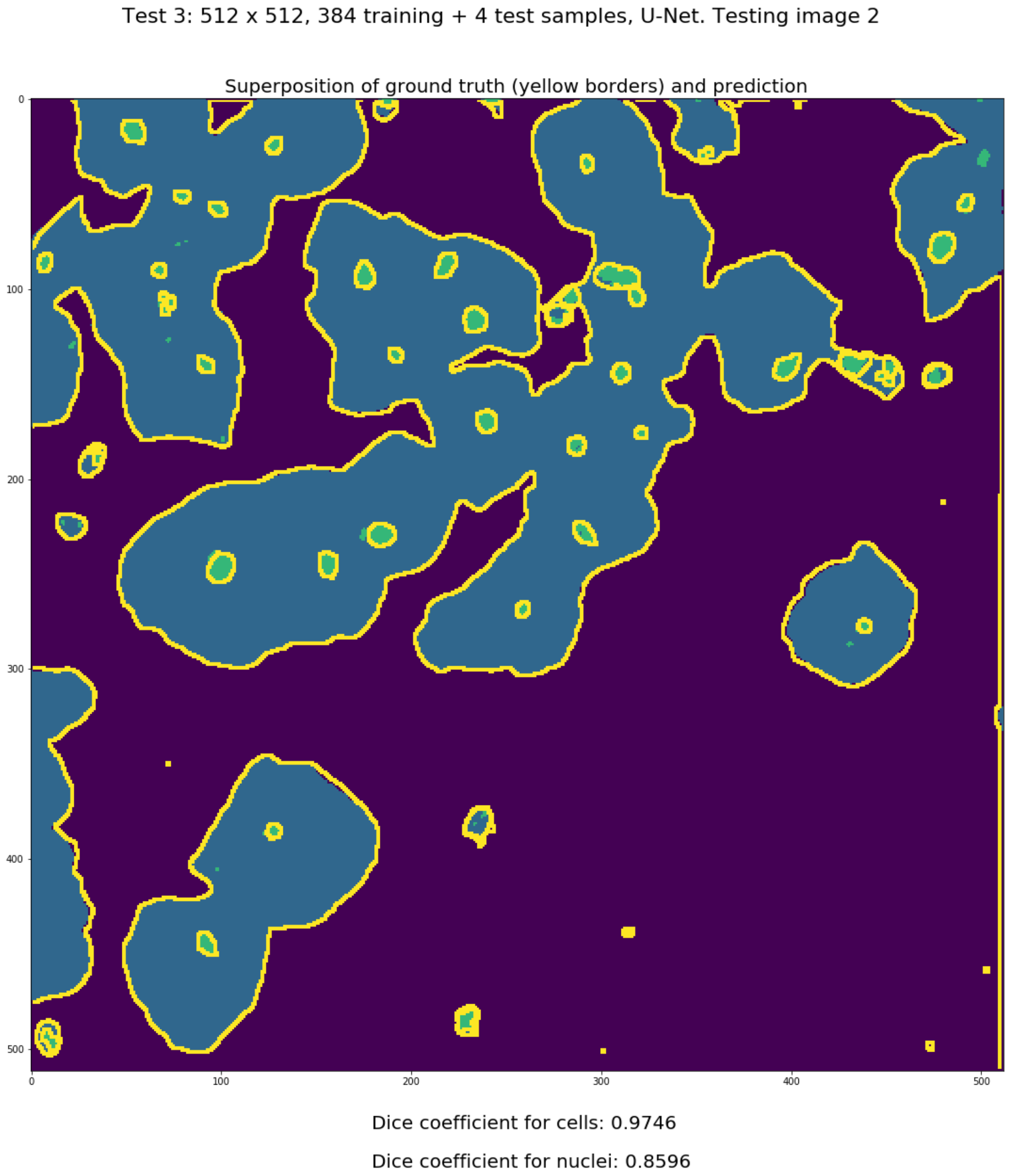

Figure 5.11: Experiment # 3 on the dataset # 3 (U-Net), test image # 2.

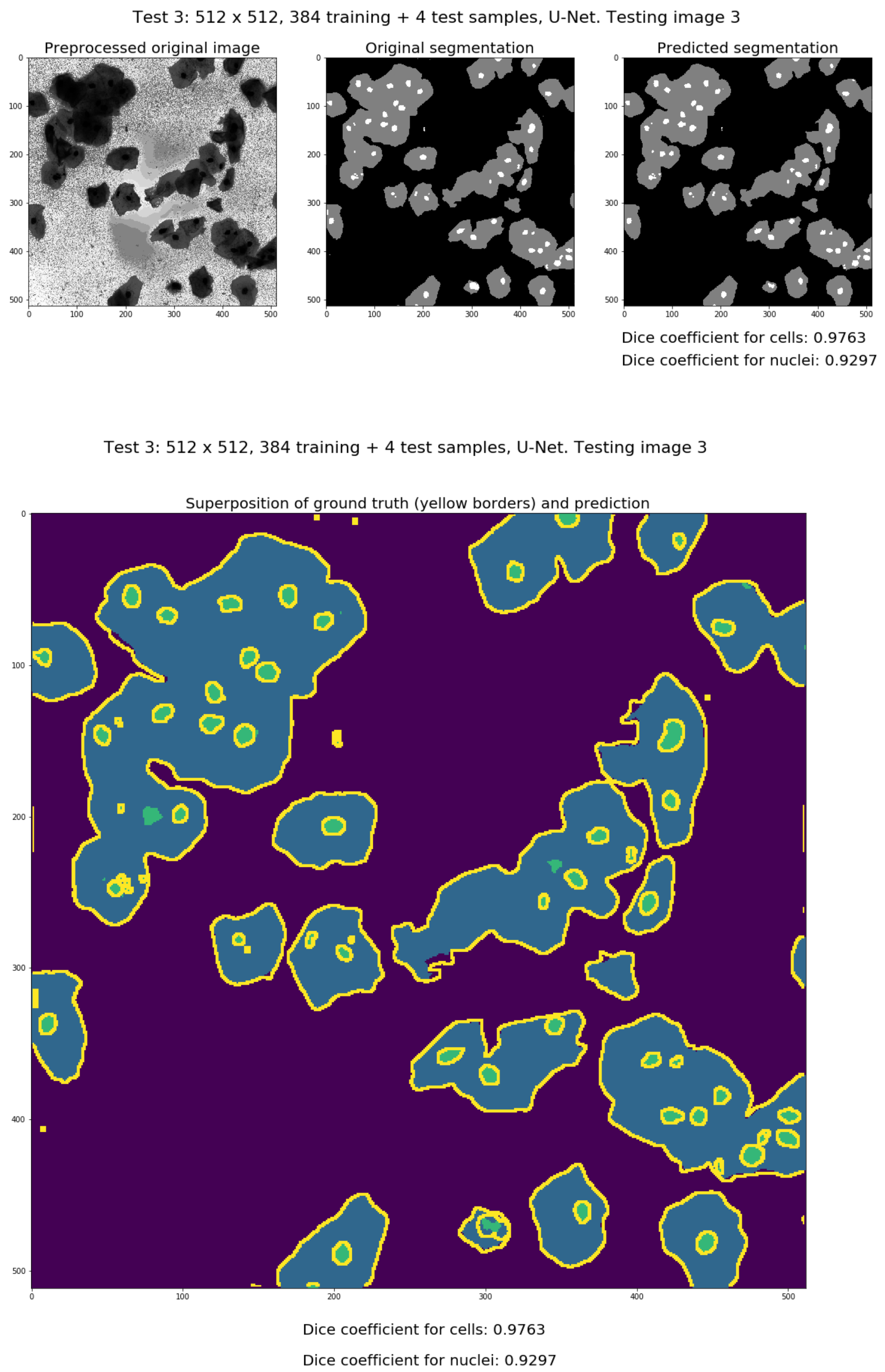

Figure 5.12: Experiment # 3 on the dataset # 3 (U-Net), test image # 3.

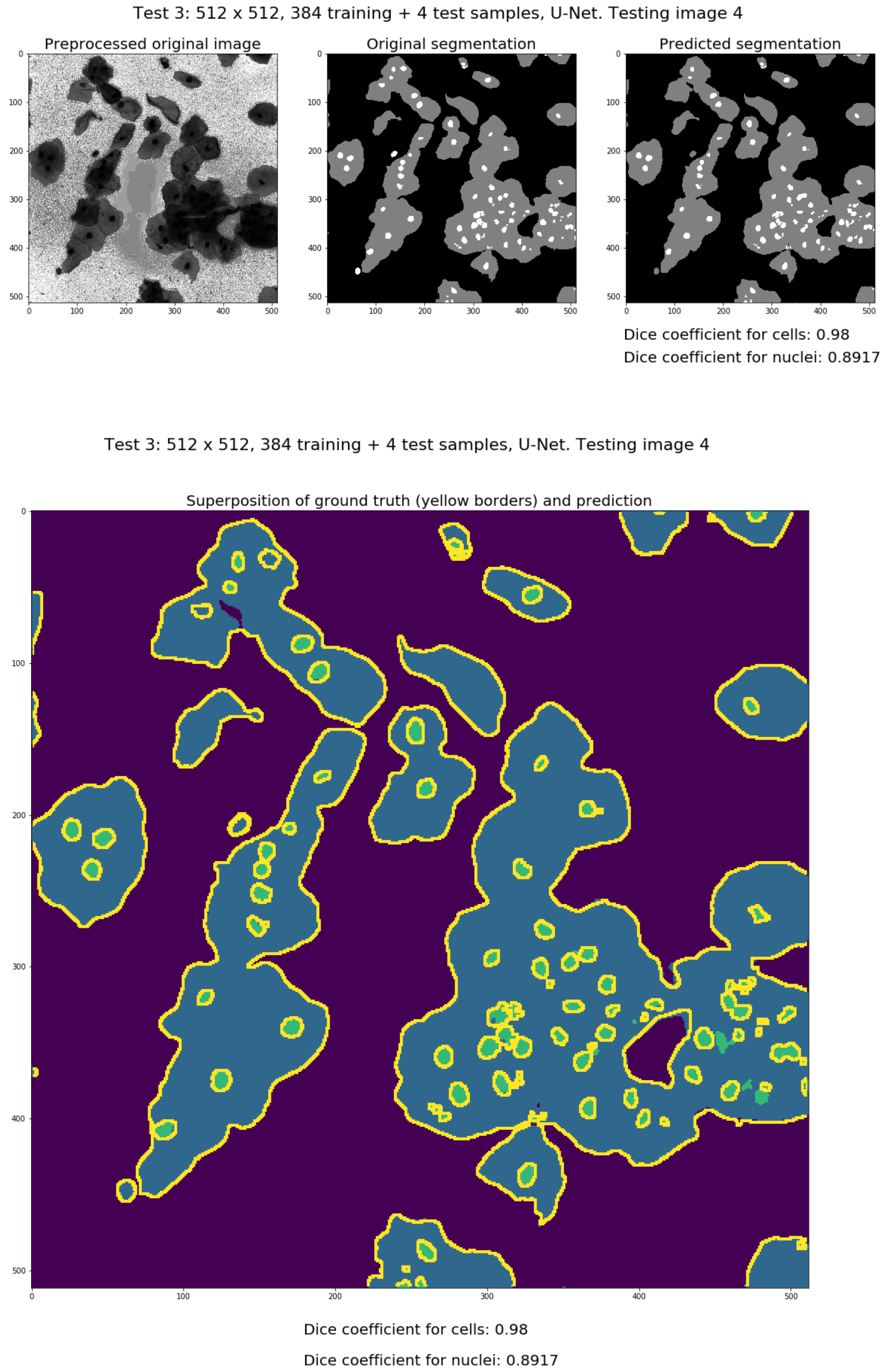

Figure 5.13: Experiment # 3 on the dataset # 3 (U-Net), test image # 4.

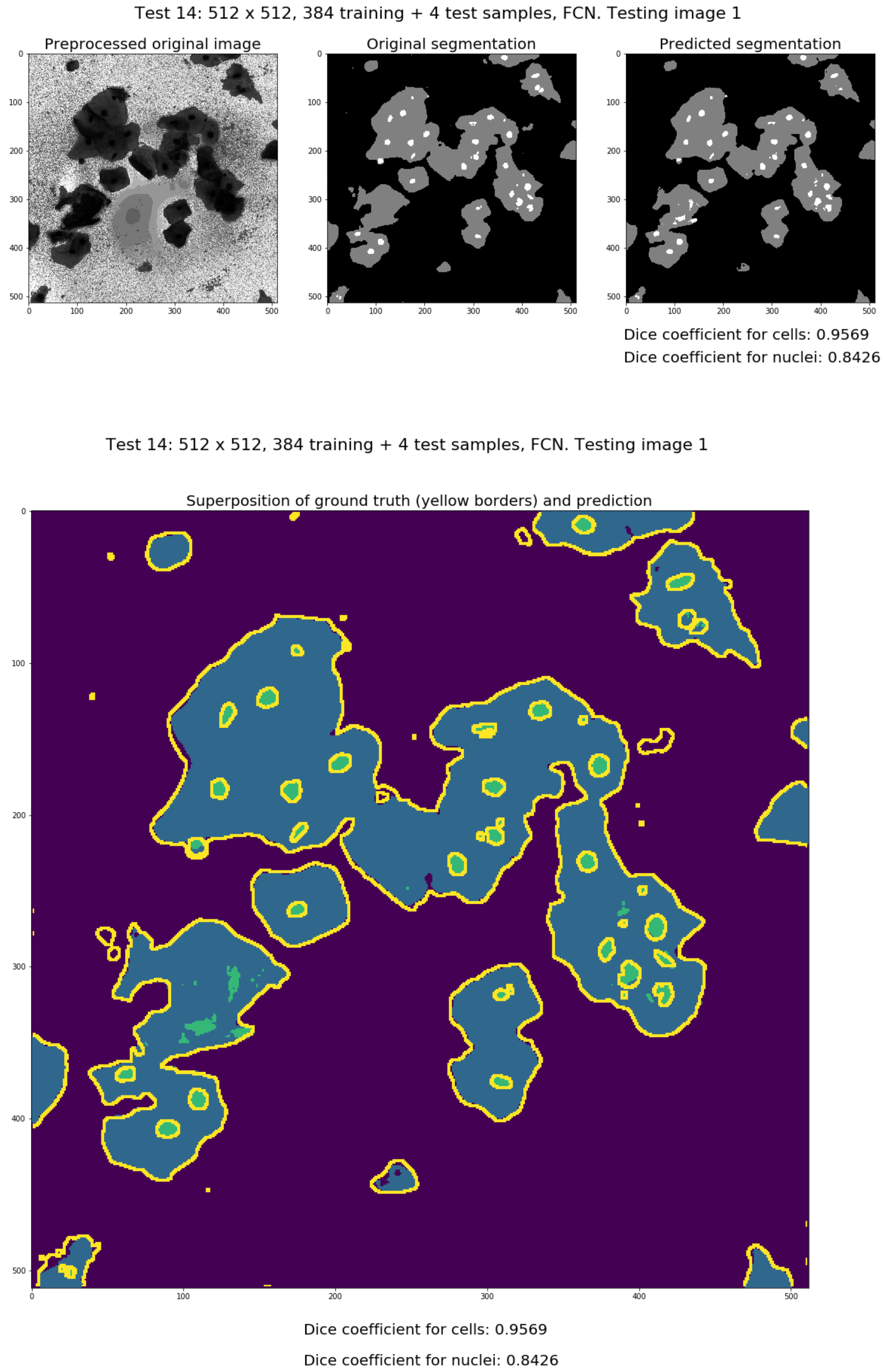

Figure 5.14: Experiment # 14 on the dataset # 3 (FCN), test image # 1.

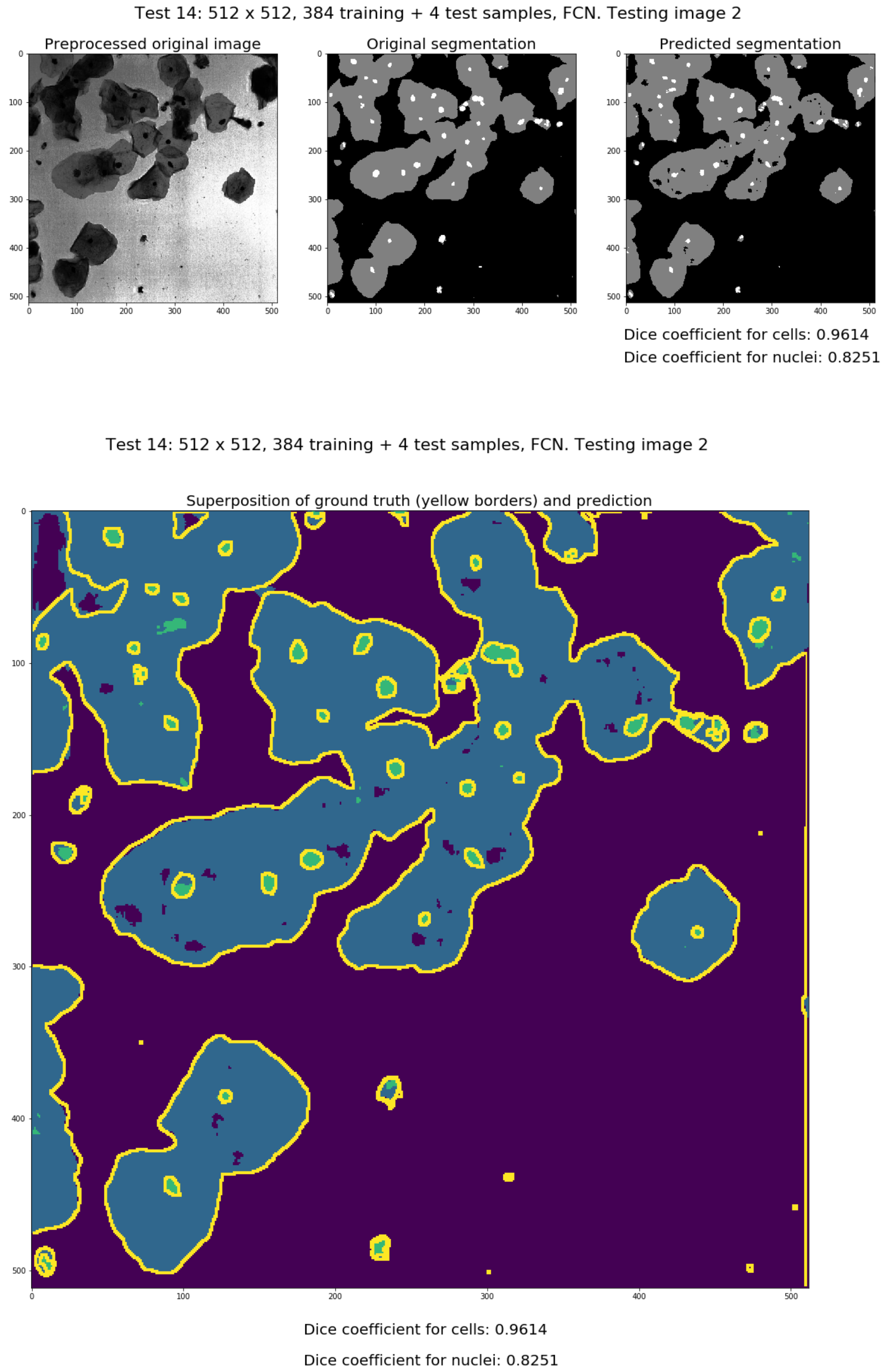

Figure 5.15: Experiment # 14 on the dataset # 3 (FCN), test image # 2.

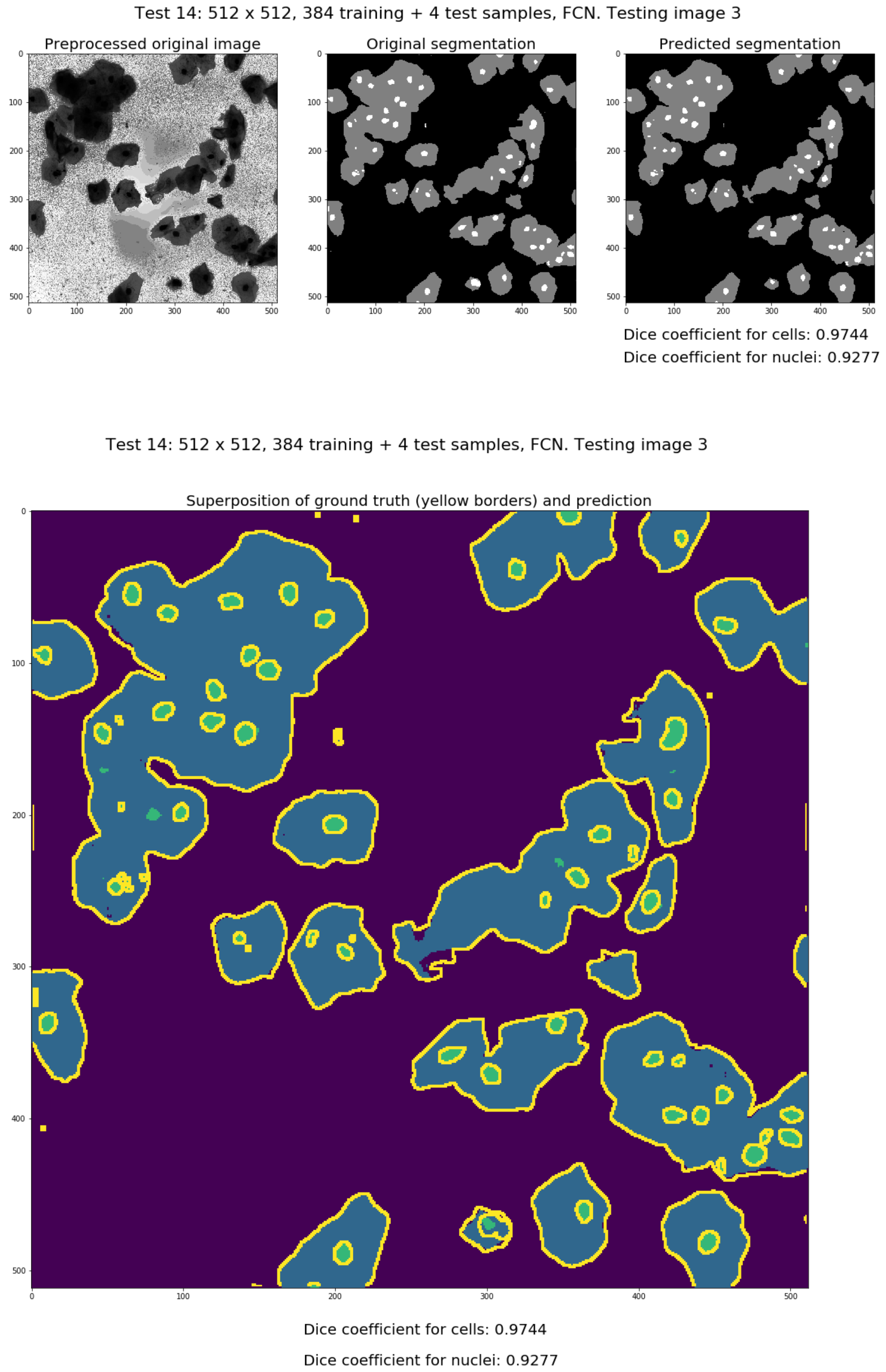

Figure 5.16: Experiment # 14 on the dataset # 3 (FCN), test image # 3.

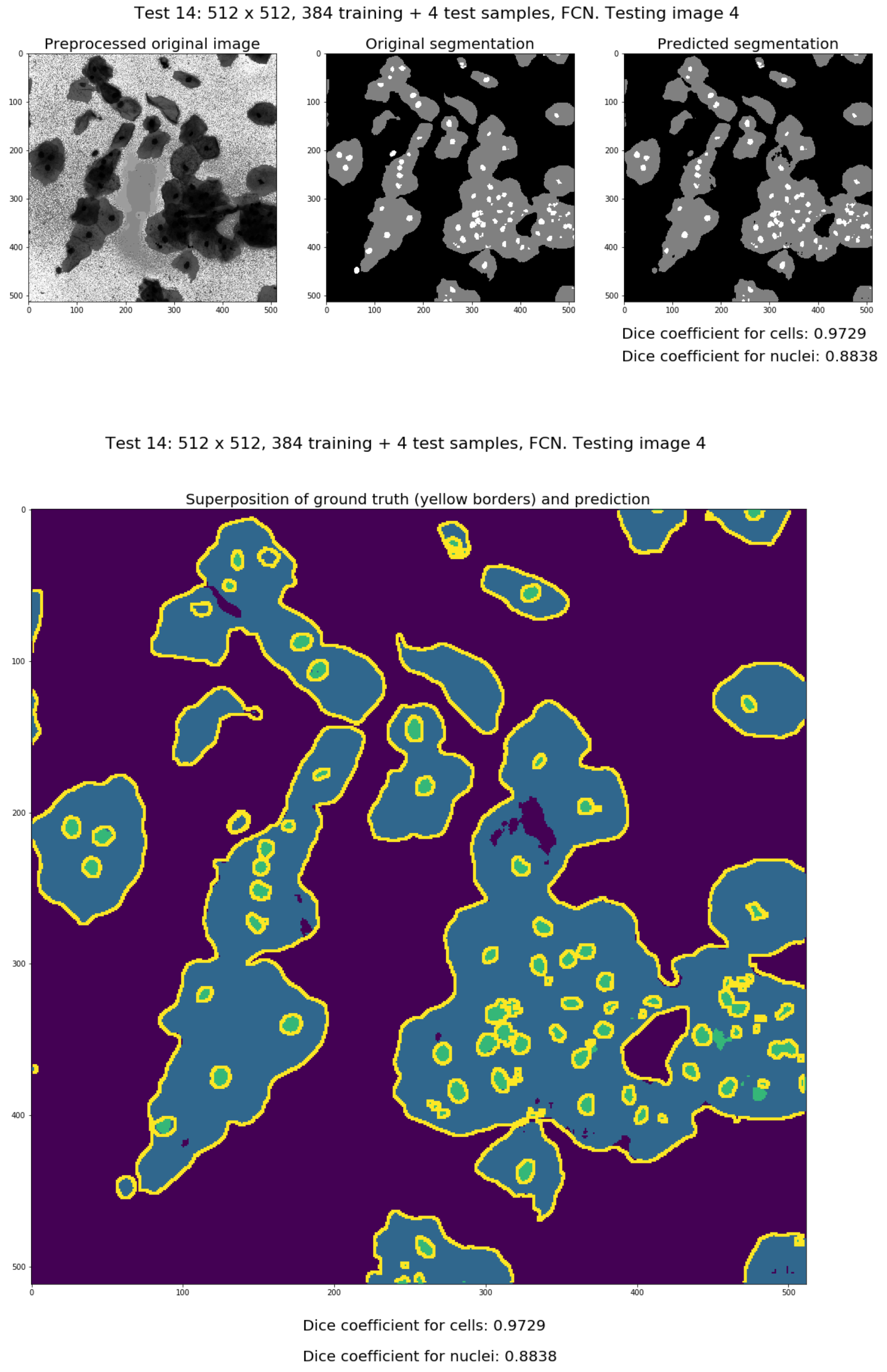

Figure 5.17: Experiment # 14 on the dataset # 3 (FCN), test image # 4.

## 5.2 BRATS dataset

All evaluations of trained networks have been summarized in the tables 5.2 and 5.3.
The Dice scores are lower than the best of known results [11], however it was easy predictable, since we used the two models with the parameters identical to the first task – segmentation of cells and nuclei. Our task was to find out how the networks, which worked well with light microscopy images, would be working with BRATS images.

The table 5.2 has the results of evaluation after training on the following sets:

- ax_2: set of all axial non-empty slices from two patients (264 total);
- ax_15: set of 270 images generated from axial slices from 15 patients;
- ax_270: set of 270 axial slices from 270 patients;
- cor_2: set of all coronal non-empty slices from two patients (347 total);
- cor_15: set of 270 images generated from coronal slices from 15 patients;
- cor_270: set of 270 coronal slices from 270 patients;
- sag_2: set of all saggital non-empty slices from two patients (289 total);
- sag_15: set of 270 images generated from saggital slices from 15 patients;
- sag_270: set of 270 saggital slices from 270 patient.

The results achieved by using both FCN and U-Net networks with the same architecture as for light microscopy images and with the best hyper-parameters. For U-Net:

- batch size: 1;
- conv. layers: depth 16/32/64/128/256 with kernel size (3, 3);
- deconv. layers: 128/64/32/16 with kernel size (2, 2);
- final layer with kernel size 3;
- optimizer: Adam (with initial learning rate 0.001);

and the same for FCN, except:

- conl. layers: depth 16/32/64/32/16;
- final layer with kernel of size (1, 1);
- optimizer Adamax (with initial learning rate 0.002).

Table 5.2: The results of evaluation of the CNNs on the BRATS dataset.

| # | Data set | Samples | FCN | | | | | U-Net | | | | |
|---|---|---|---|---|---|---|---|---|---|---|---|---|
| | | | Test acc. | Dice NCR | Dice ED | Dice ET | Avg. Dice | Test acc. | Dice NCR | Dice ED | Dice ET | Avg. Dice |
| 1 | ax_2 | 264 | 0.9771 | 0.3438 | 0.3989 | 0.3522 | 0.3650 | 0.9683 | 0.0000 | 0.0000 | 0.0000 | 0.0000 |
| 2 | ax_15 | 270 | 0.9805 | 0.4101 | 0.5986 | 0.4203 | 0.4763 | 0.9788 | 0.3112 | 0.5668 | 0.4214 | 0.4331 |
| 3 | ax_270 | 270 | 0.9836 | 0.6005 | 0.6791 | 0.5174 | 0.5990 | 0.9683 | 0.0000 | 0.0000 | 0.0000 | 0.0000 |
| | Avg. axial | | 0.9804 | 0.4515 | 0.5589 | 0.4300 | 0.4801 | 0.9718 | 0.1037 | 0.1889 | 0.1405 | 0.1444 |
| 4 | cor_2 | 347 | 0.9740 | 0.0000 | 0.0000 | 0.0000 | 0.0000 | 0.9740 | 0.0000 | 0.0000 | 0.0000 | 0.0000 |
| 5 | cor_15 | 270 | 0.9830 | 0.3582 | 0.6206 | 0.4125 | 0.4638 | 0.9740 | 0.0000 | 0.0000 | 0.0000 | 0.0000 |
| 6 | cor_270 | 270 | 0.9861 | 0.5377 | 0.6291 | 0.4474 | 0.5381 | 0.9740 | 0.0000 | 0.0000 | 0.0000 | 0.0000 |
| | Avg. coronal | | 0.9810 | 0.2986 | 0.4166 | 0.2866 | 0.3339 | 0.9740 | 0.0000 | 0.0000 | 0.0000 | 0.0000 |
| 7 | sag_2 | 289 | 0.9714 | 0.3521 | 0.3091 | 0.4634 | 0.3749 | 0.9639 | 0.0000 | 0.0000 | 0.0000 | 0.0000 |
| 8 | sag_15 | 270 | 0.9715 | 0.2733 | 0.5124 | 0.3847 | 0.3901 | 0.9639 | 0.0000 | 0.0000 | 0.0000 | 0.0000 |
| 9 | sag_270 | 270 | 0.9740 | 0.4849 | 0.5171 | 0.5040 | 0.5020 | 0.9639 | 0.0000 | 0.0000 | 0.0000 | 0.0000 |
| | Avg. saggital | | 0.9723 | 0.3701 | 0.4462 | 0.4507 | 0.4223 | 0.9639 | 0.0000 | 0.0000 | 0.0000 | 0.0000 |
| | Avg. all | 0.2705 | 0.9779 | 0.3734 | 0.4739 | 0.3891 | 0.4930 | 0.9699 | 0.0346 | 0.0630 | 0.0468 | 0.0481 |

Table 5.3: The results of evaluation of the CNNs on the BRATS dataset:

| Patient | Projection | Samples | | Dice NCR | Dice ED | Dice ET | Avg. Dice |
|---|---|---|---|---|---|---|---|
| 1 | Axial | 134 | | 0.6254 | 0.3626 | 0.2921 | 0.4267 |
| 2 | Axial | 143 | | 0.4645 | 0.3582 | 0.3421 | 0.3883 |
| | Avg. axial | | | 0.5450 | 0.3604 | 0.3171 | 0.4075 |
| 1 | Coronal | 165 | | 0.3523 | 0.4098 | 0.3075 | 0.3565 |
| 2 | Coronal | 169 | | 0.3937 | 0.4700 | 0.2488 | 0.3708 |
| | Avg. coronal | | | 0.3730 | 0.4399 | 0.2782 | 0.3637 |
| 1 | Saggital | 148 | | 0.4279 | 0.2997 | 0.2944 | 0.3407 |
| 2 | Saggital | 134 | | 0.2917 | 0.3390 | 0.3143 | 0.3150 |
| | Avg. saggital | | | 0.3598 | 0.3194 | 0.3044 | 0.3278 |
| | Avg. all | | | 0.4259 | 0.3732 | 0.2999 | 0.3663 |

The test set for the table 5.2 consists of 15 slices from 15 patients.

The table 5.3 has the results of evaluation after the training on the following training sets:

- ax_270: set of 270 axial slices from 270 patients;
- cor_270: set of 270 coronal slices from 270 patients;
- sag_270: set of 270 saggital slices from 270 patient,

and for each case the testing is done over the sets of all appropriate non-empty slices from the two patients separately.

The results of some relatively good predictions are presented in the following figures:

- figure 5.18: predictions for two axial slices from some of 15 patients, after training on 270 axial slices of 270 patients;
- figure 5.19: predictions for two coronal slices from some of 15 patients, after training on 270 coronal slices of 270 patients;
- figure 5.20: predictions for two saggital slices from some of 15 patients, after training on 270 saggital slices of 270 patients;
- figures 5.21 – 5.22: the example of 5 consecutive predicted axial slices for test data from all scans of one patient.

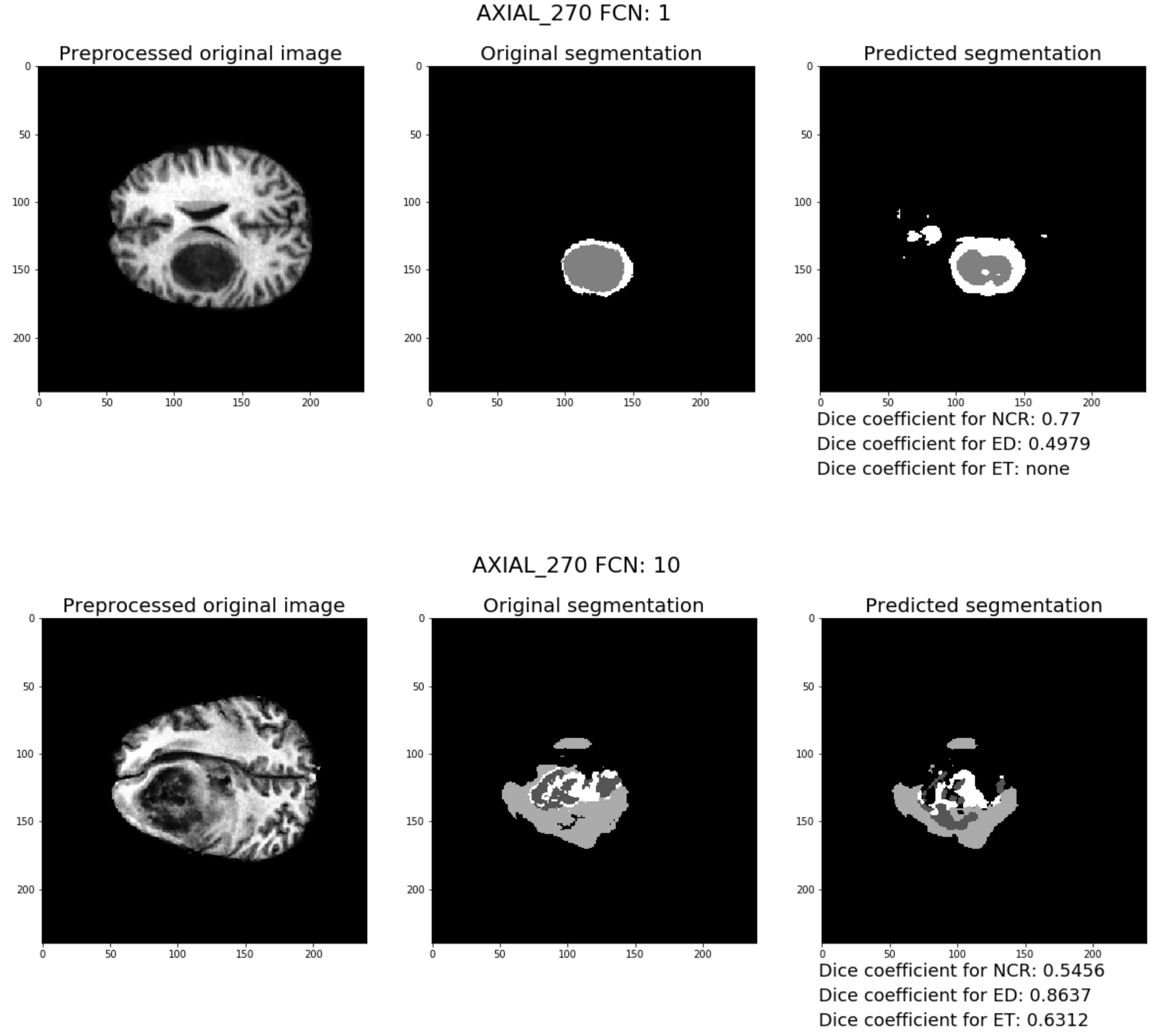

Figure 5.18: Predictions for two axial slices from some of 15 patients, after training on 270 axial slices of 270 patients

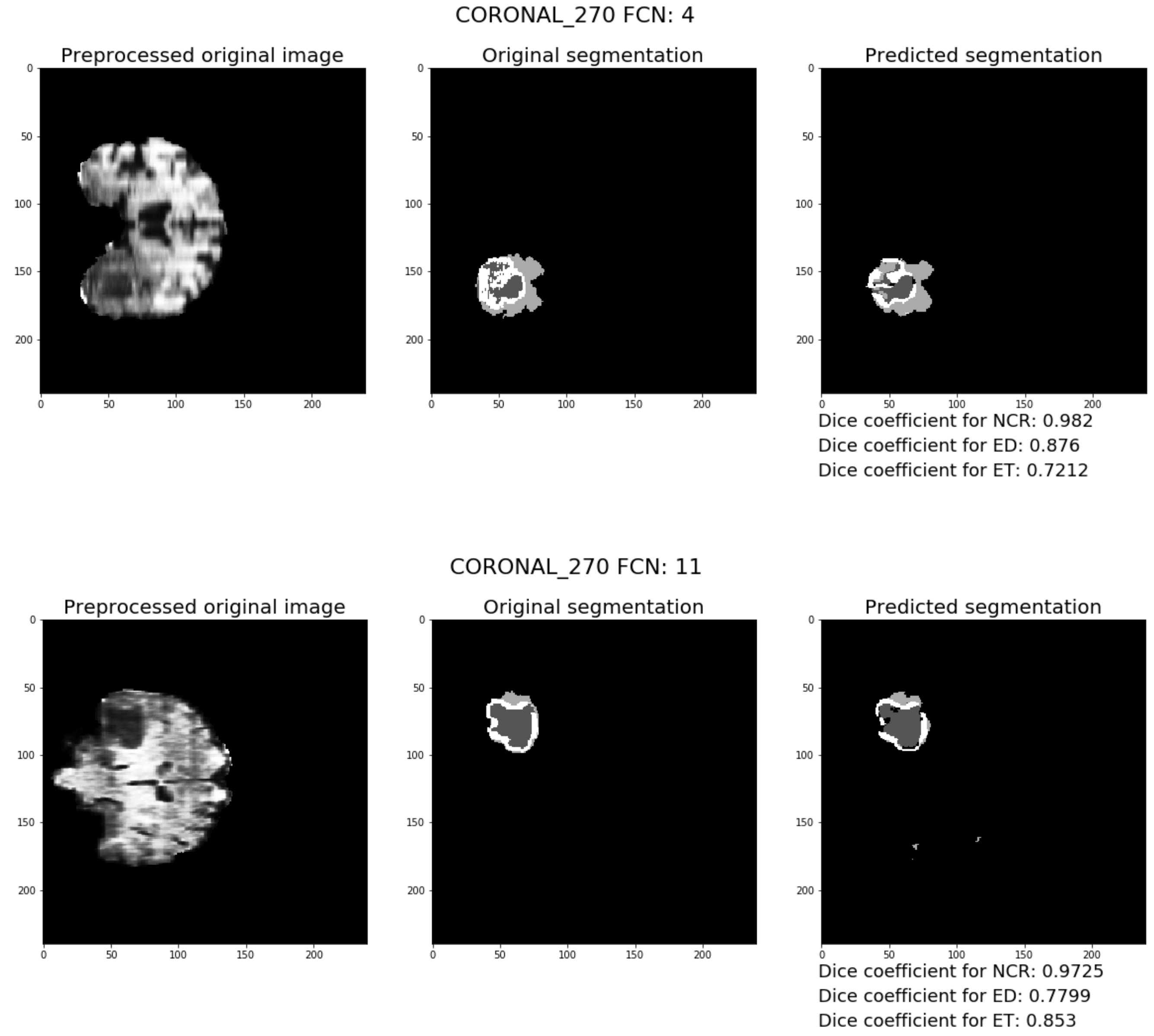

Figure 5.19: Predictions for two coronal slices from some of 15 patients, after training on 270 coronal slices of 270 patients.

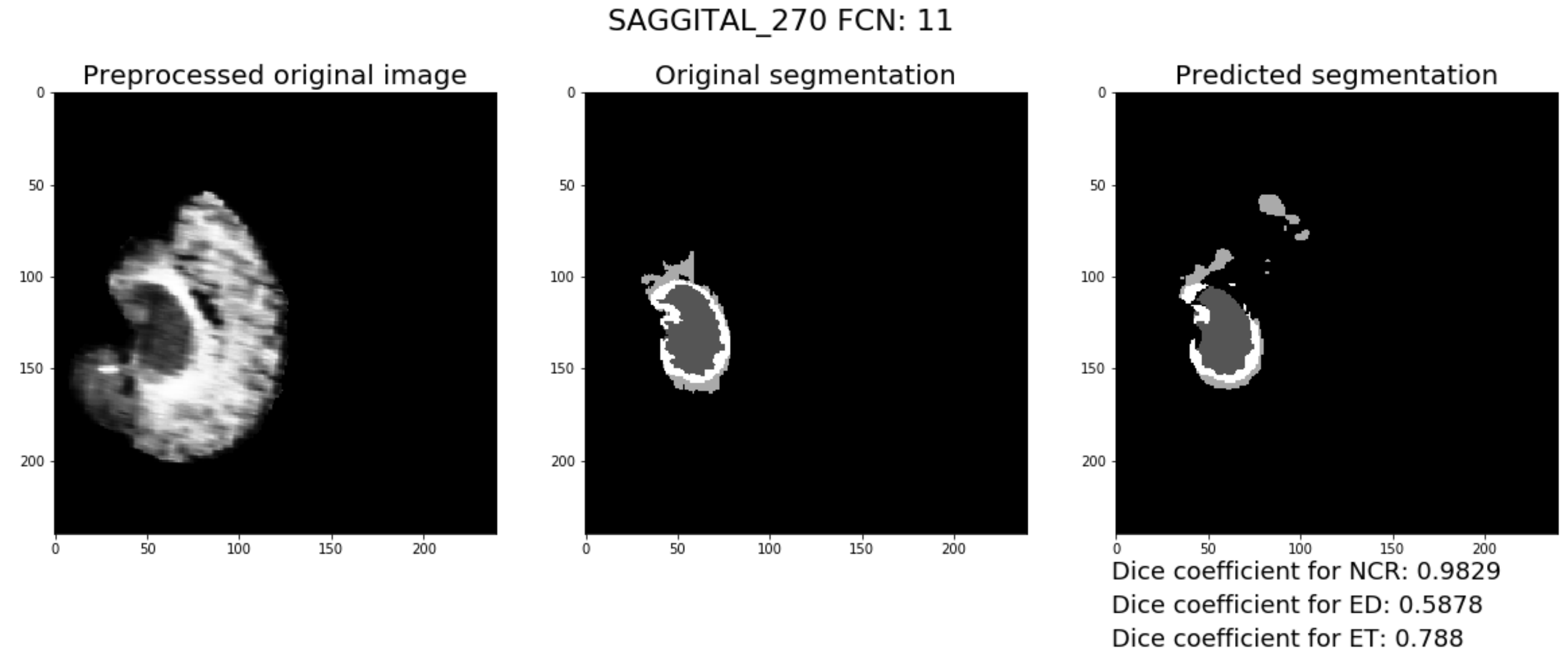

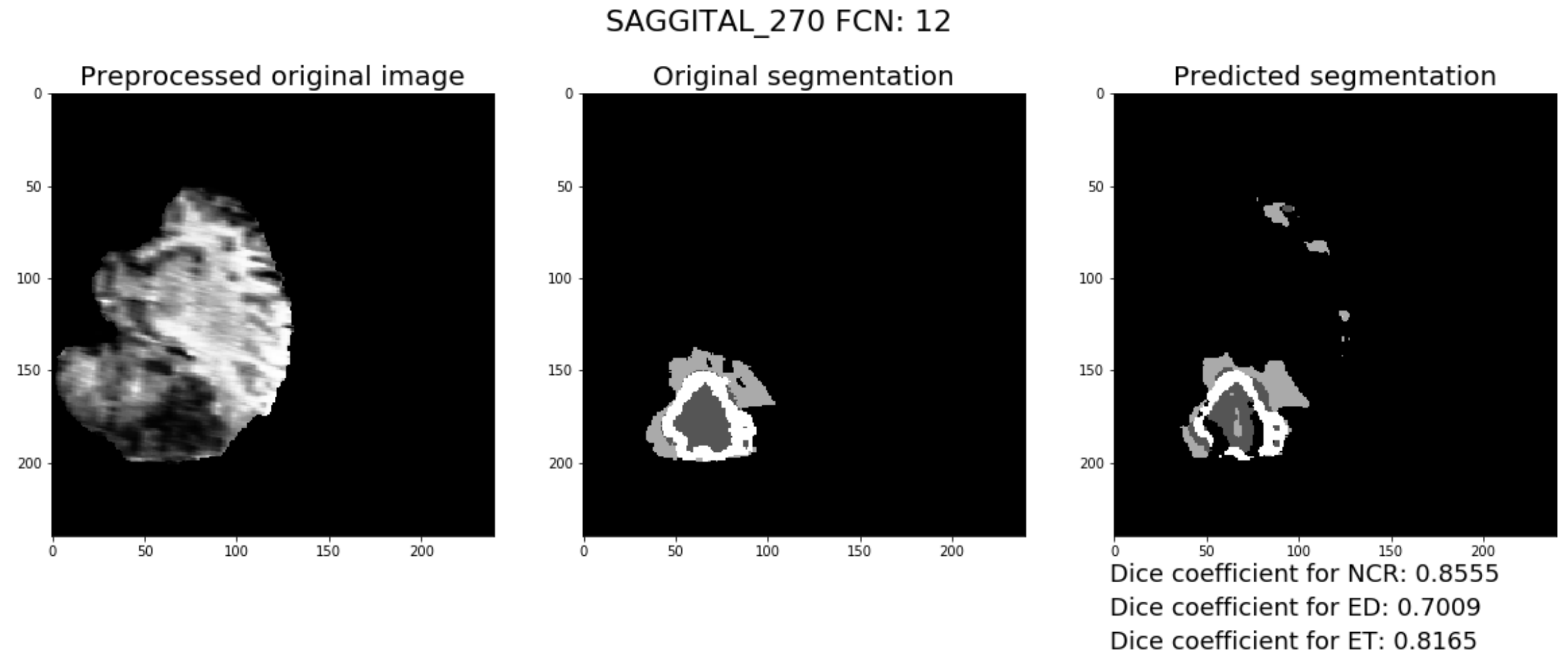

Figure 5.20: Predictions for two saggital slices from some of 15 patients, after training on 270 saggital slices of 270 patients

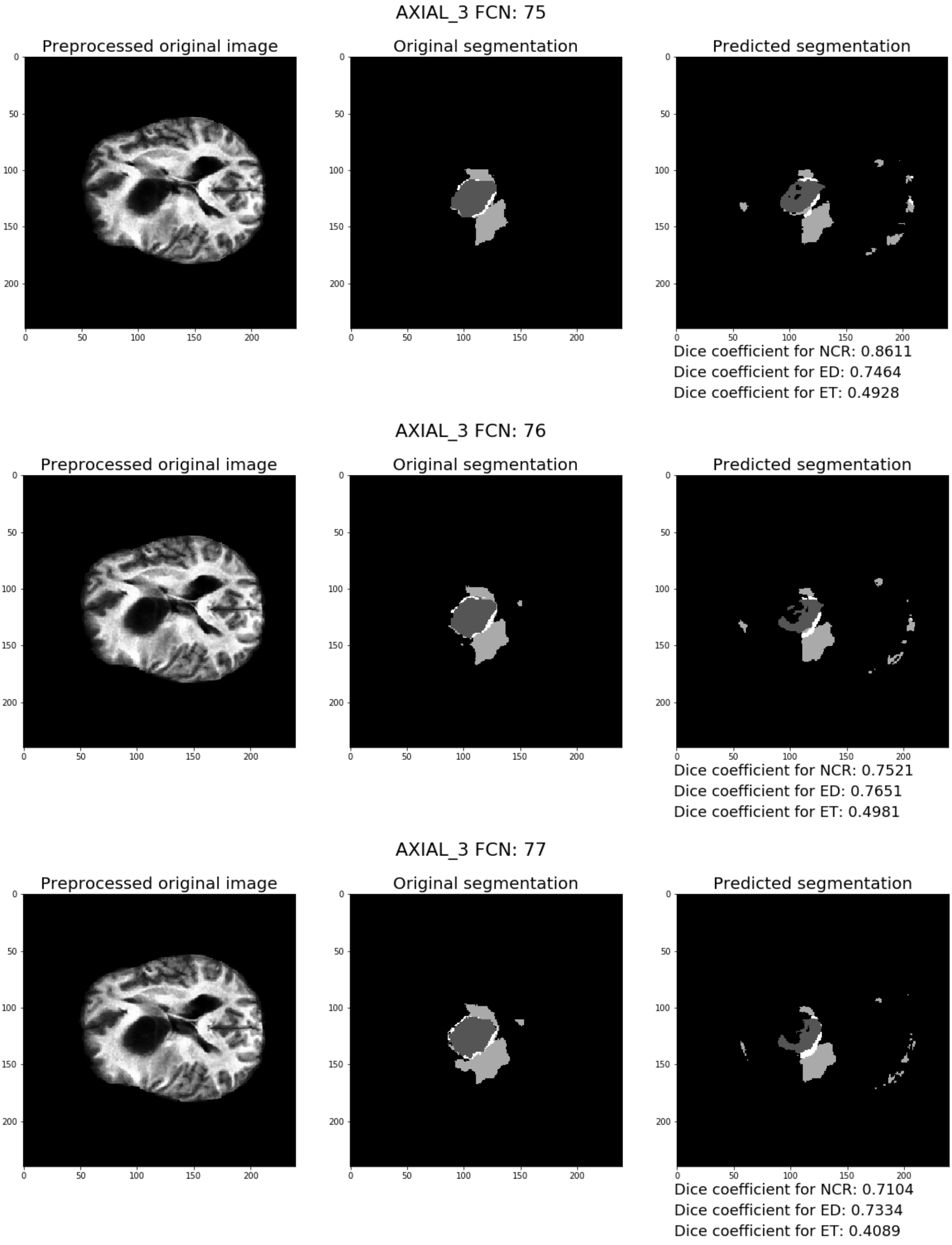

Figure 5.21: The example of 5 consecutive predicted axial slices for test data from all scans of one patient.

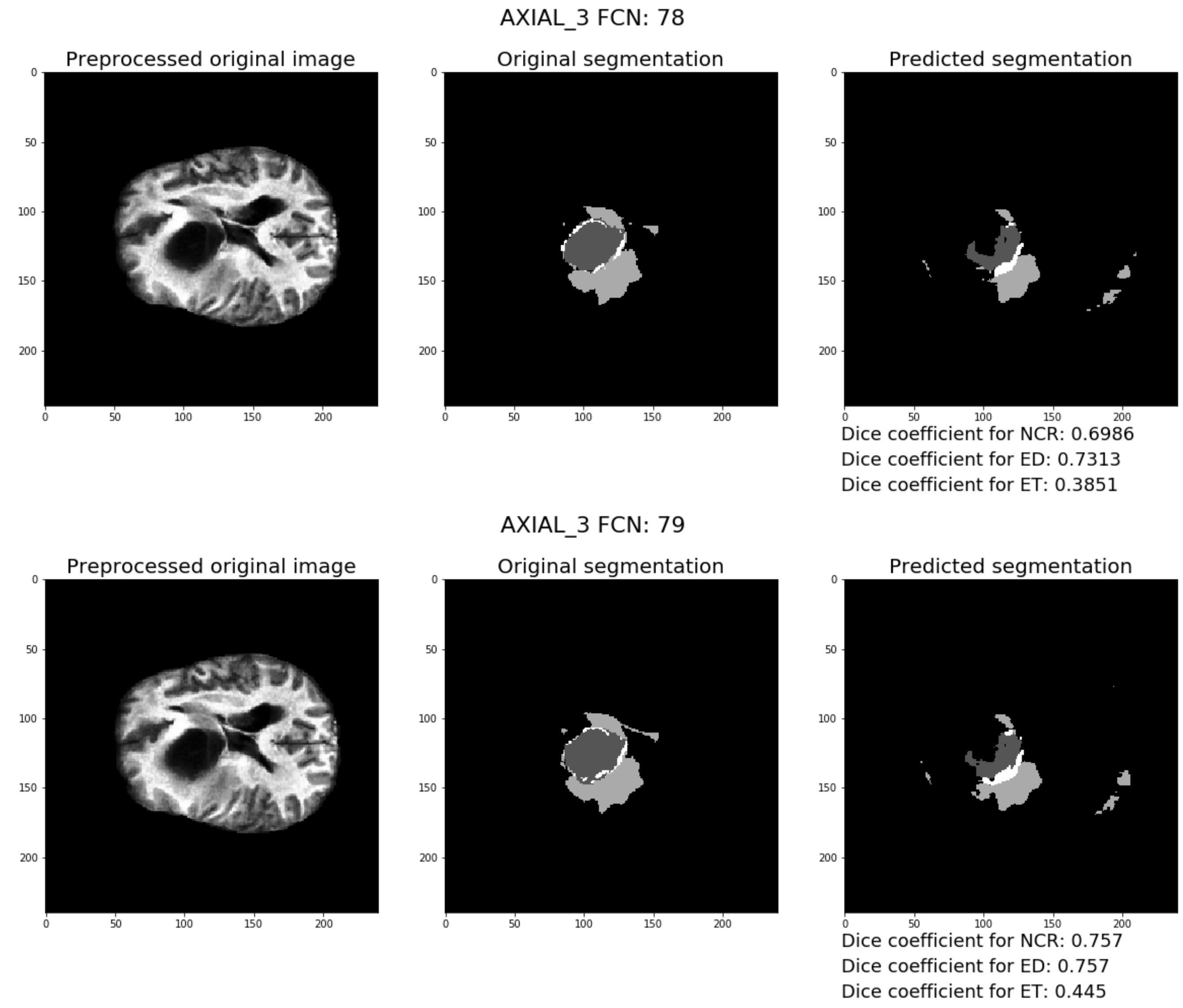

Figure 5.22: The example of 5 consecutive predicted axial slices for test data from all scans of one patient.

# 6 Conclusion

Conclusion on the light microscopy dataset:

1) On high resolutions U-Net architecture works better, than FCN. Especially, U-Net works significantly better on 1024 x 1024 original resolution. However, on 256 x 256 resolution, FCN worked better than U-Net.

2) It seems that there is no regularity in how works histogram equalization – in some cases it helps, in others – doesn't help.

3) Cross-validation was checked on 12 original training images and it helped.

4) Optimizer Adam works better than others with U-Net architecture, while optimizer Adamax works better with FCN (and sometimes Adam also works well with FCN). The suitable starting learning rates for them are respectively 0.001 and 0.002, with decaying up until 4e-5 and 8e-5.

5) The best combinations of hyperparameters are:

- kernel sizes for convolutional layers: 3 and 5 with U-Net, and 5 with FCN;
- kernel size for deconvolutional layers: 2
- kernel sizes for output convolutional layers: 1 and 3

6) The best result with U-Net over all datasets is 0.8815 for Dice score for nuclei and 0.9725 for cells, for dataset number 3 – 384 x 512 x 512 (experiment # 3).

The best result with U-Net for dataset 96 x 1024 x 1024 is only slightly worse: 0.8755 Dice score for nuclei and 0.9736 for cells (experiment # 11), which can be considered best overall bearing in mind that 1024 x 1024 is the original resolution, while 512 x 512 is 4 times less by area.

7) The best result with FCN over all datasets is 0.8698 for Dice score for nuclei and 0.9664 for cells (experiment # 14).

The best result with FCN for dataset 96 x 1024 x 1024 is significantly worse: 0.7733 Dice score for nuclei and 0.96 for cells (experiment # 30), so it is difficult to say which case is better for FCN - 024 x 1024 or 512 x 512, bearing in mind that 1024 x 1024 is the original resolution.

8) The results with both type of networks are considerably worse on 256 x 256 dataset, which can probably be explained by high loss of valuable structure information during downscaling from 1024 x 1024 to 256 x 256.

9) As a general conclusion: it is better to work with U-Net and original resolution of the images. Also, training time for U-Net were usually better, which is another argument in favor of U-Net.

Conclusion on BRATS dataset:

1) The dataset generated from 270 patients data worked best in all cases. As a general result, it is better to have less individual data from a large number of patients, than to have a data from fewer patients and augment them to a large quantity.

2) The dataset generated from all non-empty slices of two patients worked worse at all, which confirmes the first conclusion.

3) Except from axial slices, U-Net architecture didn't work at all, which confirms the general approach, when a set of different types of architectures are applied to one problem in order to find which one works out better.

4) The results for axial slices are better than for coronal and saggital, which can be explained by larger original resolution of axial projection (240 x 240 vs 240 x 155).

5) The results of test data evaluation on two full 3d scans of two patients are a little lower than for test data generated from single slices, which can be explained by the fact that the network used single slices for training and not the whole 3d volumes of patients' scans.

## Bibliography

[1] A. Krizhevsky, I. Sutskever, and G. E. Hinton, “Imagenet classification with deep convolutional neural networks.” in NIPS, 2012.

[2] J. Long, E. Shelhamer, and T. Darrell, “Fully convolutional networks for semantic segmentation,” CVPR, 2015.

[3] Ronneberger, O., Fischer, P., Brox, T.: U-Net: Convolutional Networks for Biomedical Image Segmentation. In: Miccai. (may 2015) 234–241

[4] J. Long, E. Shelhamer, and T. Darrell, “Fully convolutional networks for semantic segmentation,” CVPR, 2016.

[5] Fidon, L., Li, W., Garcia-Peraza-Herrera, L.C., et al.: Generalised Wasserstein Dice Score for Imbalanced Multi-class Segmentation using Holistic Convolutional Networks. (2017)

[6] Menze BH et al. "The Multimodal Brain Tumor Image Segmentation Benchmark (BRATS)", IEEE Transactions on Medical Imaging 34(10), 1993-2024 (2015)

[7] Bakas S, Akbari H, Sotiras A, Bilello M, Rozycki M, Kirby JS, Freymann JB, Farahani K, Davatzikos C. "Advancing The Cancer Genome Atlas glioma MRI collections with expert segmentation labels and radiomic features", Nature Scientific Data, (2017)

[8] http://www.braintumorsegmentation.org

[9] https://www.med.upenn.edu/sbia/brats2017.html

[10] https://en.wikipedia.org/wiki/Histogram_equalization

[11] Xiaomei Zhao, Yihong Wu, Guidong Song, Zhenye Li, Yazhuo Zhang, Yong Fan: A deep learning model integrating FCNNs and CRFs for brain tumor segmentation Medical Image Analysis 43 :98-111, 2018.